\documentclass[
aps,prb,reprint,
superscriptaddress, 
twocolumn, %onecolumn,
amsmath,amssymb,
longbibliography
]{revtex4-2}

\usepackage{graphicx} % Required for inserting images
\usepackage[usenames,dvipsnames]{xcolor}
\usepackage[
    bookmarksnumbered,
    pdfpagelabels=true,
    plainpages=false
    ]{hyperref}
\usepackage[capitalize]{cleveref}
\usepackage[english,russian,french]{babel}
\definecolor{cblue}{RGB}{55,126,184}
\hypersetup{
    colorlinks=true,
    linkcolor=cblue,
    citecolor=cblue,
    urlcolor=cblue
    }

\usepackage{comment}
\usepackage{array}
\usepackage{amsmath, amsfonts, amsthm, enumerate}
\usepackage{amssymb}

\usepackage{cancel}
\usepackage{physics}
\usepackage{appendix}
\usepackage{xcolor}
\usepackage{tikz}
\usetikzlibrary{shapes,snakes,arrows,chains}
\usetikzlibrary{positioning}

\usetikzlibrary{backgrounds}
\usetikzlibrary{patterns}
\pgfdeclarelayer{background}
\pgfdeclarelayer{foreground}
\pgfsetlayers{background,main,foreground}

\usepackage[left=2cm,right=2cm,top=2cm,bottom=1.5cm]{geometry}
\usepackage[export]{adjustbox}

\makeatletter
\renewcommand*\env@matrix[1][\arraystretch]{%
  \edef\arraystretch{#1}%
  \hskip -\arraycolsep
  \let\@ifnextchar\new@ifnextchar
  \array{*\c@MaxMatrixCols c}}
\makeatother

\newcommand{\beq}{\begin{equation}}
\newcommand{\eeq}{\end{equation}}
\newcommand{\bi}{{\bar{i}}}
\newcommand{\bj}{{\bar{j}}}

\newcommand{\half}{\frac{1}{2}}
\newcommand{\bq}{{\bar{q}}}
\newcommand{\bfq}{\mathbf{q}}
\newcommand{\bft}{\mathbf{t}}

\newcommand{\calD}{\mathcal{D}}

\newcommand{\SM}{\textcolor{blue}{\textbf{[SM]}}}

\renewcommand{\thesection}{\roman{section}}

\makeatletter
\renewcommand{\p@subsection}{}
\renewcommand{\p@subsubsection}{}
\makeatother

\newcommand{\thistitle}{Algebraic Fusion in a (2+1)-dimensional  Lattice Model with Generalized Symmetries}

\begin{document}

\title{\thistitle}

\author{Chinmay Giridhar}
\affiliation{Department of Physics and Astronomy, Rice University, Houston, TX 77005, USA}
\author{Philipp Vojta} \affiliation{Department of Physics, Washington University in St. Louis, MO 63130 USA}
\author{Zohar Nussinov}
\affiliation{Department of Physics, Washington University in St. Louis, MO 63130 USA}
\affiliation{Institut fur Physik, Technische Universität Chemnitz, 09107 Chemnitz, Germany
}
\affiliation{Department of Physics and Quantum Centre of Excellence for Diamond and Emergent Materials (QuCenDiEM),
Indian Institute of Technology Madras, Chennai 600036, India}

\affiliation{LPTMC, CNRS-UMR 7600, Sorbonne Universit\'e, 4 Place Jussieu, 75252 Paris cedex 05, France}

\author{Gerardo Ortiz}
\affiliation{Department of Physics, Indiana University, Bloomington, IN 47405, USA}
\affiliation{Institute for Advanced Study, Princeton, NJ 08540, USA}
\author{Andriy H. Nevidomskyy}
\affiliation{Department of Physics and Astronomy, Rice University, Houston, TX 77005, USA}
\affiliation{Rice Center for Quantum Materials and Advanced Materials Institute, Rice University, Houston, TX 77005, USA}
\affiliation{Division of Condensed Matter Physics and Materials Science, Brookhaven National Laboratory, Upton, NY 11973-5000, USA}

\date{\today}
\begin{abstract}
%The conventional Landau paradigm of classifying phases of matter by the associated (anti)unitary symmetries has recently been augmented with the notion of (inherently non-(anti)unitary)  non-invertible symmetries. These symmetries appear at %are found at self-dual points of certain lattice models including the celebrated (1+1)D transverse field Ising model. In these systems, the topological defects that are associated with the duality form a fusion algebra (rather than a group). By contrast, much less is known about the existence of such non-invertible symmetries in (2+1)D. In this work, we develop the formalism of non-invertible self-duality in the Ising plaquette model in a transverse field. We show that the use of bond-algebraic automorphisms, combined with the so-called half-gauging procedure, provides insight into the structure of the NSO, which we express as a sequential quantum circuit. The associated duality defects are constrained by the subsystem symmetries of the model, resulting in restricted mobility. We study the fusion rules of these duality defects with the defects of the subsystem symmetries. In constructing the non-invertible duality transformation, we verify explicitly that it acts as a projective unitary on the physical Hilbert space, thus satisfying a recent generalization of Wigner's theorem to non-invertible symmetries.

The notion of quantum symmetry has recently been extended to include reduced-dimensional transformations and algebraic structures beyond groups. Such generalized symmetries lead to exotic phases of matter and excitations that defy Landau’s original paradigm. Here, we develop an algebraic framework for systematically deriving the fusion rules of topological defects in higher-dimensional lattice systems with non-invertible generalized symmetries, and focus on a (2+1)-dimensional quantum Ising plaquette model as a concrete illustration. We show that bond-algebraic automorphisms, when combined with the so-called half-gauging procedure, reveal the structure of the non-invertible duality symmetry operators, which can be explicitly represented as a sequential quantum circuit. The resulting duality defects are constrained by the model’s rigid higher symmetries (lower-dimensional subsystem symmetries), leading to restricted mobility. We establish the fusion algebra of these defects. Finally, in constructing the non-invertible duality transformation, we explicitly verify that it acts as partial isometry on the physical Hilbert space, thereby satisfying a recent generalization of Wigner’s theorem applicable to non-invertible symmetries.
\end{abstract}

\maketitle

\paragraph*{Introduction.}
In physics, symmetries are typically defined within the framework of group theory,  where each symmetry operator has an inverse \cite{Asher-Peres}. However, it has long been recognized that continuum theories in two spacetime dimensions can exhibit symmetries that transcend %the 
this conventional group structure. Instead, symmetry operators or their associated defects may form more complex algebraic structures, such as a {\it fusion category}, allowing for elements with non-trivial kernels -- that is, {\it non-invertible} symmetries. The canonical example is the two-dimensional Ising conformal field theory (CFT)~\cite{frohlich_KW-CFT_2004}, whose fusion rules encompass both the global $\mathbb{Z}_2$ symmetry operators and more subtle {\it topological defect lines} \cite{TDL-Chang2019} extending along the time direction. In certain rational CFTs, these defect lines are known as Verlinde lines~\cite{verlinde1988}, and generalizations exist for broader classes of CFTs~\cite{TDL-Chang2019}. More recently, it has been shown that non-invertible symmetry operators (NSOs) can also be defined in certain quantum (1+1)-dimensional {\it lattice} models. This allows for precise, non-perturbative statements about the full theory, not just its long-wavelength behavior~\cite{aasen-defects2020,MPO-lootens2023, boundaries-choi2023,lootens2024,seiberg-LSM,seiberg_majorana2024,S3-chatterjee2024,LSM-pace2024,ebisu2025,cao2025}. The transverse-field Ising chain model (TFIC)~\cite{NO2010} is the most studied example, exhibiting a quantum analogue of the classical Kramers–Wannier duality~~\cite{kramers-wannier1941,NO2010}. It is important to note that such dualities are generally maps between {\it different} models and only become genuine symmetries at special, fine-tuned self-dual points in parameter space.

The task of constructing explicit, lattice-specific expressions for the NSOs is highly nontrivial. It involves manipulating inherently non-local objects and carefully defining their associated movement and fusion actions in a way that faithfully reproduces the expected fusion algebra~\cite{seiberg-LSM}. These NSOs can often be represented using the formalism of matrix product operators \cite{MPO-lootens2023}, alternatively described in terms of sequential quantum circuits~\cite{circuits-chen2024}, thus providing useful computational and conceptual tools. A further subtlety lies in the fact that NSOs often require an enlargement of the Hilbert space of the original lattice model. For example, translation operators appearing in fusion expressions can add or remove lattice sites, thereby modifying the Hilbert space dimension~\cite{seiberg-LSM}. 
Moreover, any such NSO must conform to Wigner’s theorem \cite{peres1995quantum}, which stipulates that all physical symmetries must be implemented in a way that preserves transition probabilities between any two states.
%\textcolor{red}{Possible minor rewording- Moreover, the analysis  may be bolstered by Wigner’s theorem \cite{peres1995quantum}, which stipulates that all physical (Hilbert space preserving) symmetries must be implemented in a way that they leave the transition probabilities between any two  states invariant.} 
These considerations highlight the deep challenges involved in defining the actions performed by non-invertible symmetries %symmetry actions that are
in a manner that is consistent both algebraically and physically. 
%within a lattice framework.

%The task of finding an explicit, lattice-specific expressions for the NSO is a highly nontrivial one -- it  requires manipulating non-local operators and carefully defining the movement and fusion operators in order to verify the fusion algebra~\cite{seiberg-LSM}. The resulting NSO can be cast in the language of matrix product operators~\cite{MPO-lootens2023}, which can also be phrased in the context of sequential quantum circuits~\cite{circuits-chen2024}.  The other troubling feature of the NSO is that they require enlarging the Hilbert space of the original lattice model, for instance translation operators that appear in the fusion expressions are subtle in that they add or remove sites, thus altering the size of the Hilbert space~\cite{seiberg-LSM}. 

In light of these challenges, this paper introduces a unified algebraic framework to construct and manipulate topological defects in quantum lattice theories. This leads to the explicit derivation of the relevant NSOs and their fusion properties by utilizing the {\it bond-algebraic} method -- a powerful approach developed earlier by two of us~\cite{PhysRevB.79.214440,Cobanera2010,Cobanera2011} %to 
that enables the study of dualities in both classical and quantum systems. % on equal footing. 
A central notion is that of homomorphisms -- here taken to be linear automorphisms -- of bond algebras, which admit implementation via unitary operators acting as isometries on the Hilbert space. In the current work, we demonstrate the power of this approach with a compelling example in a (2+1)D theory that harbors a non-invertible symmetry -- especially relevant given that the vast majority of prior studies focused on either (1+1)D~\cite{aasen-defects2020,boundaries-choi2023,lootens2024,seiberg-LSM,seiberg_majorana2024,S3-chatterjee2024,LSM-pace2024} or (3+1)D lattice theories~\cite{D4-choi2022,D4-kaidi2022,D4-koide2022,D4-koide2023,D4-triality-choi2023,D4-gorantla2024}. We note that dualities in 2+1 D theories have been studied in \cite{ebisu2025,D3-cao2023}. The example that  we consider %is paradigmatic in that the theory has
exhibits paradigmatic rigid %1-form symmetries~
generalized symmetries  \cite{Nussinov-PNAS2009,Nussinov-Annals2009} that are prevalent in compass type and many other theories (these symmetries are also commonly referred to as subsystem symmetries, particularly when they arise in the context of celebrated theories harboring dynamically constrained particles -- fractons \cite{Vijay2016,Gromov2024}). We carefully construct the relevant NSOs and derive their fusion rules, including the fracton defect operators. 
From a mathematical perspective, the NSO is realized as the composition of an invertible unitary transformation and a non-invertible projection onto a specified global symmetry sector, in accordance with the constraints set by a {\it generalized symmetry} extension of Wigner’s theorem \cite{GWT}. %{\textcolor{red}{that we introduce in the current work}}. 
The lessons learned from the bond-algebraic perspective %are powerful and we hope 
can be applied to other examples of non-invertible symmetries beyond one spatial dimension.

%\an{
%\paragraph{\textbf{Internal note.}} Gerardo: in my introduction above, I aimed to cover the following points: 
%\begin{enumerate}
%    \item The non-invertible structure of the operators is not new and has been much discussed in \textit{continuum} CFTs -- indeed, this is where the structure of fusion categories was first developed by formal mathematical physicists (and independently, by mathematicians -- for instance Ising CFT is an example of the Tambara--Yamagami category~\cite{TY})
%    \item The novelty and recent excitement, in particular as advertised by S.-H. Shao and N. Seiberg, is in the precise \textit{lattice} expressions for the non-invertible symmetry operators.
%    \item Emphasize the difficulty of obtaining lattice-regularized expressions for the non-invertible NSOs, and explain how \underline{our approach}, based on bond algebra toolkit, is powerful in obtaining such expressions and the fusion rules.
%    \item Underscore two advances we made by studying the Xu--Moore model: (i) it is (2+1)d, unlike most known examples and (ii) the model has rigid 1-form symmetries, a.k.a. subsystem symmetries, and the associated fractons.
%\end{enumerate}
%}

\paragraph*{Xu--Moore model: Symmetries and duality.}
The Xu--Moore model (a.k.a. the Ising plaquette model) \cite{Xu-Moore-PRL,Xu-Moore-NPB} is defined in terms of the Pauli $X$ and $Z$ operators on the vertices of a square lattice. The sites are labeled by coordinates $q = (q_x,q_y)$, $q_{x,y} \in \mathbb{Z}$, %where 
with lattice unit vectors $\hat x \equiv (1,0)$ and $\hat y \equiv (0,1)$ %, with 
where $1 \leq q_{x/y} \leq L_{x/y}$. With a coupling strength $g$, we define
\beq
\label{eq.XM}
H_\text{XM}[g] = -\sum_{q} \left( g^{-1} Z_{q}Z_{q+ \hat x}Z_{q+ \hat y}Z_{q+ \hat x+ \hat y} + g X_q \right).
\eeq

\noindent
The model has an extensive number of commuting (invertible) subsystem symmetries formed by the products of Pauli operators along individual rows/columns, %generated by 
\beq
\hat{\eta}^{\text{row}}_j = \prod_{q : q_y = j} X_{q};\quad
\hat{\eta}^{\text{col}}_i = \prod_{q : q_x = i} X_{q}.
\label{eq.eta}
\eeq
On a lattice of $L_x \times L_y$ sites, there are $L_x + L_y - 1$ %independent 
independent subsystem symmetries~\cite{nussinov-ortiz2023}.
%
%\paragraph*{Bond algebra and the duality.} 
%\textcolor{red}{include "D" in above section, mention that it is constant of motion and can be lifted into a symmetry by enlarging}
It was realized early on~\cite{Xu-Moore-PRL,Xu-Moore-NPB,Nussinov-Fradkin-PRB} that one can perform a duality transformation $D$ that maps the two terms in Eq.~\eqref{eq.XM} into one another: $DH(g) = H(g^{-1})D$. %At the coupling 
When $g=1$, the Xu--Moore model is self-dual, and the transformation $D$ becomes a constant of motion. Just as is the case with the quantum duality of the TFIC with periodic boundary conditions (pbc), it turns out that the operator $D$ lacks an inverse, rendering the duality \textit{non-invertible}. Note that this result is sensitive to the boundary conditions -- the duality becomes invertible (unitary) with open boundaries~\cite{XM-dualities-maity2025}.  Here we offer a perspective on non-invertibility based on the bond algebra~\cite{Cobanera2011}. 
We begin with a simple demonstration of this non-invertibility via a simple proof by contradiction. Towards that end, let us choose the bond algebra $\mathcal{A}$ to be the von Neumann algebra generated by the local terms 
(bonds) whose sum forms the Hamiltonian  
%of 
$H_\text{XM}$, i.e.,
\beq
\mathcal{A} := \{Z_{q}Z_{q+\hat x}Z_{q+ \hat y}Z_{q+ \hat x+ \hat y}; X_{q} \}
\label{eq:naive_bond_algebra_xm}.
\eeq
Now suppose that there %is 
exists an invertible duality transformation $\Phi$ that maps $Z_{q}Z_{q+ \hat x}Z_{q+ \hat y}Z_{q+ \hat x+ \hat y} \stackrel{\Phi}{\longrightarrow} X_{q}$ and, conversely, maps $X_q \stackrel{\Phi}{\longrightarrow} Z_{q} Z_{q- \hat x} Z_{q- \hat y} Z_{q- \hat x- \hat y}$. If it existed, such a transformation would%be
, by %the 
definition, be an automorphism of the bond algebra. %Consider now a system with, say, periodic boundary conditions (p.b.c.). 
Then, with pbc, the product of all plaquette $Z$ operators in any row or column equals to identity, and acting by the algebra automorphism $\Phi$ yields, schematically, 
\beq
%\Phi(\mathbb{I})
\Phi(1)=\Phi\bigg(\prod_{p \in  \substack{\text{row,}\\ \text{col}}}  [ZZZZ]_p\bigg) = \Phi\bigg(\prod_{q \in \substack{\text{row,}\\ \text{col}}} X_q\bigg) \neq %\mathbb{I}
1,
\label{eq.auto}
\eeq
resulting, since any automorphism must map the identity into itself, in the said contradiction underlying this proof. Thus, such an invertible duality $\Phi$ cannot exist within the above confines. 
This non-invertibility may, however, be remedied. In the following, we show that a unitary automorphism can be constructed, at the cost of {\it extending} the Hilbert space of the problem \cite{GWT}.

\paragraph*{Enlarging the Hilbert space.}
Adding auxiliary 
(gauge-like) degrees of freedom, acting outside %of 
the physical Hilbert space $\mathcal{H}$, introduces additional generators into the bond algebra. % of the theory. %The 
Our goal is to find the minimal number of such %auxiliaries 
auxiliary %gauge-like %degrees of freedom 
operators
%in such as fashion as to admit
necessary to achieve a bond-algebraic automorphism. %A theorem by von Neumann \textcolor{red}{[citation needed]} 
This then guarantees \cite{vonNeumann1949,KadisonRingrose} that such an automorphism is unitarily implementable %in the 
on an extended Hilbert space $\tilde{\mathcal{H}} \supset {\mathcal{H}}$. %Finally, the
After implementing the duality 
unitarily within  $\tilde{\mathcal{H}}$, the non-invertible duality is obtained %upon 
by performing a non-invertible %projecting 
projection $\widehat{P}_{+}:~ \tilde{\mathcal{H}} \to \mathcal{H}$
back into the original physical 
state space. %sector 

%Before turning to the Xu--Moore model, to illustrate the core concept, we consider 
%As 
%as a concrete example, %consider 
%the much studied quantum duality of the TFIC, where it suffices to enlarge the Hilbert space by a single site, $\tilde{\mathcal{H}} = \mathbb{C}^2 \otimes \mathcal{H}_\text{TFIC}$, and require that the automorphism maps the additional %ancilla 
%auxiliary Pauli $\sigma^z$ operator on this site as follows: $\Phi(\sigma_z) = X_1X_2\ldots X_L$.
Before turning to the Xu--Moore model, we illustrate the core concept using the TFIC. To achieve a bond-algebraic duality for the TFIC, it suffices to enlarge its Hilbert space by a single site,
$\tilde{\mathcal{H}} = \mathbb{C}^2 \otimes \mathcal{H}_\text{TFIC}$, and to require that the automorphism maps the additional auxiliary Pauli $\sigma^z$ operator on this site as
$\Phi(\sigma^z) = X_1 X_2 \ldots X_L$.
This, combined with the requirement $\Phi(Z_L \sigma_z Z_1) = X_1$ is indeed sufficient to guarantee the existence of a \textit{unitary} operator $\mathcal{U}_\Phi$ that implements the automorphism on 
the %this 
enlarged Hilbert space $\tilde{\mathcal{H}}$.
%(the additional product space happens to be trivial, i.e. $\mathbb{C}$).
It is indeed only upon {\it projecting} onto the original Hilbert space of the model that $\widehat{D}=\mathcal{U}_\Phi \widehat{P}_{+}$ becomes a NSO \cite{GWT}.
The constant of motion $D$ can be obtained as $D \equiv \widehat P_{+} \widehat D$.

%This begs the question -- by 
By how much should the Hilbert space of %the Xu--Moore model 
a system, such as the Xu--Moore model, with NSOs be enlarged in order to allow for %the 
a unitary bond algebraic automorphism %implementing 
implementation of the desired duality? Below, we show that the bond-algebraic considerations % provide a powerful tool to 
readily answer this question. 

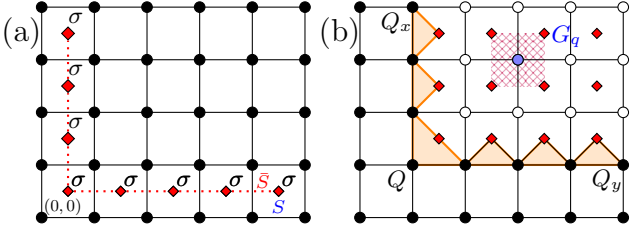
\begin{figure}[tb]
\centering
\setlength{\tabcolsep}{2pt} % reduce spacing between columns
\renewcommand{\arraystretch}{0} % tighten vertical spacing
%\begin{tabular}{@{}c@{}c@{}}
% --- Left figure ---
\begin{tikzpicture}[scale=0.7]

% Define colors
\definecolor{myblue}{RGB}{0,0,255}

% Draw the lattice
\foreach \i in {0,1,2,3,4,5} {
    \foreach \j in {0,1,2,3,4} {
        % Draw the nodes
        \node[circle, draw, fill=black, minimum size=4pt, inner sep=0pt] at (\i, \j) {};
        % Draw links (horizontal)
        \ifnum\j<4 \draw (\i, \j) -- (\i, \j+1); \fi
        % Draw links (vertical)
        \ifnum\i<5 \draw (\i, \j) -- (\i+1, \j); \fi
    }
}
\usetikzlibrary{backgrounds}
% Color the bottommost row nodes
%\foreach \i in {0,1,2,3,4,5} {
   % \node[regular polygon, regular polygon sides=4, draw, fill=blue!50, minimum size=2pt, inner sep = 1.5pt] at (\i,0) {};
%}
% extra sites bottom
\foreach \i in {1,2,3,4,5} { 
    \node[diamond, draw, fill=red, minimum size=4pt, inner sep=0pt] at (\i-0.5, 0.5){};
    \node at (\i-0.3, 0.7) {$\sigma$};
}
\foreach \i in {1,2,3,4,5} { 
    \node[diamond, draw, fill=red, minimum size=4pt, inner sep=0pt] (d\i) at (\i-0.5, 0.5) {};
    \node at (\i-0.3, 0.7) {$\sigma$};
}

% Now you can reference d1, d2, … safely
\begin{pgfonlayer}{background}
    \draw[red, dotted, thick] (d1) -- (d2) -- (d3) -- (d4) -- (d5);
\end{pgfonlayer}

% Color the leftmost column nodes
%\foreach \j in {0,1,2,3,4} {
   % \node[regular polygon, regular polygon sides=4, draw, fill=blue!50, minimum size=2pt, inner sep = 1.5pt] at (0, \j) {};
%}
% extra sites left, with names
\foreach \j in {2,3,4} {
    \node[diamond, draw, fill=red, minimum size=5pt, inner sep=0pt] (l\j) at (0.5, \j-0.5) {};
    \node at (0.65, \j-0.2) {$\sigma$};
}

% Dotted red line through them, behind everything
\begin{pgfonlayer}{background}
    \draw[red, dotted, thick] (d1)--(l2) -- (l3) -- (l4);
\end{pgfonlayer}

% Shaded interface bands and labels
%\filldraw[fill = orange!20, draw = orange,thick, opacity=0.5] (-0.2,-0.2) -- ++(0,4.4) -- ++(0.4,0) -- ++(0,-4.4);
%\node[orange] at (0.5,4.1) {$S^{y}$};
%\filldraw[fill = red!20, draw = red,thick, opacity=0.5] (-0.2,0.2) -- (5.2,0.2) -- (5.2,-0.2) -- (-0.2,-0.2) -- cycle;
\node at (4.5, 0.2) {\scriptsize\textcolor{blue}{$S$}};
\node at (4.2, 0.7) {\scriptsize\textcolor{red}{$\bar{S}$}};

%\node at (-0.2,0.2) {$q^*$};

%\draw node[scale=0.7] at (3.35,1.85) {$q+e_x$};
%\draw node[scale=0.7] at (3.4,2.8) {$q+e_x+e_y$};
%draw node[scale=0.7] at (2.1,1.8) {$q$};

%\draw node[scale=0.7] at (-0.5,0.2) {$(0,0)$};
\draw node[scale=0.7] at (0.5-0.1,0.5-0.3) {$(0,0)$};
\node at (-0.385,3.5) {\large (a)};

\end{tikzpicture}
\hspace{-0.2cm}
% --- Right figure ---
\begin{tikzpicture}[scale=0.7, background layer/.style={}, foreground layer/.style={}]

% Define colors
\definecolor{myblue}{RGB}{0,0,255}
\definecolor{myred}{RGB}{255,0,0}
\definecolor{myfill}{RGB}{240,220,130} % light gold
\definecolor{mygold}{RGB}{200,160,0}

% Grid size
\def\Nx{5}
\def\Ny{4}
% Red region corner
\def\Rx{1}
\def\Ry{1}

% --- Draw the lattice ---
\foreach \i in {0,...,\Nx} {
    \foreach \j in {0,...,\Ny} {
        \begin{pgfonlayer}{foreground}
        \ifnum\i>\Rx
            \ifnum\j>\Ry
                \node[circle, draw, fill=white, minimum size=4pt, inner sep=0pt] at (\i, \j) {};
            \else
                \node[circle, draw, fill=black, minimum size=4pt, inner sep=0pt] at (\i, \j) {};
            \fi
        \else
            \node[circle, draw, fill=black, minimum size=4pt, inner sep=0pt] at (\i, \j) {};
        \fi
        \end{pgfonlayer}
        % Links
        \ifnum\j<\Ny \draw (\i, \j) -- (\i, \j+1); \fi
        \ifnum\i<\Nx \draw (\i, \j) -- (\i+1, \j); \fi
    }
}

% ============================================================
% === Unified corner triangle at the L junction (merged) =====
% ============================================================

\begin{pgfonlayer}{background}
  % Fill the merged triangle
  \filldraw[orange!20, draw=orange, thick]
    (\Rx, \Ry) -- (\Rx+1, \Ry) -- (\Rx, \Ry+1) -- cycle;
\end{pgfonlayer}

\begin{pgfonlayer}{foreground}
  % Red diamond at the center of the merged triangle
  \node[regular polygon, regular polygon sides=4, rotate=45,
        draw, fill=myred, minimum size=4pt, inner sep=0pt]
        at (\Rx+0.5, \Ry+0.5) {};
\end{pgfonlayer}

% ============================================================
% === Horizontal edge triangles (skip the first one) =========
% ============================================================

\foreach \i in {\Rx+2,\Rx+3,\Rx+4} {
    \begin{pgfonlayer}{foreground}
    \node[regular polygon, regular polygon sides=4, rotate=45,
          draw, fill=myred, minimum size=4pt, inner sep=0pt]
          at (\i-0.5, \Ry+0.5) {};
    \end{pgfonlayer}

    \begin{pgfonlayer}{background}
    \filldraw[orange!20, draw = orange, thick]
        (\i-1, \Ry) -- (\i, \Ry) -- (\i-0.5, \Ry+0.5) -- cycle;
    \draw (\i-1, \Ry) -- (\i-0.5, \Ry+0.5);
    \draw (\i, \Ry) -- (\i-0.5, \Ry+0.5);
    \end{pgfonlayer}
}

% ============================================================
% === Vertical edge triangles (skip the first one) ===========
% ============================================================

\foreach \j in {\Ry+2,\Ry+3} {
    \begin{pgfonlayer}{foreground}
    \node[regular polygon, regular polygon sides=4, rotate=45,
          draw, fill=myred, minimum size=4pt, inner sep=0pt] at (\Rx+0.5, \j-0.5){};
    \end{pgfonlayer}

    \begin{pgfonlayer}{background}
   \filldraw[orange!20, draw = orange, thick]
    (\Rx, \j-1) -- (\Rx, \j) -- (\Rx+0.5, \j-0.5) -- cycle;
    \end{pgfonlayer}
}
%--- Fill ancilla decorations in gauged region ---
 ERROR 1
\foreach \i in {3,4,5} 
{
    \foreach \j in {3,4} 
        {
        \begin{pgfonlayer}{foreground}
            \node[draw=black, fill=myred, diamond, minimum size=4 pt, inner sep=0 pt] at (\i-0.5, \j-0.5){};
        \end{pgfonlayer} 
        }
}
% --- Blue square with cross-hatched fill ---
\begin{pgfonlayer}{background}
\fill[pattern=crosshatch, pattern color=purple, opacity=0.6]
    (2.5,2.5) rectangle (3.5,3.5);
\end{pgfonlayer}

% --- Blue center site and label ---
\begin{pgfonlayer}{foreground}
\node[circle, draw, fill=blue!50, minimum size=4pt, inner sep=0pt] at (3,3) {};
\node at (3.4,3) [above right=2pt, text=blue] {$G_q$};
\end{pgfonlayer}

% --- Corner and projections ---
\begin{pgfonlayer}{foreground}
  \node at (\Rx-0.3,\Ry-0.3) {$Q$};
  \node at (\Rx-0.3,\Ny-0.3) {$Q_x$};
  \node at (\Nx-0.3,\Ry-0.3) {$Q_y$};
\end{pgfonlayer}

\node at (-0.385,3.5) {\large (b)};
\end{tikzpicture}

%\end{tabular}

\caption{ %\textcolor{red}{To be consistent with the text we might wish to explicitly denote $(0,0)$ and $(\bar 0,\bar 0)$ in the figure.} 
(a) Xu-Moore model with sites on vertices of a periodic 5 $\times$ 4 lattice. Red diamonds indicate the new sites extending the bond algebra so as to permit the construction of an automorphism. $S$ and $\bar{S}$ are the half-gauging procedure. (b) The original, ungauged lattice sites (black) and the gauged sites on the dual lattice (red diamonds) are separated by the interface forming the duality defect. The Gauss law %, 
operator $G_q$ %, 
of the gauged Xu-Moore is shaded in purple.}
\label{fig:XM_combined}
\end{figure}

Since a relation of the form of Eq.~\eqref{eq.auto} %holds 
appears for each individual row/column, %one 
we need to %add 
{\it introduce an additional site for each %such 
row and column}. 
%coupled to one plaquette term in this row/column. 
We thus require $\mathcal{N} = L_x + L_y -1$ extra sites on our square lattice.
Without loss of generality, we choose these $\mathcal{N}$ auxiliary sites to lie along the left and lower boundaries of our lattice, denoted by $S$, as shown in Fig.~\ref{fig:XM_combined}(a).
 %To form the torus, we identify this lattice edge with the opposite, uncolored boundary. 
%The boundary $S$ contains $L_x+L_y-1$ sites. 
We denote by $\bar{S}$ 
%= \{\bar{q} \equiv q+\frac{\hat{x}}{2}+\frac{\hat{y}}{2}|q \in S\}$~\footnote{Throughout, we denote $\bar q = (\bar q_x,\bar q_y) \equiv (q_x + \frac{1}{2},q_y + \frac{1}{2})$.} 
the set of dual sites immediately above and to the right of sites in $S$ -- these are the auxiliary sites 
shown with red symbols in Fig.~\ref{fig:XM_combined}, the horizontal edge denoted by $S^{x}$ and the vertical by $S^{y}$. %(note that the overlap $S^{x} \cap S^y = \{ q^* = (0,0) \}$). 
The extended Hilbert space is thus $\tilde{\mathcal{H}}_\text{XM} = \mathbb{C}^{2\mathcal{N}} \otimes \mathcal{H}_\text{XM}$.
Each auxiliary site is coupled to the respective plaquette of the original lattice, thereby deforming 
the bond algebra $\cal A$ in Eq.(\ref{eq:naive_bond_algebra_xm}) to
\begin{eqnarray}
    \label{eq:true_bond_algebra_xm}
    \mathcal{A}' =\{ Z_{q}Z_{q+\hat x}Z_{q+ \hat y}Z_{q+ \hat x+ \hat y}\sigma^z_{q}; X_{q} , q \in S\}\\\nonumber
    \cup\{ Z_{q}Z_{q+\hat x}Z_{q+ \hat y}Z_{q+ \hat x+ \hat y}; X_{q} , q \notin S\}.
\end{eqnarray}

In order to avoid inconsistencies of the kind illustrated in Eq.~\eqref{eq.auto}, the automorphism must map the boundary as follows:
%the extra site Pauli operators $\sigma^z$ into a product over plaquettes on the horizontal and vertical edge as follows:
\begin{equation}
  \Phi_{\sf XM}(Z_{q} Z_{q+ \hat x} Z_{q+ \hat y} Z_{q+ \hat x+ \hat y} \sigma^{z}_{q})  
  =  X_{q}  \ , \ q \in S,
\label{eq:XM-auto}
\end{equation}
and 
\begin{equation}
  \Phi_{\sf XM}(X_{q+\hat x + \hat y})  
  =  Z_{q} Z_{q+ \hat x} Z_{q+ \hat y} Z_{q+ \hat x+ \hat y} \sigma^{z}_{q}  \ , \ q \in S~.
\label{eq:XM-auto-second}
\end{equation}
Equation~\eqref{eq:XM-auto} implies that each auxiliary operator $\sigma^z_q$ ($q \neq (0,0)$) is mapped onto a subsystem symmetry $\hat{\eta}^{\text{row/col}}$.
%We infer from Eq.~\eqref{eq:XM-auto} that every extra site $\sigma^z_{q}$ is mapped onto a subsystem symmetry $\hat \eta^{\text{row/col}}$. 
The case of the corner $\sigma^z_{(0, 0)}$ is more subtle  (Sec.~\ref{sec:bond_algebras_torus} in \SM %\ref{sec:unitary_action_on_projectors}
), but it too maps to a symmetry of the periodic Xu-Moore lattice.
% The way the \textcolor{blue}{extra sites} enter on the l.h.s. are indicative of gauging -- indeed, we show how to obtain the desired duality transformation in an alternative fashion, using so-called half-gauging procedure~\SM.
%\textcolor{blue}{The minimal coupling to additional sites is unitarily equivalent to specific sectors of the globally gauged theory  - this is the Xu-Moore analogue of the "partial gauging" procedure in the 1D TFIC ~\SM.}

\paragraph*{Non-invertible Symmetry.} 
The algebra automorphism is implemented on the enlarged Hilbert space by the unitary $\mathcal{U}$. This operator is explicitly constructed for a finite $L_x \times L_y$ periodic lattice in Sec.~\ref{sec:unitary_construction_torus} of \SM using a sequence of unitaries. Here, we quote the final result for the NSO implementing the duality:
%one and two-qubit gates,
%% Make a note about the importance of explicitly constructing the D operator!
\beq
\widehat{D} = \mathcal{U} \ \widehat{P}_{+}, 
\label{eq.D-operator}
\eeq
where the explicit form of the unitary $\mathcal{U}$ is given in the \textit{End Matter} for the infinite cylinder limit and in the \SM for the Torus and the non-invertible operator  $\widehat{P}_{+}$ denotes the projector into the untwisted sector of the theory, 
\begin{align}
    \widehat{P}_{+} = \prod_{q \in S } \left( \frac{1+\sigma^{z}_{q}}{2}  \right)
    %\equiv \prod_{i=1}^{L_x} \left(\frac{1+\sigma^{z\text{col}}_{i-\half}}{2} \right)\prod_{j=2}^{L_y}  \left(\frac{1+\sigma^{z\text{row}}_{j-\half}}{2}\right).\\
     = \prod_{q\in S_x} \left( \frac{1+\sigma^{z}_{q}}{2} \right)  \prod_{q'\in S_y} \left( \frac{1+\sigma^{z}_{q'}}{2} \right). 
\end{align}
% \begin{equation}
%    D_{KW} = \widehat P_{+} \mathcal{U} \widehat P_{+} = \widehat P_{+} \widehat{D}\ .
% \end{equation}
% Using the following key property of $\mathcal{U}$ : 
% \begin{equation}
%     \mathcal{U} ~ \widehat{P}_{\eta} ~{\mathcal{U}}^{\dagger} = \mathcal{U}^2  \widehat{P}_{+} ({\mathcal{U}}^{\dagger})^2 = \widehat P_{+} \  \ (\text{see} \ \SM)
%     \label{eqn:projector_phys_eta_duality}
% \end{equation}
Crucially, the projector $\widehat{P}_{+}$ only acts on the auxiliary ($\sigma$) degrees of freedom, while acting as the identity on vectors in the original Hilbert space -- as required by the generalized Wigner theorem~\cite{GWT}. Combined with \eqref{eq.D-operator} which asserts that  
%\begin{equation}
$\widehat{D}^{\dagger} \widehat{D} = \widehat P_{+}$,
%\end{equation}
this implies that transition probabilities between arbitrary  physical states $|\alpha \rangle$ and $|\beta \rangle$ remain invariant under the duality. That is, $|\langle \widehat{D}  \alpha | \widehat{D} \beta \rangle|^2 = |\langle \alpha | \widehat{D}^{\dagger} \widehat{D} | \beta \rangle |^2= |\langle \alpha | \beta \rangle|^2$ \footnote{If one were to completely project the action of $\mathcal{U}$ onto the physical sector, i.e. $D := \hat{P}_{+} \hat D = \mathcal{U} \hat P_{\eta}  \hat P_{+} = \hat P_{+} \hat P_{\eta} \ \mathcal{U}$ (using the fact that $\mathcal{U}^2 \hat P_{+} \left( \mathcal{U^{\dagger}} \right)^2 = \hat{P}_{+}$, see \SM (\ref{sec:bond_algebras_torus}\ref{sec:unitary_action_on_projectors})), we would see that it fails to conserve probabilities in the physical sector, as $D^{\dagger} D = \hat P_{\eta} \hat P_{+}$.}.
%of 

%\anc{Make this a footnote?(If one were to completely project the action of $\mathcal{U}$ onto the physical sector, i.e. $D := \widehat P_{+} \widehat D = \mathcal{U} \widehat P_{\eta}  \widehat P_{+} = \widehat P_{+} \widehat P_{\eta} \ \mathcal{U}$ (using the property that $\mathcal{U}^2 \widehat P_{+} \left( \mathcal{U^{\dagger}} \right)^2 = \widehat{P}_{+}$, see \SM (\ref{sec:bond_algebras_torus} \ref{sec:unitary_action_on_projectors}) ), we would see that it fails to conserve probabilities in the physical sector, as $D^{\dagger} D = \widehat P_{\eta} \widehat P_{+}$.).}

%It is easy to see that 

Despite preserving probabilities in the physical sector, the non-invertibility of $\widehat D$ in the enlarged Hilbert space is still manifest in this sector. %This is seen by examining the 
Indeed, the action of $\widehat D^2$ %on the local terms 
(see End Matter) may be expressed as 
\beq
\widehat{D}^2 \rightsquigarrow \widehat{P}_{\eta} \,T_{(1,1)}, 
\eeq
where $T_{l}$ denotes translation by a vector $l$ and $\widehat{P}_{\eta}$ is the projector onto the subspace where all subsystem symmetries of Eq.~(\ref{eq.eta}) have %unit 
eigenvalues equal to $1$.
%
% $U_D$ is the unitary operator (involving the aforementioned Hadamard and C^{\sf x} gates, see \SM), 
% $T_{\bft}$ is the translation operator by a vector $\bft$.
%These relations may, alternatively, be obtained 
%by examining the defect operators. 
%These operator relations may be understood by studying their associated defect terms in the Xu-Moore Hamiltonian.
An alternative approach to seeing this (adopted by many authors) is to examine the 
defect operators, to 
which we turn our attention 
to next. After finding the defect operators, we will determine their fusion rules.

\begin{figure}[tb]
\centering
\begin{tabular}{cc}
    % --- Top-left ---
    \resizebox{0.22\textwidth}{!}{%
        \begin{tikzpicture}
% Define colors
\definecolor{myblue}{RGB}{0,0,255}

% Grid size
\def\Nx{4}
\def\Ny{4}

% --- Draw links first ---
\foreach \i in {0,...,\Nx} {
    \foreach \j in {0,...,\Ny} {
        \ifnum\j<\Ny
            \draw (\i, \j) -- (\i, \j+1);
        \fi
        \ifnum\i<\Nx
            \draw (\i, \j) -- (\i+1, \j);
        \fi
    }
}

% --- Plaquette fills ---
\fill[pattern=north east lines, pattern color=blue, opacity=0.9]  (1,2) -- (2,2) -- (2,3) -- (1,3) -- cycle;
\fill[pattern=north east lines, pattern color=blue, opacity=0.9]  (1,1) -- (2,1) -- (2,2) -- (1,2) -- cycle;

% --- Blue wavy line ---
\draw[myblue, line width = 1, decorate, decoration={snake, amplitude=1.5pt, segment length=6pt}] 
    (\Nx,2) -- (2,2);

% --- Black sites on top ---
\foreach \i in {0,...,\Nx} {
    \foreach \j in {0,...,\Ny} {
        \node[circle, draw, fill=black, minimum size=4pt, inner sep=0pt] at (\i, \j) {};
    }
}

% Labels
% Larger labels with white square background
\node[draw=none, fill=white, inner sep=1pt] at (1.5, 2.5) {\Large $f$};
\node[draw=none, fill=white, inner sep=1pt] at (1.5, 1.5) {\Large $f$};
\node at (-0.35,3.5) {\Large (a)};
\end{tikzpicture}
    }
    &
    % --- Top-right ---
   \resizebox{0.22\textwidth}{!}{%
\begin{tikzpicture}
% Define colors
\definecolor{myblue}{RGB}{0,0,255}
\definecolor{myred}{RGB}{255,0,0}
\definecolor{myfill}{RGB}{240,220,130}

% Grid size
\def\Nx{4}
\def\Ny{4}
\def\Rx{1}
\def\Ry{2}

% --- Draw links first ---
\foreach \i in {0,...,\Nx} {
  \foreach \j in {0,...,\Ny} {
    \ifnum\j<\Ny
      \draw (\i, \j) -- (\i, \j+1);
    \fi
    \ifnum\i<\Nx
      \draw (\i, \j) -- (\i+1, \j);
    \fi
  }
}

% ============================================================
% === Unified corner triangle at the L junction (merged) =====
% ============================================================
\begin{pgfonlayer}{background}
  \filldraw[orange!20, draw=orange, thick]
    (\Rx, \Ry) -- (\Rx+1, \Ry) -- (\Rx, \Ry+1) -- cycle;
\end{pgfonlayer}

% Red diamond at merged apex
\begin{pgfonlayer}{foreground}
  \node[regular polygon, regular polygon sides=4, rotate=45,
        draw, fill=myred, minimum size=5pt, inner sep=0pt]
        at (\Rx+0.5, \Ry+0.5) {};
\end{pgfonlayer}

% ============================================================
% === Horizontal triangles (skip the first one) ==============
% ============================================================
\foreach \i in {\Rx+2,\Rx+3} {
  \begin{pgfonlayer}{background}
    \filldraw[orange!20, draw = orange,thick]
      (\i-1, \Ry) -- (\i, \Ry) -- (\i-0.5, \Ry+0.5) -- cycle;
  \end{pgfonlayer}
  \begin{pgfonlayer}{foreground}
    \node[regular polygon, regular polygon sides=4, rotate=45,
          draw, fill=myred, minimum size=5pt, inner sep=0pt]
          at (\i-0.5, \Ry+0.5) {};
  \end{pgfonlayer}
}

% ============================================================
% === Vertical triangles (skip the first one) ================
% ============================================================
\foreach \j in {\Ry+2} {
  \begin{pgfonlayer}{background}
    \filldraw[orange!20, draw = orange,thick]
      (\Rx, \j-1) -- (\Rx, \j) -- (\Rx+0.5, \j-0.5) -- cycle;
  \end{pgfonlayer}
  \begin{pgfonlayer}{foreground}
    \node[regular polygon, regular polygon sides=4, rotate=45,
          draw, fill=myred, minimum size=5pt, inner sep=0pt]
          at (\Rx+0.5, \j-0.5) {};
  \end{pgfonlayer}
}

% Load this in your preamble
\usetikzlibrary{patterns}

% --- Patterned blue regions ---
\fill[pattern=north east lines, pattern color=blue, opacity=0.9] 
    (3,2) -- (4,2) -- (4,3) -- (3,3) -- cycle;

\fill[pattern=north east lines, pattern color=blue, opacity=0.9] 
    (2,2) -- (3,2) -- (3,3) -- (2,3) -- cycle;

% --- Blue wavy line ---
\draw[myblue, line width=1, decorate, decoration={snake, amplitude=1.5pt, segment length=6pt}] 
  (3,0) -- (3,2);

% --- Now draw all sites on top ---
\foreach \i in {0,...,\Nx} {
  \foreach \j in {0,...,\Ny} {
    \ifnum\i>\Rx
      \ifnum\j>\Ry
        \node[circle, draw, fill=white, minimum size=4pt, inner sep=0pt] at (\i, \j) {};
      \else
        \node[circle, draw, fill=black, minimum size=4pt, inner sep=0pt] at (\i, \j) {};
      \fi
    \else
      \node[circle, draw, fill=black, minimum size=4pt, inner sep=0pt] at (\i, \j) {};
    \fi
  }
}
% --- Red dual sites on top ---
\foreach \i in {3,4} {
  \foreach \j in {4} {
    \node[draw=black, fill=myred, diamond, minimum size=5pt, inner sep=0pt] at (\i-0.5, \j-0.5) {};
  }
}
% --- Labels (unchanged) ---
\node at (2.7, 1.2) {\Large $\eta$};
\node at (3.3, 2.6) {\Large $\sigma^z$};
\node at (2.3, 2.6) {\Large $\sigma^z$};
\node at (-0.35,3.5) {\Large (b)};
\end{tikzpicture}
}
\end{tabular}
\caption{(a) A pair of fractons fuse into a horizontal $\eta$-defect: $f \times f = \eta^{\text{row}}$. The two fracton membranes combine to form a rigid string. (b) A vertical $\eta$-defect fuses with the duality defect. The $\hat \eta$ operator and $\tilde{Z}$ operator pair combine to form the fusion operator in Eq. (\ref{eqn:EtaDualityFusionOps}).}
%(lower left) The duality defect absorbs a fracton, $\mathcal{D}\otimes f = \mathcal{D}$.  (lower right) The fusion of two duality defects, $\mathcal{D}\otimes \mathcal{D}.}
\label{fig:defects}
\end{figure}
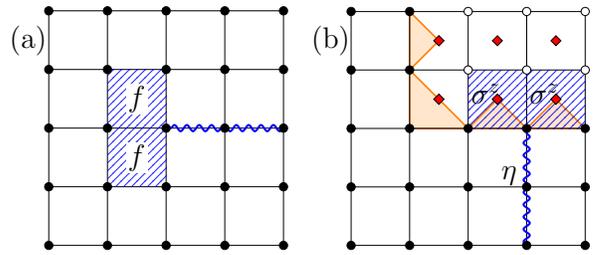

\paragraph*{Duality defect.} 
Extending earlier %studied 
partial-gauging studies of the TFIC \cite{seiberg-LSM} (see End Matter for a summary), %  for the TFIC 
we 
apply %a  %Based on the 
%partial-gauging %procedure recipe that was earlier applied to TFIC %described in 
%\cite{seiberg-LSM} (see End Matter for a summary),  
%in procedure for the TFIC described in \cite{seiberg-LSM} (see End Matter for a summary), 
%we perform 
an analogous procedure to %a subsystem of 
the Xu--Moore model. The %analog of the 
Gauss law is now implemented by  (see Fig. \ref{fig:XM_combined}(b)),
\beq \label{eq:GaugedXMGaussLaw}
G_q = X_q \sigma^z_{q} \sigma^z_{q-\hat x} \sigma^z_{q- \hat y} \sigma^z_{q- \hat x- \hat y}.  
\eeq
%Note that 
%The $\sigma^z$ degrees of freedom %live reside on the sites of the dual lattice (red diamonds) along the interface.
%To  avoid cumbersome notation, we 
%labeled the positions of the  auxiliary $\sigma$ sites by $q$ (rather than $\bq$ of the dual lattice).
%Carrying out an analogous 
The partial-gauging procedure %on a subsystem 
now leads to a 
duality %symmetry 
defect of restricted mobility   
 that resides on %the 
 the {\it geometrically extended} subsystem boundary (shaded region in Fig.~\ref{fig:XM_combined}). %between the two regions. 
 %Unlike the TFIC, %here 
 %the symmetry defect that results for the Xu--Moore model is spatially extended %(see shaded region in Fig.~\ref{fig:XM_combined}) 
 %and %possesses 
 %of restricted mobility. %We demonstrate (see Sec.~\ref{sec:gauging}\ref{sec:partial_gauging} in \SM)  that the 
The Hamiltonian terms (see Sec.~\ref{sec:gauging}\ref{sec:partial_gauging} in \SM) 
%in the Hamiltonian 
%that are 
associated with this %responsible for the duality 
defect are %as follows: %\anc{Chinmay, this structure mimics what you have in Eq. (37) of the \texttt{working\_doc}, where I used $\widehat{X}$ instead of $X'$ notation}
\begin{align}
&\Delta H^{(D)}_{Q} = g^{-1} \sigma^x_{Q} Z_{Q}Z_{Q+\hat x}Z_{Q+ \hat y} \nonumber \\
& + \sum_{j > Q_x} \sigma^x_{(j,  Q_y)} Z_{(j, Q_y)} Z_{(j+1, Q_y)} \nonumber \\ 
& + \sum_{j > Q_y} \sigma^x_{(Q_x, j)} Z_{(Q_x, j)} Z_{(Q_x, j+1)}. \label{eq.D-defect} 
%= &-\sum_{j=Q_y}^{L_y} Z_{Q_x,j} Z_{Q_x,j+1} \widehat{X}_{\bar{Q_x},\bj} \nonumber\\
% &- \sum_{i=Q_x}^{L_x} Z_{i,Q_y} Z_{i+1,Q_y} \widehat{X}_{\bj,\bar{Q_y}} 
\end{align}
%, and 
The structure of each term in Eq.~\eqref{eq.D-defect} %corresponds to 
is %that of 
captured by the shaded triangles appearing in Figs.~\ref{fig:XM_combined}, %or Fig. 
\ref{fig:defects}. %The total number of terms is equal to the length of the interface $(l_x+l_y-1)$. 
%Extending 
Propelling the subsystem interface all of the way to the origin, i.e.,  %taking 
setting $Q_x=Q_y=0$ %would result in 
leads to the geometry %depicted 
appearing in Fig. \ref{fig:XM_combined} (a). %, where bulk extra sites can be traced out. 
%On the finite lattice, the boundary of the gauged region has length $L_x+L_y-1$, and the $X$ operators become the $\hat{\sigma}_x$ operators on the extra sites as we discussed earlier. 

%On the infinite cylinder ($L_y \to \infty$) geometry, we have the option of extending $Q_y \to -\infty $ which realizes an `infinite line' defect  (constructed by partial gauging, see \SM): 
%\begin{equation}
%    \Delta H^{(D)}_{\text{cyl},Q_x} = \sum^{\infty}_{j=-\infty} \sigma^x_{(\bar Q_x, \bar j)} Z_{(Q_x, j)} Z_{(Q_x, j+1)}.
%\label{eq.infinte_line_D_defect}
%\end{equation}
%These infinite line defects admit explicit unitary realizations of mobility and fusion with similar defects (End Matter). On the finite torus, bond-algebraic arguments guarantee the existence of a unitary realizing the duality automorphism on the extended theory and analogues of the cylinder unitaries exist, yet an interpretation of the intermediate ``defect Hamiltonians" in this case is more delicate (see \SM).

\paragraph*{Defects of subsystem symmetries.}
It is natural to consider the defects %of 
associated with 
%(i.e., whose presence generally restores) 
the subsystem symmetries ~\cite{Nussinov-PNAS2009,Nussinov-Annals2009}, which we call ``$\eta$-defects'' following the convention of Ref.~\onlinecite{seiberg-LSM} in analogy to the TFIC. The subsystem symmetry defects are domain walls within a given row, with the parent Hamiltonian
\begin{eqnarray}
H^\text{row}_{(\bi,j)} &=& Z_{(i,j)}Z_{(i+1,j)}Z_{(i,j-1)}Z_{(i+1,j-1)} \nonumber \\ 
&+& Z_{(i,j+1)}Z_{(i+1,j+1)}Z_{(i,j)}Z_{(i+1,j)} \label{eq.eta-hor}\\ 
 &-& \sum_{q \neq (i,j),(i,j+1)} Z_q Z_{q+\hat x} Z_{q+\hat y} Z_{q+\hat x + \hat y} - g \sum_{q} X_{q}.\nonumber
\end{eqnarray}
The defect is centered between two columns of the original lattice, at  $\bi \equiv i+1/2$. 
%The 
An analogous expression can be written for the defect Hamiltonian of the vertical $\eta$-defect $H^\text{col}_{i,\bj}$. 
On the infinite lattice, $\mathbb{Z}^2$, the corresponding state is obtained from the ground state of the Xu--Moore model by acting with a line defect operator, 
\begin{eqnarray}
\widehat{\eta}^\text{row}_{\bi,j} = \prod_{k>\bi} X_{k,j} .
\end{eqnarray}
Similarly, the defect Hamiltonian, $H^\text{row}_{(\bi,j)}$, can be realized as $H^\text{row}_{(\bi,j)}=\widehat \eta^\text{row}_{\bi,j}H_{\sf XM} \widehat \eta^\text{row}_{\bi,j}.$
%% MAKE SURE TO LABEL \eta defects by "row/col", not "hor/vert"
One can move these defects within a given row (column) by acting with a local $X$ operator. The motion in the perpendicular direction cannot, however, be enacted by a local operator. 

Note that these $\eta$-defects create a pair of defect $Z$-plaquettes, as is evident from Eq.~(\ref{eq.eta-hor}). 
%\textcolor{red}{Can one create a single defective plaquette centered at a site $(\bi,\bj)$ of the dual lattice? The answer is yes, by acting with a membrane operator defined by a product of $\hat{X}$ as follows}
On an infinite lattice with open boundary conditions (with sites numbered $i\in (-\infty,\infty)$ in both directions), a defect on a single plaquette can be created by acting on a ground state with a membrane product 
%of $\hat{X}$ as follows
% \beq
% |f_{i,j}\rangle = \prod_{k>\bi}\prod_{l>\bj} X_{k,l} \gs.
% \label{eq.fracton}
% \eeq
\beq
\widehat{f}_{i,j} = \prod_{k>\bi}\prod_{l>\bj} X_{k,l}.
\label{eq.fracton}
\eeq
%\textcolor{red}{Here we have considered open boundary conditions for simplicity.} 
%Note that in 
In a system with PBC, such an operator would instead create four defect plaquettes situated on the corners of the membrane.
This resembles the fracton models in (3+1)D, such as the X-cube model \cite{Vijay2016}, where a quartet of fractons is similarly created by a membrane operator. 
While fracton phases of matter can only be rigorously defined in spatial dimensions $d\geq 3$, we shall nevertheless call the defects defined in Eq.~\eqref{eq.fracton} \textit{fractons}. An isolated fracton is immobile in a sense that %the movement 
its motion cannot be %enacted with 
generated by a local operator.

\paragraph*{Defect fusion rules.} At the level of product of operators, %the following statement is true:
\begin{eqnarray} 
\widehat{f}_{\bi,\bj-1} \widehat{f}_{\bi,\bj} = \widehat{\eta}^{\text{row}}_{\bi,j}; \quad \widehat{f}_{\bi-1,\bj} \widehat{f}_{\bi,\bj} = \widehat{\eta}^{\text{col}}_{i,\bj}.
\end{eqnarray}
%which 
This can be %interpreted 
viewed as a fusion rule for the corresponding fracton symmetry defects. We thus write, %written 
schematically,
%as 
\begin{eqnarray} 
f_{\bi,\bj-1}\times f_{\bi,\bj} = \eta^{\text{row}}_{\bi,j}; \quad f_{\bi-1,\bj}\times f_{\bi,\bj} = \eta^{\text{col}}_{i,\bj} .
\end{eqnarray}
The physical interpretation of these relations is that
a pair (dipole) of fractons on nearby vertical or horizontal plaquettes is nothing but an $\eta$-defect 
\footnote{Borrowing from the terminology of X-cube model, one may call these $\eta$-defects planons, as their motion is restricted in the plane perpendicular to the dipole moment, in this case along the $\eta$-line direction.} defined earlier in Eq.~\eqref{eq.eta-hor}, and we could say that a pair of fractons fuse into an $\eta$-defect, as illustrated  in Fig.~\ref{fig:defects}(a).
%
%OPERATOR RELATIONS BELOW
Of course when taken on the same plaquette, two fractons fuse into the (defect-free) identity: $f_{\bi,\bj} \times f_{\bi,\bj} = 1$. Similarly, 
\beq
\eta^{\text{row}} \times \eta^{\text{row}} = 1; \quad \eta^{\text{col}} \times \eta^{\text{col}} = 1.
\eeq
Here, we have suppressed the indices marking  the position of the $\eta$-defects (being topological, these defects are translationally invariant).

%\begin{figure}[tb]
%\includegraphics[width=0.48\textwidth]{fig-defects}
%\caption{Defect fusion. (a) A pair of fractons fuse into a horizontal $\eta$-defect: $f \otimes f = \eta^{\text{row}}$. 
%(b), (c) The duality defect `absorbs' a fracton or an $\eta$-defect, respectively: $\mathcal{D}\otimes f = \calD$, $\calD\otimes \eta = \mathcal{D}$. 
%(d) Fusion of two duality defects that differ by one row, see text. 
%}
%\label{fig:defects}
%\end{figure}

%First replacement tikz fig
%First replacement tikz fig

We now turn to the fusion rules involving the duality defect $\mathcal{D}$. The form of the defect Hamiltonian is obtained by partial gauging a portion of the lattice and the defect term corresponds to the perimeter of the chosen region - Eq.~\eqref{eq.D-defect} is the defect term for a rectangular region.
%
%\{ [CG] - I realize that the discussion in this section is actually completely general because of the later notation we use when talking about duality defect fusion. I think we can take out this statement - the line defect Equation will beintroduced in the end matter. \}}
This duality defect can be fused with a fracton resulting again in a duality defect, shown schematically in Fig.~\ref{fig:defects}(a).  To see this, note that a Hamiltonian defect of a fracton ($+(ZZZZ)$, schematically) commutes with the parent Hamiltonian of the duality defect in Eq.~\eqref{eq.D-defect}, except at the corners of the $X$-membrane. 
%\{ [CG] : I removed the operator statement of $Df = fD = D$ as it is not technically correct for fractons, a closed \textit{operator} fusion algebra is formed only if we consider D and the subsystem symmetry operators (not defect creation operators). \} }
%Another way to see this is by explicitly acting with the the membrane operator, creating a fracton, on $\widehat{D}$ in Eq.~\eqref{eq.D-operator}, which yields, at the operator level, $\widehat{D} \widehat f = \widehat f\widehat{D} = \widehat{D}$.
%
Similarly, one can fuse the $\eta$ line defect with the duality defect, as illustrated in Fig.~\ref{fig:defects}(b). 
\begin{comment}
We write the most general fusion operators that achieve this in the \SM but generically they take the form (for a vertical $\eta$-defect, see Fig.~\ref{fig:defects}):
\begin{eqnarray}
    \lambda(\eta^\text{col},\mathcal{D}) = (\prod_{j\in\eta^\text{col}} X_j)\sigma^z_L\sigma^z_R
\end{eqnarray}
where the product of $X$ operators has support on a rigid line (wavy blue in Fig.~\ref{fig:defects}) connecting the $\eta$-string to the duality defect (the two $\sigma^z$ operators shaded in  Fig.~\ref{fig:defects}b).
\end{comment}
%
Defect fusion can be written in terms of the Hilbert space fusion rules involving the corresponding defects:
\beq
\label{eq:Dxf}
\calD \times f = \calD; \quad \calD\times \eta^{\text{row}} = \calD; \quad 
\calD\times \eta^{\text{col}} = \calD.
\eeq

%The statement above implies that fractons and $\eta$-defects may be annihilated by $\mathcal{D}$ via the application of a \textit{unitary} fusion operator. 
Note that the Hilbert space of the $\cal{D}$-defect plus an $\eta$-defect is unitarily connected (without gauging/lifting the space) to the Hilbert space which hosts only a $\cal{D}$-defect. 

The most interesting process is the fusion of two duality defects, i.e. $\mathcal{D} \times \mathcal{D}$.
%, \an{where the symbol $\times$ now denotes the Hilbert-space representation fusion, akin to the operator product expansion in CFT. 
We next formulate the meaning of this fusion in a Hamiltonian-specific fashion.
%{\color{blue} Fusion of  duality defects associated with a subsystem of the XM lattice that excludes only its {\it boundaries} (i.e., omits plaquettes at its edge) will produce a defect term along those boundaries. Such a defect is unitarily equivalent to a minimal gauging of the boundary plaquettes.}
% (i.e., the smallest set of gauge operators needed to enforce consistency of the bond algebra at the subsystem boundaries).}
The Hilbert space $\mathcal{H}_{\mathcal{D}; \mathcal{D}}$ of two $\cal{D}$ defects is precisely the lifted space we have constructed with the extra $\mathcal{N}$ sites in $\bar S$. On a torus, $\mathcal{N}=L_x+L_y-1$ whereas on an infinite cylinder with $L_y\to \infty$, $\mathcal{N}$ is the cardinality of the integers). Thus, 
\beq
\mathcal{H}_{\mathcal{D}; \mathcal{D}} \simeq \mathbb{C}^{2 \mathcal{N}} \otimes \mathcal{H} 
%\otimes_{q\in S} \mathcal{H}_q  
%\otimes \mathcal{H}_1 \otimes \ldots \otimes \mathcal{H}_\mathcal{N} 
\simeq \mathcal{H}  \oplus \mathcal{H}_{\perp}.
\eeq
where $\mathcal{H}_{\perp}$ contains the additional copies of the original Hilbert space in each of the $\mathcal{N}$ twist sectors. Label the quantum configurations in these sectors by $a \in \left \{ 0,1, \cdots 2^{\mathcal{N}} - 1 \right \}$ where $a=0$ denotes the physical (untwisted) sector~\footnote{Note that the set of possible values for $a$ is uncountable on the infinite cylinder in spite of the suggestive notation used here}.
%i.e. there exists a unitary  operation taking a pair of line defects into the sum over all possible twisted sectors at the boundary. 

Henceforth, we focus on the case of the infinite cylinder due to the simplicity in describing and constructing unitary realizations of fusion and movement operators (see \textit{End Matter}). 
%for the `infinite line' duality defect which takes the form (taking $L_y\to\infty$ in Eq.~\eqref{eq.D-defect}),
%\begin{equation}
%    \Delta H^{(D)}_{\text{cyl},Q_x} = \sum^{\infty}_{j=-\infty} \sigma^x_{( Q_x,  j)} Z_{(Q_x, j)} Z_{(Q_x, j+1)}.
%    \label{eq.infinte_line_D_defect}
%\end{equation}}
The fusion rule of two $\mathcal{D}$-defects becomes
\begin{eqnarray}
\label{eq:DxD_fusion}
\mathcal{D} \times \mathcal{D} %= \mathcal{T^{-}} \mathcal{P}_{+} 
= 
\bigotimes_{q = 1}^{\mathcal{N}} (1\oplus\eta^\text{row}_{q})~. 
%&\mathcal{D} \otimes \mathcal{D} = \left[ \bigotimes_{\text{row} \ L_x>j>0} (1 \oplus f_{\bar j,\bar 0}) \right] \left[ \bigotimes_{\text{col} \ L_x>k>0} (1 \oplus f_{\bar 0,\bar k}) \right] \bigotimes (1 \oplus f_{\bar 0, \bar 0}).
\end{eqnarray}
Here, what we mean on the r.h.s. when written explicitly, is a statement about the equivalence of Hamiltonians:
%operators 
%(below, $q=1\ldots \mathcal{N}$ labels the additional sites and $a$ enumerates the corresponding twist sectors)
\beq
%\lambda H_{\mathcal{D}\times \mathcal{D}} \lambda^{-1} =   \bigoplus_{a=0}^{2^\mathcal{N}-1} \left( H_{a} \otimes_{q} \ket{b^a_{q}}\bra{b^a_{q}}_{q} \right),
\lambda H_{\mathcal{D}; \mathcal{D}} \lambda^{-1} = H \otimes \widehat{P}_{+} + 
\sum_{a=1}^{2^\mathcal{N}-1} \left( H_{a} \otimes  \ket{a}\bra{a} \right),
\label{eq:D-fusion}
\eeq
where $H_a$ is a parent Hamiltonian of the $\eta$-defects in the corresponding configuration $\{a\}$ of auxiliary sites, and $\lambda$ denotes the duality defect fusion operator $\lambda\equiv\mathcal{U}^{\dagger}_s$, derived in the \textit{End Matter}.
%\beq
%\lambda H_{\mathcal{D}\otimes \mathcal{D}} \lambda^{-1} = H \otimes_{q} \ket{0}\bra{0}_q  \bigoplus_{a=1}^{2^\mathcal{N}-1} \left( H_{a} \otimes_q \ket{b^a_q}\bra{b^a_q}_q \right),
%\eeq
%Above, $b^a_q = 0 \ \text{or} \ 1$ depending upon whether the extra site $q$ is projected into the $\sigma^z = +1$or $-1$ in a given sector $a$. 
The operator $\widehat{P}_{+}$ entering above is the same as in Eq.~\eqref{eq.D-operator}, which  projects all additional sites onto $\sigma^z=+1$ sector, i.e. $\widehat{P}_{+} (\lambda H_{\mathcal{D}; \mathcal{D}} \lambda^{-1}) \widehat{P}_{+} = H \bigotimes \ket{0}\bra{0}$.

\begin{comment}
Consider the example of Fibonacci anyons
\beq
\eta \otimes \eta = 1 \oplus \eta
\eeq
written explicitly, this means
\beq
\lambda H_{\eta \otimes \eta} \lambda^{-1} = H_\mathbb{I}\otimes |0\rangle\langle0| + H_\eta \otimes |1\rangle\langle 1|
\eeq
Is it not true that
\beq
\eta \otimes 1 = \eta \Longleftrightarrow \lambda H_{\eta \otimes 1} \lambda^{-1} = H_\eta
\eeq
\end{comment}

%Where the specific form of $\mathcal{P}_{+}$ depends on the boundary conditions of the theory (see \SM).  

%In the \SM we demonstrate that the mobility operators for the cylinder can be generalized to implement the duality on the finite torus -- however, the structure of the defect lines suffer local ``tears" upon movement.  \textcolor{cyan}{(need to ask Chinmay about boundary conditions - [CG] : the old version of this paragraph was written before the interpetation of "defect tears" on the torus. I will modify this section to reflect this)}.
%Consider two such defects that differ by one row, as in Fig.~\ref{fig:defects}(d) -- this is without loss of generality, since a given $\calD$-defect can always be moved to a desired position on a lattice, see \SM.

% Demonstrate that (denoting the dual-D operator $D^* \equiv  T_{(1,1)} D$ )
% \beq
% \mathcal{D} \otimes \mathcal{D}^* = 1 \oplus \eta
% \eeq

\paragraph*{Conclusions and outlook.}

We demonstrated that bond-algebraic methods~\cite{PhysRevB.79.214440,Cobanera2010,Cobanera2011}, when combined with the generalized Wigner theorem~\cite{GWT}, offer a systematic and transparent way to construct  NSOs directly on the lattice. Although these ideas can already be illustrated in the familiar setting of the TFIC, their full power becomes apparent in higher dimensions. We therefore focused on a (2+1)-dimensional system with rigid symmetries, using the Xu–Moore model as a concrete example.

Within this framework, we explicitly constructed the symmetry-defect operators and determine their fusion rules. An important lesson that emerges is that these fusion rules depend sensitively on the choice of boundary conditions (see also Ref.~\cite{XM-dualities-maity2025}), highlighting the subtle interplay between global structure and local algebraic relations. Our approach is entirely algebraic and self-contained: it does not assume the presence of an underlying conformal field theory, nor does it rely on prior knowledge of whether a fusion category exists.

These results are especially noteworthy because systems with rigid higher symmetries (``subsystem symmetries'') 
%-- unlike those with topological symmetries -- 
do not admit a topological quantum field theory description of their vacua. Nevertheless, by analyzing the fusion of defect operators, one can still extract meaningful information about the universality class of the critical phase.

Finally, we showed that the non-invertible duality can be naturally expressed as a sequential quantum circuit. This perspective not only clarifies the structure of the duality but also points toward possible applications in quantum information and quantum computing. More broadly, our work provides a foundation for exploring non-invertible dualities in higher-dimensional lattice systems.

%We have demonstrated that the bond-algebraic considerations, when taken together with the generalized Wigner theorem~\cite{GWT}, provide a powerful framework to derive an explicit lattice-resolved form of the non-invertible NSO. While also applicable to the well studied case of the Ising chain, in this work we apply these tools to the richer 2+1D case with subsystem symmetries, exemplified by the Xu--Moore model. We derive the symmetry defect operators and their fusion rule, which do depend sensitively on the choice of the boundary conditions. The derivation is self-contained in that it is agnostic of any knowledge of the underlying conformal field theory (if one is present) or \textit{a priori} information about the corresponding fusion category. Our formulation of the duality as a sequential quantum circuit lends itself to possible applications in quantum computing.  The present work lays the foundation for future applications of non-invertible dualities in higher-dimensional systems. 

%\textcolor{red}{Mention that no prior knowledge of underlying CFT of Xu-Moore model is required for Bond-Algebraic derivation of fusion rules.} We should close with a discussion about NSO and Wigner's theorem fixing their general form. Discuss the role of boundary conditions. We need to mention the calculation of dual variables. We need to define topological charges. On the finite torus, bond-algebraic arguments allow for analogues of the cylinder unitaries, yet an interpretation of the intermediate ``defect Hamiltonians" is more delicate (see \SM).

\begin{acknowledgements}
The authors thank P. Gorantla for insightful discussions. This work was principally supported by the Department of Energy under the Basic Energy Sciences award no. DE-SC0025047. (A.H.N.). G.O. gratefully acknowledges support from the Institute for Advanced Study. Part of this work was initiated at the Aspen Center for Physics, which is supported by National Science Foundation grant PHY-2210452.
\end{acknowledgements}

\section*{End Matter}

\subsubsection*{Non-invertibility of $\widehat D$}

The NSO $\widehat{D}$ acts projectively as a half-translation on a transverse field term $X_q$, Let $B_q \equiv Z_q Z_{q+\hat x}Z_{q+\hat y}Z_{q+\hat x + \hat y}$.
\begin{align}
& \widehat D^{\dagger} X_{q + \hat x + \hat y} \widehat D = B_q \widehat P_{+} \label{eqn:bare_X_map} \\
& \widehat D^{\dagger} B_q \widehat D = 
\left \{
\begin{array}{ll}
   X_q \widehat P_{+} & q \in S \\ \\
   \left( \prod_{n=1}^{L_a-1} X_{q+n \hat a} \right) \widehat P_{+}  & q \in S^{a'} \ \ \nonumber
\end{array} 
\right. \\
& (a,a'=x,y \ ; a \neq a') \nonumber
\label{eqn:bare_B_map}
\end{align}
%where we introduced another 
Define the projector %, this time acting on the original Hilbert space:
\begin{equation}
    \widehat{P}_{\eta} =  \prod_{a \in S}\left( \frac{1+\Phi^{-1}_{\sf XM}(\sigma^{z}_a)}{2} \right) = \mathcal{U} \ \widehat{P}_{+} \ \mathcal{U}^{\dagger}
\end{equation}
which operates non-trivially within  the original (physical) Hilbert space %\footnote{This equation is true even for bare plaquette terms at the \textit{boundary} because any ancilla sites inserted alongside $B_q$ in Eq.(\ref{eqn:bare_plaquette_map}) produce the same result, due to the projectors $\widehat P_+$.}. 
Upon conjugation by $\widehat D^2$ the local terms transform as :
% $T_{\bft}$ is the translation operator by a vector $\bft$ and we express the action of  : 
\begin{align}
    & (\widehat D^{\dagger})^2 B_q \widehat D^2 = B_{q-\hat x - \hat y} \widehat P_{\eta} \widehat P_{+} = (\widehat P_{\eta} T^{\dagger}_{(1,1)}) B_q (\widehat P_{\eta} T^{\dagger}_{(1,1)})^{\dagger} \widehat P_{+}, \nonumber \\
    & (\widehat D^{\dagger})^2 X_q \widehat D^2 = X_{q - \hat x - \hat y} \widehat P_{\eta} \widehat P_{+
    } = (\widehat P_{\eta} T^{\dagger}_{(1,1)}) X_q (\widehat P_{\eta} T^{\dagger}_{(1,1)})^{\dagger} \widehat P_{+}, \label{eqn:D_squared_anomaly}
\end{align}

which act non-invertibly on physical states owing to the projector $\widehat P_{\eta}$.

\subsubsection*{A brief review of the TFIC duality}

To
elucidate the structure of the %reveal the structure of %the %duality defect terms, 
%defects associated with the 
duality, we perform a so-called {\it partial-gauging} procedure, in which gauge degrees of freedom are introduced into a {\it portion} of the lattice of co-dimension 1 at a fixed time slice. 
%It may be useful to 
%We briefly remind the reader the essence of this procedure in the 1+1 D Ising model, following \cite{seiberg-LSM}: 
The essentials of this procedure mirror those in the TFIC \cite{seiberg-LSM}.
In the TFIC, one introduces $\mathbb{Z}_2$ gauge fields (Pauli operators) $\{\tilde Z_\bq\}$ on %the 
lattice links and imposes the Gauss law constraint by projecting into the gauge symmetry sector $\tilde{Z}_{\bq-1} \sigma^x_q \tilde{Z}_{\bq} = 1$. Performing this on a sub-system, say to the right of the site $Q$ on an infinite chain, creates a {\it duality defect}  at the interface $q=Q$. Choosing gauge invariant Pauli operators $\widehat X, \widehat Z$ both in the bulk and, crucially, also in the boundary (by dressing the matter fields $\sigma^{x/z}$ appropriately with gauge fields $X,Z$ so as to preserve the Pauli algebra and commute with gauge symmetries) %results in the respective defect Hamiltonian of the form
leads to the defect Hamiltonian 
\begin{align} 
%H^{(D-\text{TFI})}_{\bar{Q}} &= H_{\text{TFI}}(g,q<Q)|_{\sigma^x,\sigma^z} + H_{\text{TFI}}(g^{-1},q>Q)|_{\widehat X, \widehat Z} \nonumber \\
H^{(D-\text{TFI})}_{\bar{Q}} &= H_{\text{TFI}}(g,q<Q) + H_{\text{TFI}}(g^{-1},q>Q) \nonumber \\
&- g^{-1}\widehat{X}_{Q} \sigma^z_Q - g\sigma^x_Q.
\end{align}
%Specializing to 
At the self-dual point $g=1$, the local term in the Hamiltonian creating the duality defect can be written in the form 
$\Delta H^{(D)}_{\text{TFI}} = -\widehat{X}_{\bar{Q}} \sigma^z_Q$
\footnote{This form is equivalent to the notation $\Delta H^{(D)}_{J,J+1} = -\hat{X}_{J+1/2}Z_J$ 
in Ref.~\cite{seiberg-LSM}, where after relabeling $\hat{X}_{J+1/2} \longrightarrow X_{J+1}$ the authors write down the duality Hamiltonian as $-Z_{J}X_{J+1}$. Applying to the case $J=L$, one obtains $H^{(D)}_{L,1}=-Z_LX_1$ quoted in Ref.~\cite{seiberg-LSM}.}.

%\begin{figure}[t]
%\includegraphics[width=0.4\textwidth]{fig-XM-gauging}
%\caption{Schematics of the half-gauging procedure. The original, ungauged lattice sites (black) and the gauged sites on the dual lattice (red circles) are separated by the interface forming the duality defect (shaded).}
%\label{fig:XM-gauging}
%\end{figure}

\subsubsection*{Structure of the Unitary : Infinite Cylinder}

Consider infinite line defects on the infinite cylinder ($L_y \to \infty$ in Eq.~\eqref{eq.D-defect}) described by the defect term : 
\begin{equation}
    \Delta H^{(D)}_{\text{cyl},Q_x} = \sum^{\infty}_{j=-\infty} \sigma^x_{( Q_x,  j)} Z_{(Q_x, j)} Z_{(Q_x, j+1)}.
    \label{eq.infinte_line_D_defect}
\end{equation}
There exist unitary realizations of fusion and movement operators for these defects, described in this section. This enables an explicit construction for the unitary $\mathcal{U}$ where the NSO $\widehat{D} = \mathcal{U} \widehat{P}_{+}$. We derive below the expression for the desired unitary in the form
\begin{equation}
\label{eq:u_structure}
\mathcal{U} := \mathcal{U}_f \,\mathcal{U}_{\text{bulk}} \,\mathcal{U}_{s} 
\end{equation}
the operator $\mathcal{U}_s$ and $\mathcal{U}_f$ only act on/near the boundaries.

The Hamiltonian upon which the NSO acts reads as follows: 
\begin{align}
\label{eqn:cylinder_hamiltonian_xm_twisted}
&\tilde H_{c}[g] = \sum^{q_x=0}_{-\infty < q_y < \infty} \left( g^{-1} Z_{q} Z_{q+\hat x } Z_{q+\hat y } Z_{q + \hat x + \hat y} \sigma^z_{\bar q } + g X_q \right) \nonumber \\
&+ \sum^{0 < q_x < L_x}_{ -\infty < q_y < \infty} \left( g^{-1} Z_{q} Z_{q+\hat x } Z_{q+\hat y } Z_{q + \hat x + \hat y} + g X_q \right).
\end{align}
such that
\begin{align}
\label{eqn:xu_moore_duality_on_cyl}
& \mathcal{U} \tilde H_{c}[g] \ \mathcal{U}^{\dagger} = \tilde H_{c}[g^{-1}].
\end{align}
Each term in the product is a unitary operator supported along $-\infty < q_y < \infty$, at a fixed $q_x$. $\mathcal{U}_s$ is interpreted as the ``start" operator, $\mathcal{U}_{k,k+1}$ the ``movement" operators and $\mathcal{U}_{f}$ the ``finish" operator. This structure of $\mathcal{U}$ closely mimics the construction of $\mathcal{U}$ carried out in \cite{seiberg-LSM} for the (1+1)D TFIC. 
\\
\\
We next fix some notation. For any local operator $A_q$, we denote the infinite product along any fixed $q_x=i$ vertical, ordered by increasing $j$ from the right by 
\begin{equation}
    \label{eq:product_notation}
    \prod^{\infty}_{j= -\infty} A_{(i,j)} \equiv \cdots A_{(i,1)} A_{(i,0)} A_{(i,-1)} \cdots \equiv \Gamma_{i}[A_q]~.
\end{equation}
The Hamiltonian of Eq.(\ref{eqn:cylinder_hamiltonian_xm_twisted}) must be unitarily equivalent to a defect Hamiltonian, with left and right defect lines $\mathcal{D}_{R/L}$ obtained by partial gauging of the portion of the original lattice located between these defect lines. Further, we find these defects to be rigid yet effectively 'movable' along $(\pm1,0)$, sweeping out a duality transformed Xu-Moore lattice with each step - the local terms are still ``half translated" along the diagonal $(\pm 1,\pm 1)$ (with signs chosen depending on how the unitaries are constructed). The operators implementing this are the explicit unitary maps between the Hilbert spaces of Hamiltonians with different relative positions of the $\mathcal{D}_R$ and $\mathcal{D}_L$ defects.

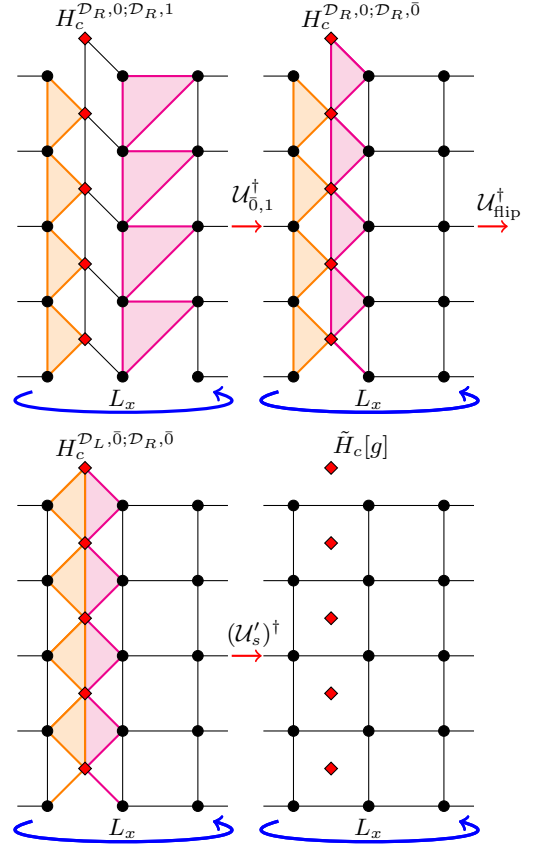
\begin{figure}[t!]
    \centering
    \begin{tikzpicture}
        \def\cols{3}
        \def\rows{3}
            \foreach \y in {-1,...,\rows}
            {
                    \draw (0.5,\y+0.5) -- ++ (0.5,-0.5);
                    \ifnum \y > -1
                        \draw (0.5+0.5,\y) -- ++(0,-1);
                        \draw (0.5,\y+0.5) -- ++(0,-1);
                        \filldraw[fill = magenta!20, draw = magenta,thick] (2,\y) -- ++ (-1,0) -- ++(0,-1) -- ++(1,1);
                    \fi
                    \ifnum \y < 3
                    \filldraw[fill = orange!20, draw = orange,thick] (0.5,\y+0.5) -- ++ (-0.5,0.5) -- ++(0,-1) -- ++(0.5,0.5);
                    \fi
            }
        %Add extra sites
                \foreach \y in {-1,...,\rows} 
                {
                    \node[draw=black, fill=red, diamond, minimum size=5pt, inner sep=0pt] 
                    at (0.5,\y+0.5){ };
                }
                \foreach \y in {-1,...,\rows} 
                {
                    \node[draw=black, fill=black, circle, minimum size=4pt, inner sep=0pt] 
                    at (1,\y){ };
                }
            \foreach \y in {-1,...,3}
            {
                \node[draw=black, fill=black, circle, minimum size=4pt, inner sep=0pt] at (0,\y) { };
                \draw (0,\y) -- ++(-0.4,0);
                \node[draw=black, fill=black, circle, minimum size=4pt, inner sep=0pt] at (2,\y) { };
                \draw (2,\y) -- ++(0.4,0);
                \ifnum \y < \rows
                \draw (2,\y) -- ++(0,1);
                \fi
            }
            \draw[blue,->,very thick] (-0.2,-1.2) to[out=195, in=345, looseness=1.5] (0.2+\cols-1,-1.2);
            \draw node at (1,-1.3) {$L_x$};
             \draw[->,red,thick] (2.45,1)--(2.85,1);
        \node at (2.75,1.4) {$\mathcal{U}_{\bar 0,1}^{\dagger}$};
        \node at (0.9,3.8) {$H^{\mathcal{D}_R,0 ; \mathcal{D}_R,1}_{c}$};
    \end{tikzpicture}
    \hspace{-2.1cm}
    \begin{tikzpicture}
        \def\cols{3}
        \def\rows{3}
            \draw[magenta,thick] (1,-1) -- ++(-0.5,0.5);
            \foreach \y in {-1,...,\rows}
            {
                    \filldraw[fill = magenta!20, draw = magenta,thick] (0.5,\y+0.5) -- ++ (0.5,-0.5);
                    \ifnum \y > -1
                       \filldraw[fill = magenta!20, draw = magenta,thick] (1,\y) -- ++(-0.5,-0.5) -- ++(0,1) -- ++(0.5,-0.5);
                    \fi
                    \ifnum \y < 3
                    \filldraw[fill = orange!20, draw = orange,thick] (0.5,\y+0.5) -- ++ (-0.5,0.5) -- ++(0,-1) -- ++(0.5,0.5);
                    \fi
            }
        %Add extra sites
                \foreach \y in {-1,...,\rows} 
                {
                    \node[draw=black, fill=red, diamond, minimum size=5pt, inner sep=0pt] 
                    at (0.5,\y+0.5){ };
                }
            \foreach \y in {-1,...,\rows}
            {
                \node[draw=black, fill=black, circle, minimum size=4pt, inner sep=0pt] at (0,\y) { };
                \draw (0,\y) -- ++(-0.4,0);
                \node[draw=black, fill=black, circle, minimum size=4pt, inner sep=0pt] at (1,\y) { };
                \draw (1,\y) -- ++(1,0);
                \node[draw=black, fill=black, circle, minimum size=4pt, inner sep=0pt] at (2,\y) { };
                \draw (2,\y) -- ++(0.4,0);
                \ifnum \y < \rows
                \draw (1,\y) -- ++(0,1);
                \draw (2,\y) -- ++(0,1);
                \fi
            }
            \draw[blue,->,very thick] (-0.2,-1.2) to[out=195, in=345, looseness=1.5] (0.2+\cols-1,-1.2);
            \draw node at (1,-1.3) {$L_x$};
            \draw[blue,->,very thick] (-0.2,-1.2) to[out=195, in=345, looseness=1.5] (0.2+\cols-1,-1.2);
            \draw[->,red,thick] (2.45,1)--(2.85,1);
        \node at (2.75,1.3) {$\mathcal{U}_{\text{flip}}^{\dagger}$};
        \node at (0.9,3.8) {$H^{\mathcal{D}_R,0 ; \mathcal{D}_R,\bar 0}_{c}$};
    \end{tikzpicture}
    \hspace{-1.8cm}
    \begin{tikzpicture}
        \def\cols{3}
        \def\rows{3}

            \draw[magenta,thick] (1,-1) -- ++(-0.5,0.5);
            \foreach \y in {-1,...,\rows}
            {
                    \ifnum \y > -1
                       \filldraw[fill = magenta!20, draw = magenta,thick] (1,\y) -- ++(-0.5,-0.5) -- ++(0,1) -- ++(0.5,-0.5);
                    \fi
                     \draw[orange,thick] (0.5,\y+0.5) -- ++ (-0.5,-0.5); 
                    \ifnum \y > -1 
                        \filldraw[fill = orange!20, draw = orange,thick] (0,\y)-- ++(0.5,-0.5) -- ++(0,1) -- ++(-0.5,-0.5);
                        
                    \fi
            }
        %Add extra sites
                \foreach \y in {-1,...,\rows} 
                {
                    \node[draw=black, fill=red, diamond, minimum size=5pt, inner sep=0pt] 
                    at (0.5,\y+0.5){ };
                }
            \foreach \y in {-1,...,\rows}
            {
                \node[draw=black, fill=black, circle, minimum size=4pt, inner sep=0pt] at (0,\y) { };
                \draw (0,\y) -- ++(-0.4,0);
                \node[draw=black, fill=black, circle, minimum size=4pt, inner sep=0pt] at (1,\y) { };
                \draw (1,\y) -- ++(1,0);
                \node[draw=black, fill=black, circle, minimum size=4pt, inner sep=0pt] at (2,\y) { };
                \draw (2,\y) -- ++(0.4,0);
                \ifnum \y < \rows
                \draw (0,\y) -- ++(0,1);
                \draw (1,\y) -- ++(0,1);
                \draw (2,\y) -- ++(0,1);
                \fi
            }
            \draw[blue,->,very thick] (-0.2,-1.2) to[out=195, in=345, looseness=1.5] (0.2+\cols-1,-1.2);
            \draw node at (1,-1.3) {$L_x$};
            \draw[blue,->,very thick] (-0.2,-1.2) to[out=195, in=345, looseness=1.5] (0.2+\cols-1,-1.2);
            \draw[->,red,thick] (2.45,1)--(2.85,1);
        \node at (2.75,1.3) {$(\mathcal{U}_{s}')^{\dagger}$};
        \node at (0.9,3.8) {$H^{\mathcal{D}_L,\bar 0 ; \mathcal{D}_R,\bar 0}_{c}$};
    \end{tikzpicture}
    \hspace{-2.1cm}
       \begin{tikzpicture}
        \def\cols{3}
        \def\rows{3}
                \foreach \y in {-1,...,\rows} 
                {
                    \node[draw=black, fill=red, diamond, minimum size=5pt, inner sep=0pt] 
                    at (0.5,\y+0.5){ };
                }
            \foreach \y in {-1,...,\rows}
            {
                \node[draw=black, fill=black, circle, minimum size=4pt, inner sep=0pt] at (0,\y) { };
                \draw (0,\y) -- ++(-0.4,0);
                \draw (0,\y) -- ++(1,0);
                \node[draw=black, fill=black, circle, minimum size=4pt, inner sep=0pt] at (1,\y) { };
                \draw (1,\y) -- ++(1,0);
                \node[draw=black, fill=black, circle, minimum size=4pt, inner sep=0pt] at (2,\y) { };
                \draw (2,\y) -- ++(0.4,0);
                \ifnum \y < \rows
                \draw (0,\y) -- ++(0,1);
                \draw (1,\y) -- ++(0,1);
                \draw (2,\y) -- ++(0,1);
                \fi
            }
            \draw[blue,->,very thick] (-0.2,-1.2) to[out=195, in=345, looseness=1.5] (0.2+\cols-1,-1.2);
            \draw node at (1,-1.3) {$L_x$};
            \draw[blue,->,very thick] (-0.2,-1.2) to[out=195, in=345, looseness=1.5] (0.2+\cols-1,-1.2);

            \node at (0.9,3.8) {$\tilde H_{c}[g]$};
    \end{tikzpicture}
    \caption{Line Defect Movement and Fusion Steps are schematically indicated. The operator $\mathcal{U}$ on the arrow indicates that the Hamiltonian on the left of the arrow is to be conjugated as $\mathcal{U}()\mathcal{U}^{-1}$ to obtain that on the right.}
    \label{fig:vertical_line_defect_fusion}
\end{figure}

Let $C^{\sf{z} / \sf{x}}_{q_1,q_2}$ be the control z/x operators respectively, acting on sites $q_1,q_2$ and $H_q$ the Hadamard operator for site $q$.
\begin{equation}
\label{eqn:fusion_operator_cylinder}
\mathcal{U}'_{s} := \Gamma_{0} \left[ C^{\sf z}_{\bar q, q} C^{\sf z}_{\bar q, q+\hat y} C^{\sf z}_{\bar q, q + \hat x} C^{\sf z}_{\bar q, q + \hat x + \hat y} H_{\bar q} \right].
\end{equation}

The defect Hamiltonian is written as $\mathcal{U}'_{s} \tilde H_{c} \mathcal{U'}^{\dagger}_{s} = H^{\mathcal{D}_L,\bar 0 ; \mathcal{D}_R,\bar 0}_{c}$. We flip $\mathcal{D}_L \to \mathcal{D}_R$ for uniformity , i.e. $\mathcal{U}_{\text{flip}} H^{\mathcal{D}_L,\bar 0 ; \mathcal{D}_R,\bar 0}_{c} \mathcal{U}^{\dagger}_{\text{flip}} = H^{\mathcal{D}_R,0 ; \mathcal{D}_R,\bar 0}_{c}$, where 
% \begin{align}
% \label{eqn:defect_hamiltonian_0}
% & \mathcal{U}'_{s} \tilde H_{c} \mathcal{U'}^{\dagger}_{s} = H^{\mathcal{D}_L,\bar 0 ; \mathcal{D}_R,\bar 0}_{c} = \nonumber  \\ & \sum_{q_x=0} \left( g^{-1} \sigma^{x}_{q} + g  X_q \sigma^{z}_{q} \sigma^z_{q - \hat y} \right) \nonumber\\
% & + \sum_{q_x = 1} \left( g^{-1} Z_{q}Z_{q+\hat x }Z_{q+\hat y }Z_{q + \hat x + \hat y} + g  X_q \sigma^z_{q - \hat x} \sigma^z_{q - \hat x - \hat y}  \right) \nonumber \\
% & + \sum_{2 \leq q_x < L_x} \left( g^{-1} Z_{q}Z_{q+\hat x }Z_{q+\hat y }Z_{q + \hat x + \hat y} + g  X_q  \right) 
% \end{align}
 
% \begin{align}
% \label{eqn:defect_hamiltonian_flipped}
% & \mathcal{U}_{\text{flip}} H^{\mathcal{D}_L,\bar 0 ; \mathcal{D}_R,\bar 0}_{c} \mathcal{U}^{\dagger}_{\text{flip}} = H^{\mathcal{D}_R,0 ; \mathcal{D}_R,\bar 0}_{c} \nonumber \\
% & = \sum_{q_x=0} \left( g^{-1} Z_q Z_{q+\hat y } \sigma^{x}_{q} + g  X_q \right) \nonumber \\
% & + \sum_{q_x = 1} \left( g^{-1} Z_{q}Z_{q+\hat x }Z_{q+\hat y }Z_{q + \hat x + \hat y} + g  X_q \sigma^z_{q - \hat x} \sigma^z_{q - \hat x - \hat y}  \right) \nonumber \\
% & + \sum_{2 \leq q_x < L_x} \left( g^{-1} Z_{q}Z_{q+\hat x }Z_{q+\hat y }Z_{q + \hat x + \hat y} + g  X_q  \right) 
% \end{align}
 
\begin{equation}
    \label{eqn:flipping_DL_to_DR_AGAIN}
    \mathcal{U}_{\text{flip}} =  \Gamma_{0} \left[ C^{\sf z}_{\bar q , q} C^{\sf z}_{\bar q , q + \hat y} \right]~.
\end{equation}
\noindent
Expressions for the defect Hamiltonians $H^{\mathcal{D}_L,\bar 0 ; \mathcal{D}_R,\bar 0}_{c}$, $H^{\mathcal{D}_R,0 ; \mathcal{D}_R,\bar 0}_{c}$ etc. are supplied in \SM (\ref{sec:unitary_construction_torus} \ref{sec:unitary_infinte_cyl}).
We now define the ``start" operator $\mathcal{U}_s$ by 
\begin{equation}
    \mathcal{U}_s :=  \mathcal{U}_{\text{flip}} \mathcal{U}'_{s}.
\end{equation}
Note that this is precisely the operator $\lambda^{\dagger}$ used in fusing the two duality defects in Eq.~\eqref{eq:D-fusion}.

To move the defect line, we conjugate by :
\begin{align}
\label{eqn:defect_movement_unitary_1}
 & \mathcal{U}_{\bar 0,1} :=  \Gamma_{1} \left[ C^{\sf z}_{q,q + \hat x + \hat y} C^{\sf z}_{q,q+\hat y } \right] \Gamma_{1} \left[ H_q \right] \Gamma_{1} \left[ C^{\sf x}_{q,q - \hat y} \right]~.
\end{align}
$\mathcal{U}_{\bar 0,1}$ effectively moves the right defect line (magenta in Fig.(\ref{fig:vertical_line_defect_fusion})) to the right, creating plaquette terms involving the physical and extra sites.
% \begin{align}
% \label{eqn:defect_motion_in_hamiltonian}
% & \left( \mathcal{U}_{\bar 0,1} \mathcal{U}_{s} \right) H_{XM,cyl} \left( \mathcal{U}_{\bar 0,1} \mathcal{U}_{s} \right)^{-1} = \sum_{q_x=0} \left( g^{-1} Z_q Z_{q+\hat y } \sigma^{x}_{q} + g  X_q \right) \nonumber \\
% & + \sum_{q_x = 1} \left( g^{-1} X_{q} + g \sigma^z_{q - \hat x} \sigma^z_{q - \hat x - \hat y} Z_q Z_{q - \hat y} \right) \nonumber \\
% & + \sum_{q_x = 2} \left( g^{-1} Z_{q}Z_{q+\hat x }Z_{q+\hat y }Z_{q + \hat x + \hat y} + g  X_{q} Z_{q - \hat x} Z_{q - \hat  x - \hat y}  \right) \nonumber \\
% & + \sum_{3 \leq q_x < L_x} \left( g^{-1} Z_{q}Z_{q+\hat x }Z_{q+\hat y }Z_{q + \hat x + \hat y} + g  X_q  \right)~.
% \end{align}
This ``movement operator" is easily generalized when \\ $k \geq 2$ : 
\begin{equation}
\label{eqn:defect_movement_unitary_general}
 \mathcal{U}_{k-1,k} := \Gamma_{k} \left[ C^{\sf z}_{q,q + \hat x + \hat y} C^{\sf z}_{q,q+\hat y } \right] \Gamma_{k} \left[ H_q \right] \Gamma_{k} \left[ C^{\sf x}_{q,q - \hat y} \right]~.
\end{equation}
Therefore,
\begin{align}
\label{eqn:defect_motion_in_hamiltonian_full_sweep}
& \left( \mathcal{U}_{L_x-2,L_x-1} \cdots \mathcal{U}_{\bar 0,1} \mathcal{U}_{s} \right) \tilde H_{c} \left( \mathcal{U}_{L_x-2,L_x-1} \cdots \mathcal{U}_{\bar 0,1} \mathcal{U}_{s} \right)^{\dagger} \nonumber \\
& = H^{\mathcal{D}_R,0;\mathcal{D}_R,L_x-1}_{c} \nonumber 
% &= \sum_{q_x = 1} \left( g^{-1} X_{q} + g \sigma^z_{q - \hat x} \sigma^z_{q - \hat x - \hat y} Z_q Z_{q - \hat y} \right) \nonumber \\
% & + \sum_{2 \leq q_x \leq L_x-1} \left( g^{-1} X_q + g  Z_{q} Z_{q - \hat y} Z_{q - \hat x} Z_{q - \hat  x - \hat y}  \right) \nonumber \\
% & + \sum_{q_x=0} \left( g^{-1} Z_q Z_{q+\hat y } \sigma^{x}_{q} + g  X_{q} Z_{q - \hat x} Z_{q - \hat  x - \hat y} \right)
\end{align}

To close the loop, we translate both the $\mathcal{D}_R$ defects to the right then perform the fusion operation. I.e.
\begin{align}
    & \mathcal{U}_{0,\bar 0} H^{\mathcal{D}_R,0;\mathcal{D}_R,L_x-1}_{c} {(\mathcal{U}_{0,\bar 0})}^{\dagger} = H^{\mathcal{D}_R,\frac{1}{2},\mathcal{D}_R,L_x-1}_{c} \\
    & \text{crucially,} \nonumber \\
    & \mathcal{U}_{L_x-1,0} H^{\mathcal{D}_R,\frac{1}{2},\mathcal{D}_R,L_x-1}_{c} {(\mathcal{U}_{L_x-1,0})}^{\dagger} = H^{\mathcal{D}_R,0;\mathcal{D}_R,\frac{1}{2}}_{c}[g^{-1}] \\
    & \text{and hence} \ (\mathcal{U}_s)^{\dagger} H^{\mathcal{D}_R,0;\mathcal{D}_R,\frac{1}{2}}_{c}[g^{-1}] \mathcal{U}_s = \tilde H_{c}[g^{-1}].
\end{align}
Defining $\mathcal{U}_f = \mathcal{U}^{\dagger}_s\, \mathcal{U}_{L_x-1,0}\, \mathcal{U}_{0,\bar 0}$, we see that $\mathcal{U}$ is obtained as claimed in Eq.\eqref{eq:u_structure}, where $\mathcal{U}_\text{bulk} = \mathcal{U}_{L_x-2,L_x-1} \cdots \mathcal{U}_{\bar 0,1}$. The final form of the NSO therefore takes on the form as in Eq.~\eqref{eq.D-operator}:
\begin{equation}
    \widehat{D} = \left(\mathcal{U}_f\, \mathcal{U}_\text{bulk}\, \mathcal{U}_{s}\right) \,\widehat{P}_{+}.
\label{eqn:U_operator_cyl_explicit} 
\end{equation}
% \textcolor{blue}{The defect configurations obtained after each movement step are isomorphic to those generated by partial gauging appropriate infinite sub-regions \SM (\ref{sec:gauging} \ref{sec:defects_infinte_cyl}). As in the case of the 1+1 D TFIC, partial gauging describes all intermediate Hamiltonians in this process. We argue in \SM (\ref{sec:gauging} \ref{sec:partial_gauging}) that this does \textbf{not} carry over to the finite torus. Despite this difficulty, bond-algebraic reasoning enables an exact construction of the analogues of the unitaries making up $\mathcal{U}$, on the torus (see \SM (\ref{sec:unitary_construction_torus} \ref{sec:protein_and_potatoes}) for details). This permits an interpretation : the intermediate steps of moving these finite defects create local 'tears' in the (outer) defect line, which are essential to ensure unitary equivalence. $\mathcal{U}_f$ is qualitatively different on the torus : it 'fixes the tears' on the boundary to restore the required bond-algebraic isomorphism.}

\bibliography{refs,fusion,higher-D,bond-algebra}

\clearpage
\setcounter{equation}{0}
\setcounter{figure}{0}
\setcounter{table}{0}
\makeatletter
\renewcommand{\theequation}{S\arabic{equation}}
\renewcommand{\thefigure}{S\arabic{figure}}
\renewcommand{\bibnumfmt}[1]{[#1]}
\renewcommand{\citenumfont}[1]{#1}

\setcounter{section}{0}
\renewcommand\thesection{\Roman{section}}

\onecolumngrid

%\begin{widetext}
\begin{center}
	\Large{\textbf{Supplemental Material for ``\thistitle"}}
\end{center}
%\end{widetext}

\tableofcontents

\section{Bond Algebras and Minimal Gauging} \label{sec:bond_algebras_torus}
Consider the finite $L_x \times L_y$ Xu-Moore model with periodic boundary conditions.
%two_subfigures_________________________________________________________________
\begin{figure}[h!]
    \centering
    \begin{tikzpicture}
% Define colors
\definecolor{myblue}{RGB}{0,0,255}

% Draw the lattice
\foreach \i in {0,1,2,3,4,5} {
    \foreach \j in {0,1,2,3,4} {
        % Draw the nodes
        \node[circle, draw, fill=black, minimum size=4pt, inner sep=0pt] at (\i, \j) {};
        
        % Draw links (horizontal)
        \ifnum\j<4
            \draw (\i, \j) -- (\i, \j+1);
        \fi

        % Draw links (vertical)
        \ifnum\i<5
            \draw (\i, \j) -- (\i+1, \j);
        \fi
    }
}

% Color the bottommost row nodes
\foreach \i in {0,1,2,3,4,5} {
    \node[circle, draw, fill=myblue, minimum size=4pt, inner sep=0pt] at (\i, 0){};
}
%extra sites
% \foreach \i in {1,2,3,4,5}
% { 
% \node[diamond, draw, fill=red, minimum size=5pt, inner sep=0pt] at (\i-0.5, 0.5){};
% \node at (\i-0.7, 0.7) {$\sigma$};
% }
% Color the leftmost column nodes
\foreach \j in {0,1,2,3,4} {
    \node[circle, draw, fill=myblue, minimum size=4pt, inner sep=0pt] at (0, \j) {};
}
%extra sites
% \foreach \j in {2,3,4} {
%     \node[diamond, draw, fill=red, minimum size=5pt, inner sep=0pt] at (4.5, \j-0.5) {};
%     \node at (4.3, \j-0.3) {$\sigma$};
% }

% \draw[red] (4.8,-0.2) -- (4.8,4.2) -- (5.2,4.2) -- (5.2,-0.2) -- (4.8,-0.2);
% \node[red] at (5.5,2) {$S^{y}$};
% \draw[red] (-0.2,0.2) -- (5.2,0.2) -- (5.2,-0.2) -- (-0.2,-0.2) -- (-0.2,0.2);
% \node[red] at (-0.5,0) {$S^{x}$};

\draw node[scale=0.7] at (2.35,2.8) {$q+\hat y $};
\draw node[scale=0.7] at (3.3,2.8) {$q + \hat x + \hat y$};
\draw node[scale=0.7] at (2.2,1.8) {$q$};

%arrows_______________________________
\draw[blue,->,very thick] (0,4) to[out=195, in=165, looseness=1.5] (0,0);
\draw[blue,<-,very thick] (0,4) to[out=165, in=15, looseness=1.5] (5,4);

\end{tikzpicture}
        \caption{Xu-Moore model with periodic boundary conditions enforced by identifying opposite edges}
    \end{figure}
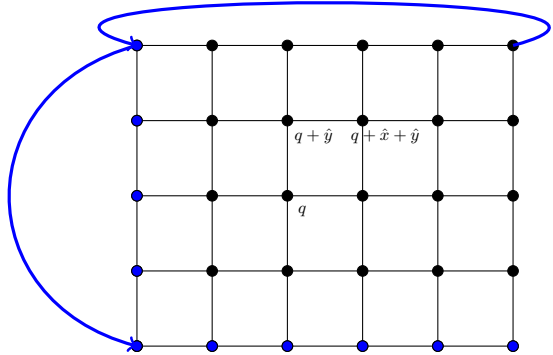
    \hspace{0.5cm}

\begin{figure}
        \centering
        \begin{tikzpicture}

% Define colors
\definecolor{myblue}{RGB}{0,0,255}

% Draw the lattice
\foreach \i in {0,1,2,3,4,5} {
    \foreach \j in {0,1,2,3,4} {
        % Draw the nodes
        \node[circle, draw, fill=black, minimum size=4pt, inner sep=0pt] at (\i, \j) {};
        
        % Draw links (horizontal)
        \ifnum\j<4
            \draw (\i, \j) -- (\i, \j+1);
        \fi

        % Draw links (vertical)
        \ifnum\i<5
            \draw (\i, \j) -- (\i+1, \j);
        \fi
    }
}

% Color the bottommost row nodes
\foreach \i in {0,1,2,3,4,5} {
    \node[circle, draw, fill=myblue, minimum size=4pt, inner sep=0pt] at (\i, 0){};
}
%extra sites
\foreach \i in {1,2,3,4,5}
{ 
\node[diamond, draw, fill=red, minimum size=5pt, inner sep=0pt] at (\i-0.5, 0.5){};
\node at (\i-0.3, 0.7) {$\sigma$};
}
% Color the leftmost column nodes
\foreach \j in {0,1,2,3,4} {
    \node[circle, draw, fill=myblue, minimum size=4pt, inner sep=0pt] at (0, \j) {};
}
%extra sites
\foreach \j in {2,3,4} {
    \node[diamond, draw, fill=red, minimum size=5pt, inner sep=0pt] at (0.5, \j-0.5) {};
    \node at (0.6, \j-0.3) {$\sigma$};
}

\filldraw[red,thick,fill=red,opacity=0.3] (-0.2,-0.2) -- ++(0,4.4) -- ++(0.4,0) -- ++(0,-4.4);
\node[red] at (-0.5,2) {$S^{y}$};
\filldraw[orange,thick,fill=orange,opacity=0.3] (-0.2,0.2) -- (5.2,0.2) -- (5.2,-0.2) -- (-0.2,-0.2) -- (-0.2,0.2);
\node[orange] at (5.5,0) {$S^{x}$};

\draw node[scale=0.7] at (3.35,1.85) {$q+\hat x $};
\draw node[scale=0.7] at (3.3,2.8) {$q + \hat x + \hat y$};
\draw node[scale=0.7] at (2.1,1.8) {$q$};

\end{tikzpicture}
        \caption{Xu-Moore model with periodic boundary conditions and extra sites coupled to plaquette terms.}
    \caption{Xu-Moore model on a finite periodic lattice with 'boundary' plaquette terms coupled to $\mathbb{Z}_2$ gauge degrees of freedom.}
    \label{fig:xu_moore_finite}
\end{figure}

Let $P$ denote the set of all sites in the finite lattice and $S^{y/x}$ denote the set of a vertical/horizontal line of sites which are identified with the corresponding opposite edge when pasting into a torus. 
\begin{equation}
\label{eqn:s_boxdot}
S = S^{y} \cup S^{x}
\end{equation}
Denote the gauge fields by $\sigma^z_{q}$ for any $q \in S$.   
\\
\\
The 'gauged' finite Xu-Moore model on a torus is thus :  
\begin{equation}
\label{eqn:def_xumoore_extended}
\Tilde{H}_{\sf XM}[g] := \sum_{q \in P \backslash S} ( g^{-1} Z_{q} Z_{q+\hat x } Z_{q+\hat y } Z_{q + \hat x + \hat y} + g X_q ) + \sum_{q \in S} ( g^{-1} Z_{q} Z_{q+\hat x } Z_{q+\hat y } Z_{q + \hat x + \hat y} \sigma^{z}_{q} + g X_q )~.
\end{equation}
The Hilbert space of this Hamiltonian is enlarged to $\mathcal{H}_{\sf XM,\text{gauged}} = \left( \prod_{|S|} \mathbb{C}^2 \right) \bigotimes \mathcal{H}_{\sf XM} $
\\
\\
Define the generators of the \textit{enlarged} Bond Algebra as  : 
\begin{equation}
\label{eqn:def_gen_enlarged_bond_algebra_xu_moore}
\Tilde{G}_{L_x L_y} :=
\left \{
\begin{array}{ll}
   Z_{q + \hat x + \hat y} Z_{q+\hat x } Z_{q+\hat y } Z_{q}  & q \in P \backslash S \\ \\
   Z_{q + \hat x + \hat y} Z_{q+\hat x } Z_{q+\hat y } Z_{q} \sigma^{z}_{q} & q \in S \\ \\
    X_q & q \in P 
\end{array}
\right.
\end{equation}
The Bond Algebra $\Tilde{A}_{L_x L_y}$ generated by these operators is thus 
\begin{equation}
\Tilde{A}_{L_x L_y} := \langle \Tilde{G}_{L_x L_y} \rangle
\end{equation}
Now we will define an automorphism on $\Tilde{A}_{L_x L_y}$ enforcing a $\tilde H_{\sf XM}[g] \leftrightarrow \tilde H_{\sf XM}[g^{-1}]$ duality on $\Tilde{H}_{\sf XM}$ by mapping the above set of generators into a different set of generators for the same Bond Algebra. Note that \textbf{$L_x x \equiv 0 \equiv L_y y$} and $S^{x} \cap S^y = \{ (0,0) \}$ (see Fig.(\ref{fig:xu_moore_finite})) : 
\\
\\
\begin{equation}
\label{eqn:def_automorphism_xu_moore}
\Phi_{\sf XM} := 
\left\{
\begin{array}{ll}
    \Phi_{\sf XM} \left( X_{q+ \hat x+ \hat y} \right) = Z_{q} Z_{q+ \hat x} Z_{q+ \hat y} Z_{q+ \hat x+ \hat y} \sigma^{z}_{q} & q \in S
    \\
    \\
    \Phi_{\sf XM} \left( X_{q + \hat x + \hat y} \right) = Z_{q} Z_{q+ \hat x} Z_{q+ \hat y} Z_{q+ \hat x+ \hat y} & q \notin S
    \\
    \\
    \Phi_{\sf XM} \left( Z_{q} Z_{q+ \hat x} Z_{q+ \hat y} Z_{q+ \hat x+ \hat y} \right) = X_q & q \notin S
    \\
    \\
    \Phi_{\sf XM} \left( Z_{q} Z_{q+ \hat x} Z_{q+ \hat y} Z_{q+ \hat x+ \hat y} \sigma^{z}_{q} \right) = X_q & q \notin S
\end{array}
\right.
\end{equation}
The above definition of $\Phi_{\sf XM}$ implies the following mapping for the ancilla sites : 
\begin{align}
    &\Phi_{\sf XM} \left( \sigma^{z}_{q} \right) = \prod^{L_x-1}_{n = 0} X_{q+n \hat x}  \ \ \ \  q \in S^{y} \backslash \{ (0,0) \} \\
    &\Phi_{\sf XM} \left( \sigma^{z}_{q} \right) = \prod^{L_y-1}_{n = 0} X_{q+n \hat y} \ \ \ \  q \in S^{x} \backslash \{ (0,0) \} \\
    &\Phi_{\sf XM} \left( \sigma^{z}_{(\bar 0 , \bar 0)} \right) = \left( \prod^{L_x-1}_{m = 1} \prod^{L_y-1}_{n = 1} X_{(m,n)} \right) X_{(0,0)} \ \ \ \ q = (0,0) 
\end{align}
\begin{figure}[h!]
\centering
\begin{tikzpicture}
\begin{scope}[xscale=-1,xshift=-4]
% Define colors
\definecolor{myblue}{RGB}{0,0,255}

% Draw the lattice
\foreach \i in {0,1,2,3,4,5} {
    \foreach \j in {0,1,2,3,4} {
        % Draw the nodes
        \node[circle, draw, fill=black, minimum size=4pt, inner sep=0pt] at (\i, \j) {};
        
        % Draw links (horizontal)
        \ifnum\j<4
            \draw (\i, \j) -- (\i, \j+1);
        \fi

        % Draw links (vertical)
        \ifnum\i<5
            \draw (\i, \j) -- (\i+1, \j);
        \fi
    }
}

% Color the bottommost row nodes
\foreach \i in {0,1,2,3,4,5} {
    \node[circle, draw, fill=myblue, minimum size=4pt, inner sep=0pt] at (\i, 0){};
}
%extra sites
\foreach \i in {1,2,3,4,5}
{ 
\node[diamond, draw, fill=red, minimum size=5pt, inner sep=0pt] at (\i-0.5, 0.5){};
}
% Color the rightmost column nodes
\foreach \j in {0,1,2,3,4} {
    \node[circle, draw, fill=myblue, minimum size=4pt, inner sep=0pt] at (5, \j) {};
}
%extra sites
\foreach \j in {2,3,4} {
    \node[diamond, draw, fill=red, minimum size=5pt, inner sep=0pt] at (4.5, \j-0.5) {};
}

\draw[magenta,thick] (-0.2,2.8) -- ++(5.4,0) -- ++(0,0.4) -- ++(-5.4,0) -- ++(0,-0.4);
\draw[magenta,->,thick] (4.5,3.5) to[out=180, in=90, looseness=1.5] (3.5,3.2);

\draw[orange,thick] (1.8,-0.2) -- ++(0,4.4) -- ++(0.4,0) -- ++(0,-4.4) -- ++(-0.4,0);
\draw[orange,->,thick] (1.5,0.5) to[out=120, in=135, looseness=1.5] (1.8,1.5);

\draw[red,thick] (5+0.2,-0.2) -- ++ (-0.4,0) -- ++ (0,0.2) -- ++ (-0.8,0.8) -- ++(-3.2,0) -- ++(0,2.4) -- ++ (3.4,0) -- ++ (0,-2.2) -- ++(0.8,-0.8) -- ++(0.2,0) -- ++(0,-0.4);
\draw[red,->,thick] (4.5,0.5) to[out=180, in=225, looseness=1.5] (4.8,-0.2);

\draw[green,thick] (-0.2,1.2) -- ++(0,-1.4) -- ++(1.4,0) -- ++(0,1.4) -- ++(-1.4,0);
\draw[green,->,thick] (0,1) to[out=180, in=135, looseness=1.5] (-0.2,0.5);

\node[circle, draw, fill=purple, minimum size=4pt, inner sep=0pt] at (0,0) {};
\node[circle, draw, fill=purple, minimum size=4pt, inner sep=0pt] at (5,0) {};
\node[circle, draw, fill=purple, minimum size=4pt, inner sep=0pt] at (0,4) {};
\node[circle, draw, fill=purple, minimum size=4pt, inner sep=0pt] at (5,4) {};

\node at (5.2,4.2) {\textcolor{purple}{$q^*$}};

\end{scope}

\draw[->,thick] (-6.3,1.7) -- ++ (-0.7,-0.7);
\node at (-6.5,2) { $\Phi_{\sf XM} \rightharpoondown T_{(-\frac{1}{2},-\frac{1}{2})}$ };

\end{tikzpicture}
\caption{Illustration of the duality mapping for the gauge degrees of freedom. Site $(0,0)$ is indicated in purple. The convention chosen here is a half-translation along the $(-1,-1)$ direction : its inverse $\Phi^{-1}_{\sf XM}$ will translate the local terms in the opposite direction. In later sections the Unitary will be constructed for $\Phi^{-1}_{\sf XM}$.}
\label{fig:correct_xm_duality_p1}
\end{figure}
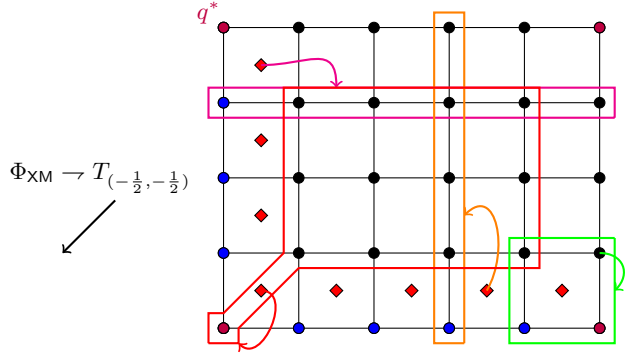
\\
The bulk duality assignment (the last two lines in Eq.(\ref{eqn:def_automorphism_xu_moore})) fixes the result of the duality on boundary plaquette terms : 
\begin{align}
\label{eqn:consistency_check_1_xm_bond_algebra_automorphism}
  & \Phi_{\sf XM}(Z_{q} Z_{q+ \hat x} Z_{q+ \hat y} Z_{q+ \hat x+ \hat y}) = \prod^{L_x-1}_{n = 1} X_{q+n \hat x} \ \ \ \ q \in S^{y} \backslash \{ (0,0) \}  \\
    & \Phi_{\sf XM}(Z_{q} Z_{q+ \hat x} Z_{q+ \hat y} Z_{q+ \hat x+ \hat y}) = \prod^{L_y-1}_{n = 1} X_{q+n \hat y} \ \ \ \ q \in S^{x} \backslash \{ (0,0) \}  \\ 
    & \Phi_{\sf XM}(Z_{(0,0)} Z_{(1,0)} Z_{(0,1)} Z_{(1,1)}) = \prod^{L_y-1}_{n = 1} \prod^{L_x-1}_{m = 1} X_{(m,n)} 
\end{align}
which follow from the expression of the above $ZZZZ$ terms as a product of bulk plaquette terms which are independent generators of $\tilde{\mathcal{A}}_{L_xL_y}$ in $\Tilde{G}_{L_x L_y}$, then distributing $\Phi_{\sf XM}$ over this product.
\\
\\
The action of  $\Phi_{\sf XM}$ on $\sigma^{z}_{q}$ is decided by Eq.(\ref{eqn:def_automorphism_xu_moore}), to ensure that :
\begin{equation}
\label{eqn:consistency_check_2_xm_bond_algebra_automorphism}
\Phi_{\sf XM}(Z_{q} Z_{q+ \hat x} Z_{q+ \hat y} Z_{q} \sigma^{z}_{q}) =  X_{q}  \ \ \ \ q \in S
\end{equation}
Which 'fixes' the inconsistency which would arise had we demanded a naive automorphism without extra sites. \textit{Note that each $\sigma^{z}_{q}$ is mapped to a symmetry of the bare Hamiltonian - due to periodic boundary conditions it is evident that even $\Phi_{\sf XM} \left( \sigma^z_{(\bar 0 , \bar 0)} \right)$ is a symmetry.}
\\
\\
$\Phi_{\sf XM}$ is a well defined automorphism on $\Tilde{\mathcal{A}}_{L_x L_y}$, which satisfies : 
\begin{equation}
\label{eqn:extended_xu_moore_duality}
\Phi_{\sf XM}(\Tilde{H}_{\sf XM}[g]) = \Tilde{H}_{\sf XM}[g^{-1}]
\end{equation}
And so $\exists \ \text{a Hilbert space} \ \mathcal{M}$  and a unitary $\mathcal{U} \in \mathcal{L}(\mathcal{H}_{\sf XM,\text{gauged}} \otimes \mathcal{M}) $ such that 
\begin{equation}
\label{eqn:unitary_implm_of_xm_duality}
\mathcal{U}^{\dagger} \left( \Tilde{H}_{\sf XM}[g] \otimes I \right) \mathcal{U}  = \Tilde{H}_{\sf XM}[g^{-1}] \otimes I
\end{equation}
Thus, the conjugation $\mathcal{U} \left( \cdot \right) \mathcal{U}^{\dagger}$ implements $\Phi^{-1}_{\sf XM}$, which is used as the canonical choice throughout the main text. 
\subsection{Explicit mapping of $\sigma^z$ under $\mathcal{U}$} \label{sec:unitary_action_on_projectors}
We show in later sections by explicit construction that $\mathcal{M} \equiv \mathbb{C}$ for the finite periodic Xu-Moore model, and thus assume it in the following discussion. Following Bond Algebraic considerations, it is evident that : 
\begin{eqnarray}
    &\mathcal{U} \ \sigma^{z}_{(\bar 0 , \bar 0)} \mathcal{U}^{\dagger} = \Phi^{-1}_{\sf XM} \left( \sigma^{z}_{(\bar 0 , \bar 0)} \right) = \left( \prod^{L_x}_{m = 2} \prod^{L_y}_{n = 2} X_{(m,n)} \right) X_{(1,1)} \\ \nonumber \\
     &\mathcal{U} \ \sigma^{z}_{(\bar 0 , \bar j)} \ \mathcal{U}^{\dagger} = \Phi^{-1}_{\sf XM} \left( \sigma^{z}_{(\bar 0 , \bar j)} \right) = \hat\eta^{\text{row}}_{j+1} \\ \nonumber \\
     &\mathcal{U} \ \sigma^{z}_{(\bar j , \bar 0)} \ \mathcal{U}^{\dagger} = \Phi^{-1}_{\sf XM} \left( \sigma^{z}_{(\bar j , \bar 0)} \right) = \hat\eta^{\text{col}}_{j+1}
\end{eqnarray}
where $(a+L_x,b+L_y) \equiv (a,b)$.

\begin{figure}[h!]
\centering
\begin{tikzpicture}
\begin{scope}[xscale=-1,xshift=-4]
% Define colors
\definecolor{myblue}{RGB}{0,0,255}

% Draw the lattice
\foreach \i in {0,1,2,3,4,5} {
    \foreach \j in {0,1,2,3,4} {
        % Draw the nodes
        \node[circle, draw, fill=black, minimum size=4pt, inner sep=0pt] at (\i, \j) {};
        
        % Draw links (horizontal)
        \ifnum\j<4
            \draw (\i, \j) -- (\i, \j+1);
        \fi

        % Draw links (vertical)
        \ifnum\i<5
            \draw (\i, \j) -- (\i+1, \j);
        \fi
    }
}

% Color the bottommost row nodes
\foreach \i in {0,1,2,3,4,5} {
    \node[circle, draw, fill=myblue, minimum size=4pt, inner sep=0pt] at (\i, 0){};
}
%extra sites
\foreach \i in {1,2,3,4,5}
{ 
\node[diamond, draw, fill=red, minimum size=5pt, inner sep=0pt] at (\i-0.5, 0.5){};
}
% Color the rightmost column nodes
\foreach \j in {0,1,2,3,4} {
    \node[circle, draw, fill=myblue, minimum size=4pt, inner sep=0pt] at (5, \j) {};
}
%extra sites
\foreach \j in {2,3,4} {
    \node[diamond, draw, fill=red, minimum size=5pt, inner sep=0pt] at (4.5, \j-0.5) {};
}

\draw[magenta,thick] (-0.2,2.8+1) -- ++(5.4,0) -- ++(0,0.4) -- ++(-5.4,0) -- ++(0,-0.4);
\draw[magenta,->,thick] (4.5,3.5) to[out=180, in=270, looseness=1.5] (3.5,3.2+0.6);

\draw[orange,thick] (1.8-1,-0.2) -- ++(0,4.4) -- ++(0.4,0) -- ++(0,-4.4) -- ++(-0.4,0);
\draw[orange,->,thick] (1.5,0.5) to[out=45, in=45, looseness=1.5] (1.8-0.6,1.5);

\draw[red,thick] (5+0.2-1,-0.2+1) -- ++ (-0.4,0) -- ++ (0,0.2) -- ++ (-0.8,0.8) -- ++(-3.2,0) -- ++(0,2.4) -- ++ (3.4,0) -- ++ (0,-2.2) -- ++(0.8,-0.8) -- ++(0.2,0) -- ++(0,-0.4);
\draw[red,->,thick] (4.5,0.5) to[out=180, in=225, looseness=1.5] (4.8-1,-0.2+1);

\node[circle, draw, fill=purple, minimum size=4pt, inner sep=0pt] at (0,0) {};
\node[circle, draw, fill=purple, minimum size=4pt, inner sep=0pt] at (5,0) {};
\node[circle, draw, fill=purple, minimum size=4pt, inner sep=0pt] at (0,4) {};
\node[circle, draw, fill=purple, minimum size=4pt, inner sep=0pt] at (5,4) {};

\node at (5.2,0.2) {\textcolor{purple}{$q^*$}};

\end{scope}

\draw[->,thick] (-6.3-0.7,1.7-0.7) -- ++ (0.7,0.7);
\node at (-6.5,2) { $\Phi_{\sf XM}^{-1} \rightharpoondown T_{(\frac{1}{2},\frac{1}{2})}$ };

\end{tikzpicture}
\caption{Action of $\mathcal{U}$ (or $\Phi^{-1}_{\sf XM}$) on the extra sites}
\label{fig:correct_xm_duality_p1}
\end{figure}
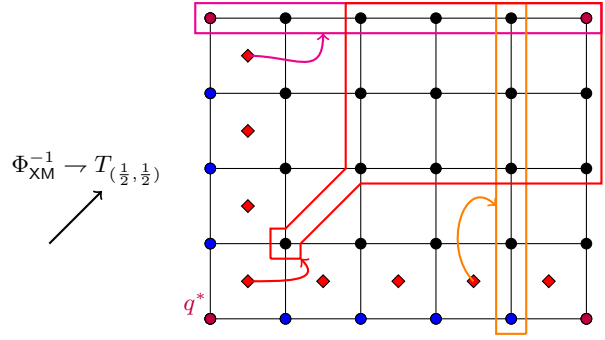

In the $(1+1)D$ TFIC we know that the unitary exchanges the ancilla site with the $\hat \eta$ symmetry operator of the theory and hence exchanges the projectors into the physical $\sigma^z = 1$ and the $\hat \eta = 1$ sectors - thus $\mathcal{U}_{\text{TFIC}}^2$ acts trivially on these two operators. \textit{In the Xu-Moore case, we show that $\mathcal{U}^2$ acts trivially on the projectors, though it does not act trivially on the ancillas or subsystem symmetries.} First, we consider : 
\begin{eqnarray}
    &\mathcal{U}^2 \ \sigma^{z}_{(\bar 0, \bar 0)} (\mathcal{U}^{\dagger})^2 = \Phi^{-1}_{\sf XM} \left( \left( \prod^{L_x}_{m = 2} \prod^{L_y}_{n = 2} X_{(m,n)} \right) X_{(1,1)}\right) \nonumber \\
    &= \sigma^{z}_{(\bar 0, \bar 0)} \left( \prod_{L_x-1 \geq q_x \geq 2} \sigma^{z}_{(\bar q_x, \bar 0)} \right) \left( \prod_{L_y-1 \geq q_y \geq 2} \sigma^{z}_{(\bar 0, \bar q_y)} \right) \\ \nonumber \\
    &\mathcal{U}^2 \ \sigma^{z}_{(\bar 0,\bar L_y-1)} \ (\mathcal{U}^{\dagger})^2 = \Phi^{-1}_{\sf XM} \left( \hat\eta^{\text{row}}_{L_y} \right) = \prod_{0 \leq q_x \leq L_x-1} \sigma^z_{(\bar q_x,\bar 0)} \\ \nonumber \\
    &\mathcal{U}^2 \ \sigma^{z}_{(\bar L_x-1,0)} \ (\mathcal{U}^{\dagger})^2 = \Phi^{-1}_{\sf XM} \left( \hat\eta^{\text{row}}_{L_y} \right) = \prod_{0 \leq q_y \leq L_y-1} \sigma^z_{(\bar 0,\bar q_y)} \\ \nonumber \\
     &\mathcal{U}^2 \ \sigma^{z}_{(\bar j,\bar 0)} \ (\mathcal{U}^{\dagger})^2 = \Phi^{-1}_{\sf XM} \left( \hat\eta^{\text{col}}_{j+1} \right) = \sigma^{z}_{(\bar j+1,0)} \ \ 1 \leq j \leq L_x - 2 \\ \nonumber \\
     &\mathcal{U}^2 \ \sigma^{z}_{(\bar 0,\bar j)} \ (\mathcal{U}^{\dagger})^2 = \Phi^{-1}_{\sf XM} \left( \hat\eta^{\text{row}}_{j+1} \right) = \sigma^{z}_{(\bar 0,\bar j+1)} \ \ 1 \leq j \leq L_y - 2
\end{eqnarray}

\begin{figure}[h!]
        \centering
        \begin{tikzpicture}

% Define colors
\definecolor{myblue}{RGB}{0,0,255}

% Color the bottommost row nodes
%extra sites
\foreach \i in {1,2,3,4,5,6}
{ 
\node[diamond, draw, fill=red, minimum size=5pt, inner sep=0pt] at (\i-0.5, 0.5){};
\node[diamond, draw, fill=red, minimum size=5pt, inner sep=0pt] at (\i-0.5, 0.5+4){};
}
% Color the leftmost column nodes
%extra sites
\foreach \j in {2,3,4,5} {
    \node[diamond, draw, fill=red, minimum size=5pt, inner sep=0pt] at (0.5, \j-0.5) {};
    \node[diamond, draw, fill=red, minimum size=5pt, inner sep=0pt] at (0.5+5, \j-0.5) {};
}

\draw[blue,very thick] (0,0) -- ++(0,4) -- ++(5,0) -- ++(0,-4) -- ++(-5,0);
\draw[blue,very thick] (5,0) -- ++(1.5,0);
\draw[blue,very thick] (5,4) -- ++(1.5,0);
\draw[blue,very thick] (0,4) -- ++(0,1.5);
\draw[blue,very thick] (5,4) -- ++(0,1.5);
\node at (-0.5,0.5) {\textcolor{blue}{$M$}};

% Draw the lattice
\foreach \i in {0,1,2,3,4,5,6} {
    \foreach \j in {0,1,2,3,4,5} {
        % Draw the nodes
        \node[circle, draw, fill=black, minimum size=4pt, inner sep=0pt] at (\i, \j) {};
        
        % Draw links (horizontal)
        \ifnum\j<5
            \draw (\i, \j) -- (\i, \j+1);
        \fi

        % Draw links (vertical)
        \ifnum\i<6
            \draw (\i, \j) -- (\i+1, \j);
        \fi
    }
}

\draw[red, thick] (2.3,4.5+0.2) -- ++(3.4,0) -- ++(0,-2.4) -- ++(-0.4,0) -- ++ (0,2) -- ++(-3,0) -- ++(0,0.4);
\draw[red,->,thick] (0.5,0.5) to[out=15, in=270, looseness=1.5] (3.5,4.3);

\draw[green, thick] (5+0.2,0.2+1) -- ++(0,4.6-1) -- ++(0.6,0) -- ++(0,-4.6+1) -- ++ (-0.6,0);
\draw[green, thick] (5+0.2,0.8-1) -- ++(0,1) -- ++(0.6,0) -- ++(0,-1);
\draw[green,->,thick] (4.5,0.5) to[out=90, in=180, looseness=1.5] (5.2,1.5);

\draw[orange, thick] (0.2+1,4.2) -- ++(5.6-1,0) -- ++(0,0.6) -- ++(-5.6+1,0) -- ++(0,-0.6);
\draw[orange, thick] (-0.2,4.2) -- ++(1,0) -- ++(0,0.6) -- ++(-1,0);
\draw[orange,->,thick] (0.5,3.5) to[out=0, in=270, looseness=1.5] (1.5,4.2);

\end{tikzpicture}
    \caption{Action of $\mathcal{U}^2$ on extra sites. The boundary $M$ indicates the extent of the system, everything outside it is periodic repetition.}
    \label{fig:ancillas_under_U^2}
\end{figure}
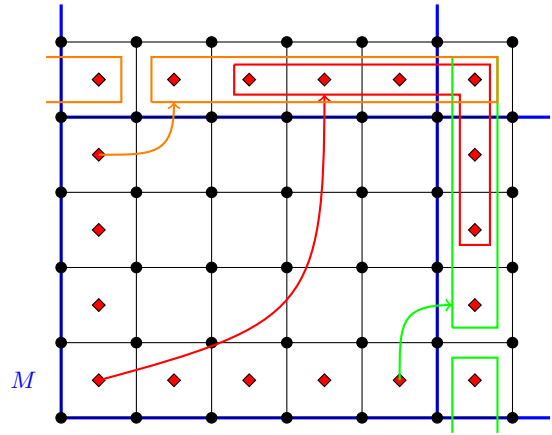

The abelian group generated by the ancillas $\{ \sigma^z_{\bar a} \}$ is preserved under conjugation by $\mathcal{U}^2$ : each of $ \mathcal{U}^2 \sigma^z_{\bar a}  (\mathcal{U}^{\dagger})^2$ is a different generator for the same group. Therefore, if we define $\widehat P_{+} = \prod_{a \in S} \left( \frac{1+\sigma^z_{\bar a}}{2}\right)$ it is evident that 
\begin{equation}
    \mathcal{U}^2 \widehat{P}_{+} (\mathcal{U}^{\dagger})^2 = \widehat{P}_{+}~.
\end{equation}

\subsection{Infinite Cylinder limit} \label{sec:extra_sites_infinite_cylinder}

The case of the infinite cylinder can be understood as the ``bulk" region of the torus with $L_y \to \infty$. Intuitively, the bottom row of extra sites is 'pushed off' to infinity. Formally, one constructs the automorphism on this infinite Bond Algebra by defining the same half-translation automorphism everywhere, as before.

\begin{equation}
\label{eqn:bond_algebra_xm_cyl}
\mathcal{G}_{\sf XM} =\{ Z_q Z_{q+\hat x } Z_{q+\hat y } Z_{q + \hat x + \hat y} , X_q \ \ \forall \ q \ ; \ \sigma^z_{(\bar 0, \bar j)}|_{-\infty < j < \infty}  \} \ \ \text{with} \ \ \mathcal{A}_{\sf XM} = \langle \mathcal{G}_{\sf XM} \rangle
\end{equation}

\begin{figure}[h!]
\centering
\begin{tikzpicture}
        \def\cols{3}
        \def\rows{2}
        % Add vertex bulges (small circles)
        \foreach \x in {0,...,\cols} 
        {
            \foreach \y in {0,...,\rows} 
            {
                \node[draw=black, fill=black, circle, minimum size=4pt, inner sep=0pt] at (\x,\y){ };
            }
        }
        %Add extra sites
            \foreach \y in {-1,...,\rows} {
                \node[draw=black, fill=red, diamond, minimum size=5pt, inner sep=0pt] 
                    at (1+0.5,\y+0.5){ };
             }

        \foreach \x in {0,...,\cols} 
        {
            \foreach \y in {-1,...,\rows} 
            {
                \draw (\x,\y) -- (\x,\y+1);
                
                \ifnum \x<\cols
                    \ifnum \y > -1
                    \draw (\x,\y) -- (\x+1,\y);
                    \fi
                \fi
            }
        }

        \node[draw = orange, thick, circle, minimum size = 7pt, inner sep=0pt] at (2,2) { };
        \draw[orange, ->, thick] (2,2) to[out = 315, in = 0, looseness=1.5] (1.9,1.5);
        \draw[orange,thick] (1.9,1.1) --++(0,0.8) --++(-0.8,0) --++(0,-0.8) --++(0.8,0);

        \node[draw = magenta, thick, circle, minimum size = 7pt, inner sep=0pt] at (1.5,0.5) { };
        \draw[magenta, ->, thick] (1.5,0.5) to[out = 0, in = 90, looseness=1.5] (1.9,0.1);
        \draw[magenta,thick] (-1,0.1) --++(5,0) --++(0,-0.2) --++(-5,0) --++(0,0.2);

        \draw[blue,->,very thick] (-0.3,-0.8) to[out=195, in=345, looseness=1.5] (0.3+\cols,-0.8);

       \foreach \y in {0,...,2}
            {
                \draw (0,\y) -- ++(-0.4,0);
                \draw (3,\y) -- ++(0.4,0);
            }
\end{tikzpicture}
\caption{Illustration of Bond Algebraic Automorphism}
\label{fig:xm_cylinder_BA_automorphism}
\end{figure}
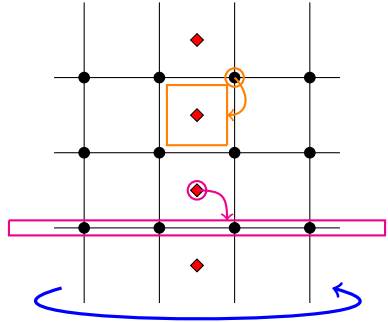

We can define the automorphism on $\mathcal{A}_{\sf XM}$ by mapping $\mathcal{G}_{\sf XM}$ to a different set of generators $\mathcal{G}'_{\sf XM}$ :

\begin{align}
\label{eqn:xm_automorphism_cyl}
&\mathcal{G}_{\sf XM} \xrightarrow{\Phi_{\sf XM}} \mathcal{G}'_{\sf XM} \ \ \text{where} \nonumber \\ \nonumber \\
& \Phi_{\sf XM}(Z_q Z_{q+\hat x} Z_{q+\hat y} Z_{q + \hat x + \hat y}) = X_q & q_x \neq 0 \\
& \Phi_{\sf XM}(Z_q Z_{q+\hat x} Z_{q+\hat y} Z_{q + \hat x + \hat y} \sigma^{z}_{q} ) = X_{q} & q_x=0 \\
& \Phi_{\sf XM}(X_q) = Z_{q - \hat x- \hat y} Z_{q- \hat y} Z_{q - \hat x} Z_{q} & q_x \neq 1 \\
& \Phi_{\sf XM}(X_q) = Z_{q - \hat x- \hat y} Z_{q- \hat y} Z_{q - \hat x} Z_{q} \sigma^z_{q - \hat x - \hat y} & q_x = 1
\end{align}
which implies that (see Fig.(\ref{fig:xm_cylinder_BA_automorphism})).
\begin{equation}
\Phi_{\sf XM}(\sigma^{z}_{q}) =  \prod^{L_x-1}_{q'_x=0} X_{(q'_x,q_y)}~.
\end{equation}

Evidently,
\begin{equation}
    \label{eqn:xm_automorphism_cyl_hamiltonian}
    \Phi_{\sf XM}(\tilde H_{\sf XM,cyl}[g]) = \tilde H_{\sf XM,cyl}[g^{-1}]~.
\end{equation}

We seek to find a unitary $\mathcal{U}$ such that 
\begin{equation}
\label{eqn:unitary_implm_of_xm_duality}
\mathcal{U}\Tilde{H}_{\sf XM, cyl}[g] \mathcal{U}^{\dagger}  = \Tilde{H}_{\sf XM, cyl}[g^{-1}] 
\end{equation}
A direct construction of $\mathcal{U}$ for the infinite cylinder is carried out in the End Matter. 

\section{Xu-Moore on Torus : Constructing the Unitary} \label{sec:unitary_construction_torus}

We describe a well-defined sequence of Unitary operators that enact the Duality map on the minimally gauged theory according to Eq.(\ref{eqn:def_automorphism_xu_moore}) by sequential action on L-shaped subsystems. We are motivated primarily by the Bond Algebraic automorphism in carrying out this construction explicitly.\subsection{Notation}
\begin{figure}[h!]
\centering
\begin{tikzpicture}
% Define colors
\definecolor{myblue}{RGB}{0,0,255}

% Draw the lattice
\foreach \i in {0,1,2,3,4,5} {
    \foreach \j in {0,1,2,3,4} {
        % Draw the nodes
        \node[circle, draw, fill=black, minimum size=4pt, inner sep=0pt] at (\i, \j) {};
        
        % Draw links (horizontal)
        \ifnum\j<4
            \draw (\i, \j) -- (\i, \j+1);
        \fi

        % Draw links (vertical)
        \ifnum\i<5
            \draw (\i, \j) -- (\i+1, \j);
        \fi
    }
}
% Color the bottommost row nodes
\foreach \i in {0,1,2,3,4,5} {
    \node[circle, draw, fill=myblue, minimum size=4pt, inner sep=0pt] at (\i, 0){};
}
%extra sites
\foreach \i in {1,2,3,4,5}
{ 
\node[diamond, draw, fill=red, minimum size=5pt, inner sep=0pt] at (\i-0.5, 0.5){};
\node at (\i-0.3, 0.7){};
}
% Color the leftmost column nodes
\foreach \j in {0,1,2,3,4} {
    \node[circle, draw, fill=myblue, minimum size=4pt, inner sep=0pt] at (0, \j) {};
}
%extra sites
\foreach \j in {2,3,4} {
    \node[diamond, draw, fill=red, minimum size=5pt, inner sep=0pt] at (0.5, \j-0.5) {};
    \node at (0.6, \j-0.3){};
}

\draw[magenta] (-0.2,-0.2) -- ++(0,4.4) -- ++(0.4,0) -- ++(0,-4) -- ++(5,0) -- ++(0,-0.4) -- ++(-5.4,0);
\node[magenta] at (0,4.5) {$L_0$};
\draw[red] (0.4,0.4) -- ++(0,3.2) -- ++(0.2,0) -- ++(0,-3) -- ++(4.2,0) -- ++(0,-0.2) -- ++(-4.4,0);
\node[red] at (0.5,3.8) {$L_{\sigma}$};

\draw[orange] (1-0.2,1-0.2) -- ++(0,4.4-1) -- ++(0.4,0) -- ++(0,-4+1) -- ++(5-1,0) -- ++(0,-0.4) -- ++(-5.4+1,0);
\node[orange] at (1,4.5) {$L_1$};
\draw[orange] (2-0.2,2-0.2) -- ++(0,4.4-2) -- ++(0.4,0) -- ++(0,-4+2) -- ++(5-2,0) -- ++(0,-0.4) -- ++(-5.4+2,0);
\node[orange] at (2,4.5) {$L_2$};
\draw[orange] (3-0.2,3-0.2) -- ++(0,4.4-3) -- ++(0.4,0) -- ++(0,-4+3) -- ++(5-3,0) -- ++(0,-0.4) -- ++(-5.4+3,0);
\node[orange] at (3,4.5) {$L_3$};
\draw[orange] (4-0.2,4-0.2) -- ++(0,4.4-4) -- ++(0.4,0) -- ++(5-4,0) -- ++(0,-0.4) -- ++(-5.4+4,0);
\node[orange] at (4,4.5) {$L_4$};;

\end{tikzpicture}
    \caption{L-shaped layers of terms indicated for a 5 x 4 site Xu-Moore model with PBC}
    \label{fig:xm_twisted_4by3}
\end{figure}
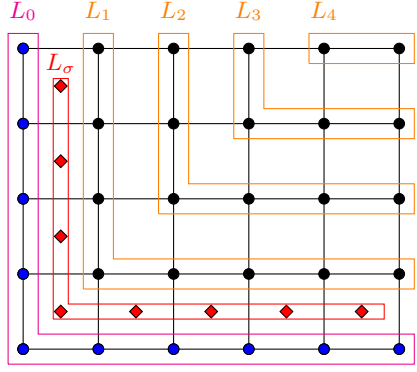

Consider an $L_x \times L_y$ site Xu-Moore Hamiltonian where, say $L_x \geq L_y \geq 3$. All sites can be organized into L-shaped subsets $L_1,L_2, \cdots L_{k_{\text{max}}}$ where $k_{\text{max}} = L_x-1$. Let $\bar q_{x/y} \equiv q_{x/y} + \frac{1}{2}$, which is used to represent the coordinates of the ancilla sites (which all have coordinates displaced from those on $L_1$ by ($\frac{1}{2}$,$\frac{1}{2}$)). \textbf{The coordinates $q_{x/y}$ are understood modulo $L_{x/y}$ respectively and start at (0,0)}.
\begin{align}
    & L_{0} := \left\{q : q_x = 0 \ \text{or} \ q_y = 0 \right\} \ , \ L_{\sigma} := \left\{\bar q : \bar q_x = \bar 0 \ \text{or} \ \bar q_y = \bar 0 \right\} \\
    & L_{1\leq k <k_{\text{max}}} := \left\{q : q_{x/y} \geq k  \right\} \cup \{ (0,k) , (k,0) \} \\
    & L_{k_{\text{max}}} := \left\{ q = (q_x,0) : q_x \geq k_{\textbf{max}} \right \} \cup \{ (0,0) \}
\end{align}
See Fig.(\ref{fig:xm_twisted_4by3}). The sequential unitaries will be supported on each subset $L_{k/\sigma}$ of sites. All members of $L_0$ are uniquely contained in some $L_{k>0}$ and the defect terms always involve sites from neighboring layers. Consequently, any unitary supported on $L_{k \neq 0}$ will modify terms belonging to $L_0$.
\begin{align}
\label{eqn:hamiltonian_xm_min_gauged_layers}
&\tilde H_{\sf XM}[g] := \sum_{k=1}^{k_{\text{max}}-1} H_k[g]
\ \ \text{where} \ \ H_0[g] = \sum_{q \in L_0} \left( g^{-1} Z_q Z_{q+\hat x } Z_{q+\hat y } Z_{q + \hat x + \hat y} \sigma^{z}_{q} + g X_q \right) \\
& H_{k>0}[g] = \sum_{q \in L_k \setminus L_0} \left( g^{-1} Z_q Z_{q+\hat x } Z_{q+\hat y } Z_{q + \hat x + \hat y} + g X_q \right)
\end{align}
Upon defect motion, we will obtain plaquette terms involving extra sites which take the following form : 
\begin{align}
    \label{eq:plaquette_terms_in_sigmas}
 &H_{\sigma}[g^{-1}] := \sum^{L_y-1}_{q_y = 1} g ~\sigma^z_{0,q_y} \sigma^z_{0,q_y+1} Z_{1,q_y} Z_{1,q_y+1} + \sum^{L_x-1}_{q_x = 1} g ~\sigma^z_{q_x,0} \sigma^z_{q_x+1,0} Z_{q_x,1} Z_{q_x+1,1} + g ~\sigma^z_{0,0} \sigma^z_{0,1} \sigma^z_{1,0} Z_{1,1} 
\end{align}
Fix the following ordering for the multiplication of a string of local operators $A_q \equiv A_{(j,k)}$ : 
\begin{align}
&\prod_{j_1 \leq j \leq j_2} A_{(j,k)} := A_{(j_2,k)} A_{(j_2-1,k)}  \cdots A_{(j_1,k)} \\
&\prod_{k_1 \leq k \leq k_2} A_{(j,k)} := A_{(j,k_2)} A_{(j,k_2-1)}  \cdots A_{(j,k_1)}
\end{align}

\subsection{Constructing the Unitary} \label{sec:protein_and_potatoes}
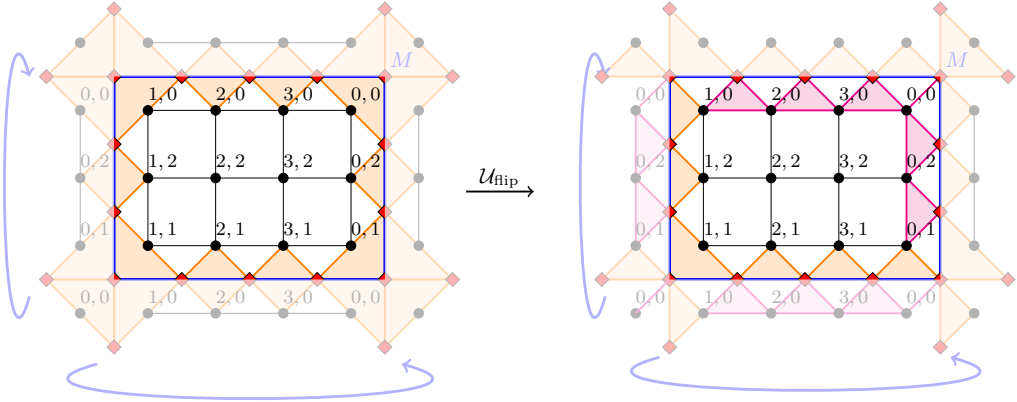
\begin{figure}[h!]
\centering
\scalebox{0.9}{

\begin{tikzpicture}

% Define colors
\definecolor{myblue}{RGB}{0,0,255}

\filldraw[fill=white, draw=white] (1,1) rectangle (5,4);

%dual lattice edges

\foreach \i in {0,1,2,3,4,5} {
    \foreach \j in {0,1,2,3,4} {
        
        % Draw links (horizontal)
        \ifnum\j<4
            \draw (\i+0.5, \j+0.5) -- ++ (0,1);
        \fi

        % Draw links (vertical)
        \ifnum\i<5
            \draw (\i+0.5, \j+0.5) -- ++ (1,0);
        \fi
    }
}

%arrows to indicate the structure gauge-invariant X variables

\foreach \i in {1,2,3,4}
{ 
\foreach \j in {1,2,3} 
{
    \ifnum \i = 1
        \ifnum \j = 1
            \filldraw[fill = orange!20, draw = orange,thick] (\i+0.5,\j+0.5) -- ++(-0.5,0.5) -- ++(0,-1) -- ++(1,0) -- ++(-0.5,0.5);
            \filldraw[fill = orange!20, draw = orange,thick] (\i,\j) -- ++(-1,0) -- ++(1,1);
        \else
            \ifnum \j=3
                \filldraw[fill = orange!20, draw = orange,thick] (\i+0.5,\j+0.5) -- ++(0.5,0.5) -- ++(-1,0) -- ++(0,-1) -- ++(0.5,0.5);
                \filldraw[fill = orange!20, draw = orange,thick] (\i,\j+1) -- ++(-1,0) -- ++(1,-1);
            \else
                \filldraw[fill = orange!20, draw = orange,thick] (\i+0.5,\j+0.5) -- ++(-0.5,0.5) -- ++(0,-1) -- ++(0.5,0.5);
                 \filldraw[fill = orange!20, draw = orange,thick] (\i,\j+1) -- ++(-0.5,-0.5) -- ++(0.5,-0.5);
            \fi
        \fi
    \else
        \ifnum \i = 4
            \ifnum \j = 1
            \filldraw[fill = orange!20, draw = orange,thick] (\i+0.5,\j+0.5) -- ++(0.5,0.5) -- ++(0,-1) -- ++(-1,0) -- ++(0.5,0.5);
            \filldraw[fill = orange!20, draw = orange,thick] (\i+1,\j) -- ++(1,0) -- ++(-1,1);
        \else
            \ifnum \j=3
                \filldraw[fill = orange!20, draw = orange,thick] (\i+0.5,\j+0.5) -- ++(-0.5,0.5) -- ++(1,0) -- ++(0,-1) -- ++(-0.5,0.5);
                \filldraw[fill = orange!20, draw = orange,thick] (\i+1,\j+1) -- ++(1,0) -- ++(-1,-1);
            \else
                \filldraw[fill = orange!20, draw = orange,thick] (\i+0.5,\j+0.5) -- ++(0.5,0.5) -- ++(0,-1) -- ++(-0.5,0.5);
                \filldraw[fill = orange!20, draw = orange,thick] (\i+1,\j+1) -- ++(0.5,-0.5) -- ++(-0.5,-0.5);
            \fi
        \fi
        \else
            \ifnum \j = 1
                \filldraw[fill = orange!20, draw = orange,thick] (\i,\j) -- ++(0.5,0.5) -- ++ (0.5,-0.5) -- ++ (-1,0);
                \filldraw[fill = orange!20, draw = orange,thick] (\i,\j) -- ++(0.5,-0.5) -- ++ (0.5,0.5) -- ++ (-1,0);
            \else
                \ifnum \j = 3
                    \filldraw[fill = orange!20, draw = orange,thick] (\i,\j+1) -- ++(0.5,-0.5) -- ++ (0.5,0.5) -- ++ (-1,0);
                    \filldraw[fill = orange!20, draw = orange,thick] (\i,\j+1) -- ++(0.5,0.5) -- ++ (0.5,-0.5) -- ++ (-1,0);
                \fi
            \fi
        \fi
    \fi
}
}

\filldraw[fill = orange!20, draw = orange,thick] (1,1) -- ++(-1,0) -- ++(1,-1) -- ++(0,1);
\filldraw[fill = orange!20, draw = orange,thick] (1,1) -- ++(1,0) -- ++(-1,-1) -- ++(0,1);

\filldraw[fill = orange!20, draw = orange,thick] (1,4) -- ++(-1,0) -- ++(1,1) -- ++(0,-1);
\filldraw[fill = orange!20, draw = orange,thick] (1,4) -- ++(1,0) -- ++(-1,1) -- ++(0,-1);

\filldraw[fill = orange!20, draw = orange,thick] (5,4) -- ++(-1,0) -- ++(1,1) -- ++(0,-1);
\filldraw[fill = orange!20, draw = orange,thick] (5,4) -- ++(1,0) -- ++(-1,1) -- ++(0,-1);

\filldraw[fill = orange!20, draw = orange,thick] (5,1) -- ++(-1,0) -- ++(1,-1) -- ++(0,1);
\filldraw[fill = orange!20, draw = orange,thick] (5,1) -- ++(1,0) -- ++(-1,-1) -- ++(0,1);

\foreach \i in {0,1,2,3,4,5}
{ 
\foreach \j in {0,1,2,3,4} {
    \node[circle, draw, fill=black, minimum size=4pt, inner sep=0pt] at (\i+0.5, \j+0.5) {};
}
}

\foreach \i in {0,1,2,3,4}
{ 
\foreach \j in {0,1,2,3} {
    \ifnum \j < 3
        \ifnum \i < 4
        \node at (\i+0.72, \j+0.72) {\footnotesize $\i,\j$};
        \else
            \ifnum \j < 3
            \node at (\i+0.72, \j+0.72) {\footnotesize $0,\j$};
            \fi
        \fi
    \else
        \ifnum \i < 4
        \node at (\i+0.72, \j+0.72) {\footnotesize $\i,0$};
        \fi
    \fi
}
}

 \node at (4+0.72, 3+0.72) {\footnotesize $0,0$};

\draw[blue,->,very thick] (0.75,-0.25) to[out=195, in=345, looseness=1.5] (5.5-0.25,-0.25);
\draw[blue,->,very thick] (-0.25,1-0.25) to[out=255, in=105, looseness=1.5] (-0.25,4.25-0.25);

\draw[draw=blue, very thick] (1,1) rectangle (5,4);
\node at (5.25,4.25) {\textcolor{blue}{$M$}};

%dual lattice sites

\foreach \i in {0,1,2,3,4,5}
{
    \node[diamond, draw, fill=red, minimum size=6pt, inner sep=0pt] at (1, \i) {};
    \node[diamond, draw, fill=red, minimum size=6pt, inner sep=0pt] at (5, \i) {};
}

\foreach \i in {0,2,3,4,6}
{
    \node[diamond, draw, fill=red, minimum size=6pt, inner sep=0pt] at (\i,1) {};
    \node[diamond, draw, fill=red, minimum size=6pt, inner sep=0pt] at (\i,4) {};
}

\draw[->,black,thick] (6.2,2.3)-- ++(1,0);
        \node at (6.7,2.5) {$\mathcal{U}_{\text{flip}}$};

\fill[white, opacity=0.7, even odd rule]
  (-1,-1) rectangle (6.2,6)   % outer full-coverage rectangle
  (1,1) rectangle (5,4);    % inner hole (highlight area)

\end{tikzpicture}

}
\hspace{-1.5cm}
\scalebox{0.9}{

\begin{tikzpicture}

% Define colors
\definecolor{myblue}{RGB}{0,0,255}

\filldraw[fill=white, draw=white] (1,1) rectangle (5,4);

%dual lattice edges

%arrows to indicate the structure gauge-invariant X variables

\foreach \i in {1,2,3,4}
{ 
\foreach \j in {1,2,3} 
{
    \ifnum \i = 1
        \ifnum \j = 1
            \filldraw[fill = orange!20, draw = orange,thick] (\i+0.5,\j+0.5) -- ++(-0.5,0.5) -- ++(0,-1) -- ++(1,0) -- ++(-0.5,0.5);
            %\filldraw[fill = orange!20, draw = orange,thick] (\i,\j) -- ++(-1,0) -- ++(1,1
            \else
                \filldraw[fill = orange!20, draw = orange,thick] (\i+0.5,\j+0.5) -- ++(-0.5,0.5) -- ++(0,-1) -- ++(0.5,0.5);
                \filldraw[fill = orange!20, draw = orange,thick] (\i+0.5+4,\j+0.5) -- ++(-0.5,0.5) -- ++(0,-1) -- ++(0.5,0.5);
                 %\filldraw[fill = orange!20, draw = orange,thick] (\i,\j+1) -- ++(-0.5,-0.5) -- ++(0.5,-0.5);
         \fi
    \else
        \ifnum \i = 4
            \ifnum \j = 1
            \filldraw[fill = orange!20, draw = orange,thick] (\i,\j) -- ++(0.5,0.5) -- ++ (0.5,-0.5) -- ++ (-1,0);
            \filldraw[fill = magenta!20, draw = magenta,thick] (\i,4) -- ++(-0.5,-0.5) -- ++ (1,0) -- ++(-0.5,0.5);
            \filldraw[fill = magenta!20, draw = magenta,thick] (\i,1) -- ++(-0.5,-0.5) -- ++ (1,0) -- ++(-0.5,0.5);
            %\filldraw[fill = orange!20, draw = orange,thick] (\i+1,\j) -- ++(1,0) -- ++(-1,1);
        \else
                \filldraw[fill = magenta!20, draw = magenta,thick] (\i+0.5,\j+0.5) -- ++(0.5,-0.5) -- ++ (-0.5,-0.5) -- ++(0,1);
                 \filldraw[fill = magenta!20, draw = magenta,thick] (\i+0.5-4,\j+0.5) -- ++(0.5,-0.5) -- ++ (-0.5,-0.5) -- ++(0,1);
        \fi
        \else
            \ifnum \j = 1
                \filldraw[fill = orange!20, draw = orange,thick] (\i,\j) -- ++(0.5,0.5) -- ++ (0.5,-0.5) -- ++ (-1,0);
                \filldraw[fill = orange!20, draw = orange,thick] (\i,\j+3) -- ++(0.5,0.5) -- ++ (0.5,-0.5) -- ++ (-1,0);
                %\filldraw[fill = orange!20, draw = orange,thick] (\i,\j) -- ++(0.5,-0.5) -- ++ (0.5,0.5) -- ++ (-1,0);
            \else
                \ifnum \j = 3
                    \filldraw[fill = magenta!20, draw = magenta,thick] (\i,\j+1) -- ++(-0.5,-0.5) -- ++ (1,0) -- ++(-0.5,0.5);
                    \filldraw[fill = magenta!20, draw = magenta,thick] (\i,\j+1-3) -- ++(-0.5,-0.5) -- ++ (1,0) -- ++(-0.5,0.5);
                    %\filldraw[fill = orange!20, draw = orange,thick] (\i,\j+1) -- ++(0.5,0.5) -- ++ (0.5,-0.5) -- ++ (-1,0);
                \fi
            \fi
        \fi
    \fi
}
}
%stuff outside M
\filldraw[fill = orange!20, draw = orange,thick] (1+0.5,4+0.5) -- ++(-0.5,0.5) -- ++(0,-1) -- ++(1,0) -- ++(-0.5,0.5);
\filldraw[fill = orange!20, draw = orange,thick] (1+0.5,0+0.5) -- ++(-0.5,0.5) -- ++(0,-1) -- ++(0.5,0.5);
\filldraw[fill = orange!20, draw = orange,thick] (5+0.5,0+0.5) -- ++(-0.5,0.5) -- ++(0,-1) -- ++(0.5,0.5);
\filldraw[fill = orange!20, draw = orange,thick] (5+0.5,1+0.5) -- ++(-0.5,0.5) -- ++(0,-1) -- ++(1,0) -- ++(-0.5,0.5);
\filldraw[fill = orange!20, draw = orange,thick] (0,1) -- ++(0.5,0.5) -- ++ (0.5,-0.5) -- ++ (-1,0);
\filldraw[fill = orange!20, draw = orange,thick] (0,4) -- ++(0.5,0.5) -- ++ (0.5,-0.5) -- ++ (-1,0);
\filldraw[fill = orange!20, draw = orange,thick] (5+0.5,4+0.5) -- ++(-0.5,0.5) -- ++(0,-1) -- ++(1,0) -- ++(-0.5,0.5);
\filldraw[fill = orange!20, draw = orange,thick] (4,4) -- ++(0.5,0.5) -- ++ (0.5,-0.5) -- ++ (-1,0);
%\filldraw[fill = magenta!20, draw = magenta,thick] (4,4) -- ++(-0.5,-0.5) -- ++ (1,0) -- ++(-0.5,0.5);
\draw[magenta,thick] (4+0.5,3+0.5) -- ++(0.5,0.5);
\draw[magenta,thick] (0.5,3+0.5) -- ++(0.5,0.5);
\draw[magenta,thick] (4+0.5,0.5) -- ++(0.5,0.5);

% \filldraw[fill = orange!20, draw = orange,thick] (1,1) -- ++(-1,0) -- ++(1,-1) -- ++(0,1);
% \filldraw[fill = orange!20, draw = orange,thick] (1,1) -- ++(1,0) -- ++(-1,-1) -- ++(0,1);

% \filldraw[fill = orange!20, draw = orange,thick] (1,4) -- ++(-1,0) -- ++(1,1) -- ++(0,-1);
% \filldraw[fill = orange!20, draw = orange,thick] (1,4) -- ++(1,0) -- ++(-1,1) -- ++(0,-1);

% \filldraw[fill = orange!20, draw = orange,thick] (5,4) -- ++(-1,0) -- ++(1,1) -- ++(0,-1);
% \filldraw[fill = orange!20, draw = orange,thick] (5,4) -- ++(1,0) -- ++(-1,1) -- ++(0,-1);

% \filldraw[fill = orange!20, draw = orange,thick] (5,1) -- ++(-1,0) -- ++(1,-1) -- ++(0,1);
% \filldraw[fill = orange!20, draw = orange,thick] (5,1) -- ++(1,0) -- ++(-1,-1) -- ++(0,1);

\foreach \i in {0,1,2,3,4,5}
{ 
\foreach \j in {0,1,2,3,4} 
{
    \node[circle, draw, fill=black, minimum size=4pt, inner sep=0pt] at (\i+0.5, \j+0.5){ };
}
}

\foreach \i in {0,1,2,3,4}
{ 
\foreach \j in {0,1,2,3} {
    \ifnum \j < 3
        \ifnum \i < 4
        \node at (\i+0.72, \j+0.72) {\footnotesize $\i,\j$};
        \else
            \ifnum \j < 3
            \node at (\i+0.72, \j+0.72) {\footnotesize $0,\j$};
            \fi
        \fi
    \else
        \ifnum \i < 4
        \node at (\i+0.72, \j+0.72) {\footnotesize $\i,0$};
        \fi
    \fi
}
}

\node at (4+0.72, 3+0.72) {\footnotesize $0,0$};

\draw[blue,->,very thick] (0.85,-0.15) to[out=195, in=345, looseness=1.5] (5+0.15,-0.15);
\draw[blue,->,very thick] (-0.25+0.3,1-0.25) to[out=255, in=105, looseness=1.5] (-0.25+0.3,4.25-0.25);

\draw[draw=blue, very thick] (1,1) rectangle (5,4);
\node at (5.25,4.25) {\textcolor{blue}{$M$}};

%dual lattice sites

\foreach \i in {0,1,2,3,4,5}
{
    \node[diamond, draw, fill=red, minimum size=6pt, inner sep=0pt] at (1, \i) {};
    \node[diamond, draw, fill=red, minimum size=6pt, inner sep=0pt] at (5, \i) {};
}

\foreach \i in {0,2,3,4,6}
{
    \node[diamond, draw, fill=red, minimum size=6pt, inner sep=0pt] at (\i,1) {};
    \node[diamond, draw, fill=red, minimum size=6pt, inner sep=0pt] at (\i,4) {};
}

\foreach \i in {1,2,3} {
    \foreach \j in {1,2} {
        
        % Draw links (horizontal)
        \ifnum\j<3
            \draw (\i+0.5, \j+0.5) -- ++ (0,1);
        \fi

        % Draw links (vertical)
        \ifnum\i<4
            \draw (\i+0.5, \j+0.5) -- ++ (1,0);
        \fi
    }
}

\foreach \j in {1,2}
{
    \ifnum\j<3
            \draw (5+0.5, \j+0.5) -- ++ (0,1);
        \fi
}

\draw[magenta, thick] (0.5,0.5) -- ++(0.5,0.5);

\fill[white, opacity=0.7, even odd rule]
  (-1,-1) rectangle (6.2,6)   % outer full-coverage rectangle
  (1,1) rectangle (5,4);    % inner hole (highlight area)

\end{tikzpicture}
}
\caption{Original theory in the figure is on a 4 $\times$ 3 periodic lattice. The blue boundary $M$ indicates the extent of the lattice, everything outside $M$ is part of the same unit being repeated, to make apparent the nature of periodic boundary conditions. The defect lines (indicated in magenta and orange) obtained by applying $\mathcal{U}_s$ to $\tilde H_{\sf XM}$.}
\label{fig:initial_fusion_step}
\end{figure}

\begin{center}
   \textbf{The $\mathcal{U}_s$ Operator} 
\end{center}
To see a connection with partial gauging, consider the action upon $\tilde H_{\sf XM}$ by $\mathcal{U}'_s$.
\begin{align}
    & \mathcal{U}'_s := \left( \prod_{\bar q \in L_{\sigma}} \left( \prod_{\langle \bar q,q \rangle} C^{\sf z}_{\bar q,q} \right) \right) \left( \prod_{\bar q \in L_{\sigma}} H_{\sigma} \right) \ \ \text{where} \ \langle \ ,\rangle \ \text{indicate nearest neighbors} \\
    &  \mathcal{U}'_s \tilde H_{\sf XM}[g]  (\mathcal{U}'_s)^{-1} = \sum_{q \in L_0} \left( g^{-1} \sigma^{x}_{q} + g X_q \left( \prod_{\langle q' , q \rangle } \sigma^{z}_{q'} \right) \right) + \nonumber \\
    &\sum_{q \in L_1 \setminus \{ (0,1),(1,0) \} } \left( g^{-1} Z_q Z_{q+\hat x } Z_{q+\hat y } Z_{q + \hat x + \hat y} + g X_q \left( \prod_{\langle q' , q \rangle} \sigma^{z}_{q'} \right) \right) + \sum_{2 \leq l \leq k_{\text{max}}-1} H_l
\end{align}
See Fig.(\ref{fig:initial_fusion_step}). The original matter fields $X,Z$ of this theory \textit{resemble the gauge invariant matter fields $\tilde X, \tilde Z$ obtained upon performing \textbf{partial} gauging (as opposed to a full gauging) on an ancilla-coupled rectangular sub-region, as in} Eq.(\ref{eqn:gauged_rectangular_region_hamiltonian}). \textit{Thus, we have found an explicit unitary that links the minimally gauged Hamiltonian to a specific gauge sector of the globally ancilla-coupled Hamiltonian}. The idea is to find a further sequence of unitaries whose combined action will map the minimally gauged Hamiltonian to its Bond-Algebraic dual (which must exist on grounds of Von Neumann's theorem).
Mirroring the discussion for the cylinder, it will be convenient to ``flip" one of these defect lines.
\begin{align}
\label{eqn:U_flip_torus}
\mathcal{U}_{\text{flip}} = \left( \prod_{1 \leq k \leq L_y-1} \left( C^{\sf z}_{(\bar 0 , \bar k),(0,k)} C^{\sf z}_{(\bar 0 , \bar k),(0,k+1)} \right) \prod_{1 \leq k \leq L_x-1} \left( C^{\sf z}_{(\bar k , \bar 0),(k,0)} C^{\sf z}_{(\bar k , \bar 0),(k+1,0)} \right) \right)  C^{\sf z}_{(\bar 0 , \bar 0) , (0,0)}
\end{align}
The inverse fusion operator $\mathcal{U}_s$ is defined as for the infinite cylinder :  
\begin{equation}
\label{eq:U_s_torus}
\mathcal{U}_s = \mathcal{U}_{\text{flip}} \mathcal{U}'_s 
\end{equation}
Upon conjugation, we obtain the defect Hamiltonian consisting of two defects ``adjacent" to each other. Schematically, we the defect lines by $H^{\mathcal{D},m}_{\text{outer}}$ to indicate the layer $L_m$ upon which the lower left layer of the outer term is supported. $H^{\mathcal{D}}_{\text{inner}}$ is \textit{always} understood to be the corresponding ``dual" defect to $H^{\mathcal{D},m}_{\text{outer}}$ at every movement step, and is thus not labeled by any further index.
The terms not involved in forming defect configurations are a part of the ``bulk". 
The following is a schematic of the Hamiltonian hosting a pair of defects which are unmoved : 
\begin{equation}
    \mathcal{U}_s \tilde H_{\sf XM}[g]  (\mathcal{U}_s)^{-1} = \textcolor{orange}{H^{\mathcal{D},\bar 0}_{\text{outer}}} + \textcolor{magenta}{H^{\mathcal{D}}_{\text{inner}}} + H^{(1)}_{\text{bulk}}
\end{equation}
where 
\begin{equation}
\label{eq:H_bulk}
    H^{(m+1)}_{\text{bulk}} = \sum^{k_{\text{max}}}_{k=m+1} H_k[g] + \sum_{q \in L_0 \setminus L_1} g X_{q} + (1-\delta_{m,0}) \left( \textcolor{red}{H_{\sigma}[g^{-1}] + \sum_{k=1}^{m-1} H_k[g^{-1}] + \sum_{q \in L_{m} \setminus L_0 } g^{-1} X_q } \right)
 \end{equation}
The terms are color-coded to match with Fig.(\ref{fig:initial_fusion_step}) (and Fig.(\ref{fig:u1_movement_result_torus})) if we let $L_x = 4$ and $L_y = 3$. 
\\
\\
Intuitively, $H^{(m+1)}_{\text{bulk}}$ denotes the sum of the $L_x - m-1 \times L_y - m-1$ ``untransformed"  XM local terms (black in Fig.(\ref{fig:u1_movement_result_torus})) \textbf{and} the ``duality-transformed" local terms (red in Fig.(\ref{fig:u1_movement_result_torus})). 
\\
\\
Explicitly : 
\begin{align}
\label{eqn:fusion_step_xm_torus}
& \mathcal{U}_s \tilde H_{\sf XM}[g]  (\mathcal{U}_s)^{-1} \nonumber \\ 
& = \left( \textcolor{magenta}{g^{-1} \sigma^x_{\bar 0 , \bar 0} Z_{0,0}} + g X_{0,0} \right) + \left( \textcolor{magenta}{g^{-1} \sigma^x_{\bar 1 , \bar 0} Z_{1,0} Z_{2,0}} + \textcolor{orange}{g X_{1,0} \sigma^z_{\bar 0 , \bar L_y-1} \sigma^z_{\bar 0 , \bar 0}}  \right) + \left( \textcolor{magenta}{g^{-1} \sigma^x_{\bar 0 , \bar 1} Z_{0,1} Z_{0,2}} + \textcolor{orange}{g X_{0,1} \sigma^z_{\bar L_x-1 , \bar 0} \sigma^z_{\bar 0 , \bar 0}} \right)  \nonumber \\ 
& + \sum_{2 \leq k \leq L_x-1} \left( \textcolor{magenta}{g^{-1} \sigma^x_{\bar k , \bar 0} Z_{k,0} Z_{k+1,0}} + g X_{k,0} \right) + \sum_{2 \leq k \leq L_y-1} \left( \textcolor{magenta}{g^{-1} \sigma^x_{\bar 0 , \bar k} Z_{0,k} Z_{0,k+1}} + g X_{0,k} \right) \\ 
& + \sum_{q \in L_1 \setminus \{ (0,1),(1,0) \} } \left( g^{-1} Z_q Z_{q+\hat x } Z_{q+\hat y } Z_{q + \hat x + \hat y} + \textcolor{orange}{g X_q \left( \prod_{\langle \bar q' , q \rangle} \sigma^{z}_{q'} \right)} \right) + \sum_{2 \leq l < k_{\text{max}}} H_l \ .\nonumber 
\end{align}
%\subsection{The Movement Operators}
\begin{center}
   \textbf{The Movement Operators} 
\end{center}
The approach behind constructing this unitary is to ``move" the outer (orange in Fig.(\ref{fig:initial_fusion_step})) defect line toward the inner (magenta) line. The structure of the unitary achieving this transformation will be constructed on the grounds that it mimic that of the cylinder \textit{along the edges}, with crucial modifications at the boundaries. The unitaries are highly non-local and are supported between two successive layers $L_k, L_{k+1}$ across which the defect line is moved. 
The movement operator to move the defect line across layers $L_m$ and $L_{m+1}$ is : 
\begin{align}
& \mathcal{U}_{m,m+1} := \mathcal{U}^{0}_m \left( \prod_{q \in L_{m+1} \setminus \{ (0,(m+1) \text{mod} L_y),((m+1) \text{mod} L_x,0) \} } C^{\sf z}_{q,q+\hat x } C^{\sf z}_{q,q+\hat y } C^{\sf z}_{q,q + \hat x + \hat y} \right) \nonumber \\
& \left( \prod_{q \in L_{m+1}} H_q \right) \left( \prod_{m+2 \leq k \leq L_y} C^{\sf x}_{(m+1,k),(m+1,k-1)} \right) \left( \prod_{m+2 \leq k \leq L_x} C^{\sf x}_{(k,m+1),(k-1,m+1)} \right) \\ \nonumber \\
& \text{where} \ \ \mathcal{U}^0_{m} = C^{\sf z}_{((m+1)\text{mod}L_x,0),((m+2)\text{mod}L_x,0)} C^{\sf z}_{(0,(m+1)\text{mod}L_y),(0,(m+2)\text{mod}L_y)} \ \ \text{if} \ m < L_y-1 \\ \nonumber \\
& \text{and} \ \ \mathcal{U}^0_{L_y-1} = I~.
\end{align}
However, the Hamiltonians obtained upon defect motion defined by these movement operators are \textbf{not} unitarily equivalent to a defect Hamiltonian obtained by partial gauging (see \ref{sec:gauging} \ref{sec:partial_gauging}) a portion of the lattice. Explicitly, this manifests in the form of additional terms which we call \textit{tears}. They are formed because $\mathcal{U}_{\bar 0,1}$ maps one of the terms of the outer defect line to a tear : $X_{s}ZZ \to X_s \sigma^z$, occurring always at the intersections of the inner and outer defect lines. On the infinite cylinder, the infinite defects lines do not intersect and hence do not suffer tears. The `defect Hamiltonian' at each stage is schematically : 
\begin{equation}
  \left(\mathcal{U}_{m,m-1} \cdots \mathcal{U}_s \right)  \tilde H_{\sf XM}[g]  (\mathcal{U}_{m,m-1} \cdots \mathcal{U}_s)^{-1} = \textcolor{orange}{H^{\mathcal{D},m}_{\text{outer}}} + \textcolor{magenta}{H^{\mathcal{D}}_{\text{inner}}} + \textcolor{cyan}{H^{m}_{\text{tears}}} + H^{(m+1)}_{\text{bulk}},  
\end{equation}
where, for $1 \leq m \leq L_y$
\begin{equation}
    \label{eqn:tear_terms}
    H^{m}_{\text{tears}} = \sum^{m}_{k=1} \left( X_{k,0} \sigma^x_{k,0} + X_{0,k} \sigma^x_{0,k} \right) + \delta_{m,L_y} \sum_{k=L_y}^{L_x-1} X_{k,0} \sigma^x_{k,0}~.
\end{equation}
% \an{where $H_{\text{bulk}}$ is a restriction of $H_{\sf XM}[g]$ to the bulk region of size $(L_x-1) \times (L_y-1)$ (unshaded region in Fig.~\ref{fig:u1_movement_result_torus}).}
Concretely, the first movement operator $\mathcal{U}_{\bar 0,1}$ moves the outer defect supported between layers $L_{\sigma}$ and $L_1$ inward, to a ``defect" supported between $L_1$ and $L_2$. It is given by : 
\begin{align}
\label{eqn:U1_movement_operator_torus}
& \mathcal{U}_{\bar 0,1} = \left( C^{\sf z}_{(1,0),(2,0)} C^{\sf z}_{(0,1),(0,2)} \right) \left( \prod_{q \in L_1 \setminus \{ (0,1),(1,0) \} } C^{\sf z}_{q,q+\hat x } C^{\sf z}_{q,q+\hat y } C^{\sf z}_{q,q + \hat x + \hat y} \right) \nonumber \\
& \left( \prod_{q \in L_1} H_q \right) \left( \prod_{2 \leq k \leq L_y} C^{\sf x}_{(1,k),(1,k-1)} \right) \left( \prod_{2 \leq k \leq L_x} C^{\sf x}_{(k,1),(k-1,1)} \right) 
\end{align}

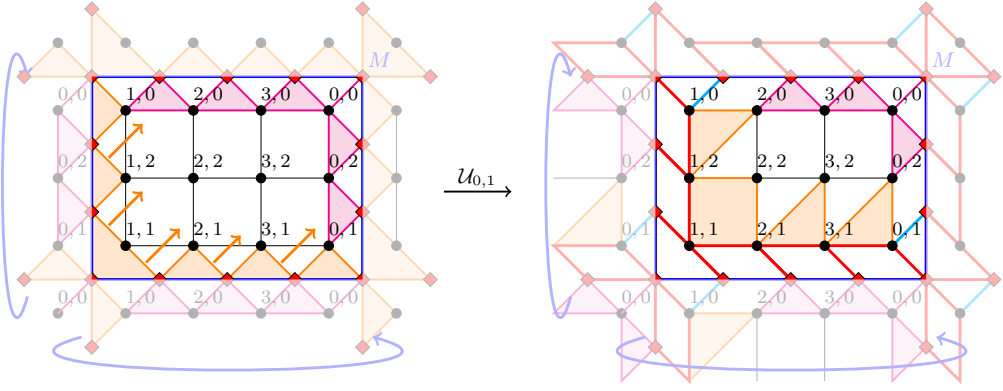
\begin{figure}[ht]
\centering
\scalebox{0.9}
{
\begin{tikzpicture}

% Define colors
\definecolor{myblue}{RGB}{0,0,255}

\filldraw[fill=white, draw=white] (1,1) rectangle (5,4);

%dual lattice edges

%arrows to indicate the structure gauge-invariant X variables

\foreach \i in {1,2,3,4}
{ 
\foreach \j in {1,2,3} 
{
    \ifnum \i = 1
        \ifnum \j = 1
            \filldraw[fill = orange!20, draw = orange,thick] (\i+0.5,\j+0.5) -- ++(-0.5,0.5) -- ++(0,-1) -- ++(1,0) -- ++(-0.5,0.5);
            %\filldraw[fill = orange!20, draw = orange,thick] (\i,\j) -- ++(-1,0) -- ++(1,1
            \else
                \filldraw[fill = orange!20, draw = orange,thick] (\i+0.5,\j+0.5) -- ++(-0.5,0.5) -- ++(0,-1) -- ++(0.5,0.5);
                \filldraw[fill = orange!20, draw = orange,thick] (\i+0.5+4,\j+0.5) -- ++(-0.5,0.5) -- ++(0,-1) -- ++(0.5,0.5);
                 %\filldraw[fill = orange!20, draw = orange,thick] (\i,\j+1) -- ++(-0.5,-0.5) -- ++(0.5,-0.5);
         \fi
    \else
        \ifnum \i = 4
            \ifnum \j = 1
            \filldraw[fill = orange!20, draw = orange,thick] (\i,\j) -- ++(0.5,0.5) -- ++ (0.5,-0.5) -- ++ (-1,0);
            \filldraw[fill = magenta!20, draw = magenta,thick] (\i,4) -- ++(-0.5,-0.5) -- ++ (1,0) -- ++(-0.5,0.5);
            \filldraw[fill = magenta!20, draw = magenta,thick] (\i,1) -- ++(-0.5,-0.5) -- ++ (1,0) -- ++(-0.5,0.5);
            %\filldraw[fill = orange!20, draw = orange,thick] (\i+1,\j) -- ++(1,0) -- ++(-1,1);
        \else
                \filldraw[fill = magenta!20, draw = magenta,thick] (\i+0.5,\j+0.5) -- ++(0.5,-0.5) -- ++ (-0.5,-0.5) -- ++(0,1);
                 \filldraw[fill = magenta!20, draw = magenta,thick] (\i+0.5-4,\j+0.5) -- ++(0.5,-0.5) -- ++ (-0.5,-0.5) -- ++(0,1);
        \fi
        \else
            \ifnum \j = 1
                \filldraw[fill = orange!20, draw = orange,thick] (\i,\j) -- ++(0.5,0.5) -- ++ (0.5,-0.5) -- ++ (-1,0);
                \filldraw[fill = orange!20, draw = orange,thick] (\i,\j+3) -- ++(0.5,0.5) -- ++ (0.5,-0.5) -- ++ (-1,0);
                %\filldraw[fill = orange!20, draw = orange,thick] (\i,\j) -- ++(0.5,-0.5) -- ++ (0.5,0.5) -- ++ (-1,0);
            \else
                \ifnum \j = 3
                    \filldraw[fill = magenta!20, draw = magenta,thick] (\i,\j+1) -- ++(-0.5,-0.5) -- ++ (1,0) -- ++(-0.5,0.5);
                    \filldraw[fill = magenta!20, draw = magenta,thick] (\i,\j+1-3) -- ++(-0.5,-0.5) -- ++ (1,0) -- ++(-0.5,0.5);
                    %\filldraw[fill = orange!20, draw = orange,thick] (\i,\j+1) -- ++(0.5,0.5) -- ++ (0.5,-0.5) -- ++ (-1,0);
                \fi
            \fi
        \fi
    \fi
}
}
%stuff outside M
\filldraw[fill = orange!20, draw = orange,thick] (1+0.5,4+0.5) -- ++(-0.5,0.5) -- ++(0,-1) -- ++(1,0) -- ++(-0.5,0.5);
\filldraw[fill = orange!20, draw = orange,thick] (1+0.5,0+0.5) -- ++(-0.5,0.5) -- ++(0,-1) -- ++(0.5,0.5);
\filldraw[fill = orange!20, draw = orange,thick] (5+0.5,0+0.5) -- ++(-0.5,0.5) -- ++(0,-1) -- ++(0.5,0.5);
\filldraw[fill = orange!20, draw = orange,thick] (5+0.5,1+0.5) -- ++(-0.5,0.5) -- ++(0,-1) -- ++(1,0) -- ++(-0.5,0.5);
\filldraw[fill = orange!20, draw = orange,thick] (0,1) -- ++(0.5,0.5) -- ++ (0.5,-0.5) -- ++ (-1,0);
\filldraw[fill = orange!20, draw = orange,thick] (0,4) -- ++(0.5,0.5) -- ++ (0.5,-0.5) -- ++ (-1,0);
\filldraw[fill = orange!20, draw = orange,thick] (5+0.5,4+0.5) -- ++(-0.5,0.5) -- ++(0,-1) -- ++(1,0) -- ++(-0.5,0.5);
\filldraw[fill = orange!20, draw = orange,thick] (4,4) -- ++(0.5,0.5) -- ++ (0.5,-0.5) -- ++ (-1,0);
%\filldraw[fill = magenta!20, draw = magenta,thick] (4,4) -- ++(-0.5,-0.5) -- ++ (1,0) -- ++(-0.5,0.5);
\draw[magenta,thick] (4+0.5,3+0.5) -- ++(0.5,0.5);
\draw[magenta,thick] (0.5,3+0.5) -- ++(0.5,0.5);
\draw[magenta,thick] (4+0.5,0.5) -- ++(0.5,0.5);

% \filldraw[fill = orange!20, draw = orange,thick] (1,1) -- ++(-1,0) -- ++(1,-1) -- ++(0,1);
% \filldraw[fill = orange!20, draw = orange,thick] (1,1) -- ++(1,0) -- ++(-1,-1) -- ++(0,1);

% \filldraw[fill = orange!20, draw = orange,thick] (1,4) -- ++(-1,0) -- ++(1,1) -- ++(0,-1);
% \filldraw[fill = orange!20, draw = orange,thick] (1,4) -- ++(1,0) -- ++(-1,1) -- ++(0,-1);

% \filldraw[fill = orange!20, draw = orange,thick] (5,4) -- ++(-1,0) -- ++(1,1) -- ++(0,-1);
% \filldraw[fill = orange!20, draw = orange,thick] (5,4) -- ++(1,0) -- ++(-1,1) -- ++(0,-1);

% \filldraw[fill = orange!20, draw = orange,thick] (5,1) -- ++(-1,0) -- ++(1,-1) -- ++(0,1);
% \filldraw[fill = orange!20, draw = orange,thick] (5,1) -- ++(1,0) -- ++(-1,-1) -- ++(0,1);

\foreach \i in {0,1,2,3,4,5}
{ 
\foreach \j in {0,1,2,3,4} {
    \node[circle, draw, fill=black, minimum size=4pt, inner sep=0pt] at (\i+0.5, \j+0.5) {};
}
}

\foreach \i in {0,1,2,3,4}
{ 
\foreach \j in {0,1,2,3} {
    \ifnum \j < 3
        \ifnum \i < 4
        \node at (\i+0.72, \j+0.72) {\footnotesize $\i,\j$};
        \else
            \ifnum \j < 3
            \node at (\i+0.72, \j+0.72) {\footnotesize $0,\j$};
            \fi
        \fi
    \else
        \ifnum \i < 4
        \node at (\i+0.72, \j+0.72) {\footnotesize $\i,0$};
        \fi
    \fi
}
}

\node at (4+0.72, 3+0.72) {\footnotesize $0,0$};

\draw[blue,->,very thick] (0.85,-0.15+0.3) to[out=195, in=345, looseness=1.5] (5+0.15,-0.15+0.3);
\draw[blue,->,very thick] (-0.25+0.3,1-0.25) to[out=255, in=105, looseness=1.5] (-0.25+0.3,4.25-0.25);

\draw[draw=blue, very thick] (1,1) rectangle (5,4);
\node at (5.25,4.25) {\textcolor{blue}{$M$}};

%dual lattice sites

\foreach \i in {0,1,2,3,4,5}
{
    \node[diamond, draw, fill=red, minimum size=6pt, inner sep=0pt] at (1, \i) {};
    \node[diamond, draw, fill=red, minimum size=6pt, inner sep=0pt] at (5, \i) {};
}

\foreach \i in {0,2,3,4,6}
{
    \node[diamond, draw, fill=red, minimum size=6pt, inner sep=0pt] at (\i,1) {};
    \node[diamond, draw, fill=red, minimum size=6pt, inner sep=0pt] at (\i,4) {};
}

\foreach \i in {1,2,3} {
    \foreach \j in {1,2} {
        
        % Draw links (horizontal)
        \ifnum\j<3
            \draw (\i+0.5, \j+0.5) -- ++ (0,1);
        \fi

        % Draw links (vertical)
        \ifnum\i<4
            \draw (\i+0.5, \j+0.5) -- ++ (1,0);
        \fi
    }
}

\foreach \j in {1,2}
{
    \ifnum\j<3
            \draw (5+0.5, \j+0.5) -- ++ (0,1);
        \fi
}

\draw[magenta, thick] (0.5,0.5) -- ++(0.5,0.5);

\draw[->,black,thick] (6.2,2.3)-- ++(1,0);
        \node at (6.7,2.5) {$\mathcal{U}_{0,1}$};

\fill[white, opacity=0.7, even odd rule]
  (-1,-1) rectangle (6.2,6)   % outer full-coverage rectangle
  (1,1) rectangle (5,4);    % inner hole (highlight area)

\draw[->,orange,very thick] (1.25,1.8) --++(0.5,0.5);
\draw[->,orange,very thick] (1.25,2.8) --++(0.5,0.5);
\draw[->,orange,very thick] (1.8,1.25) --++(0.5,0.5);
\draw[->,orange,very thick] (2.8,1.25) --++(0.4,0.4);
\draw[->,orange,very thick] (3.8,1.25) --++(0.5,0.5);

\end{tikzpicture}
}
\hspace{-1.2cm}
\scalebox{0.9}
{
\begin{tikzpicture}

% Define colors
\definecolor{myblue}{RGB}{0,0,255}

\filldraw[fill=white, draw=white] (1,1) rectangle (5,4);

%dual lattice edges

%arrows to indicate the structure gauge-invariant X variables

\foreach \i in {3,4}
{ 
    \filldraw[fill = magenta!20, draw = magenta,thick] (\i,4) -- ++(-0.5,-0.5) -- ++ (1,0) -- ++(-0.5,0.5);
     \filldraw[fill = magenta!20, draw = magenta,thick] (\i,1) -- ++(-0.5,-0.5) -- ++ (1,0) -- ++(-0.5,0.5);
}
\filldraw[fill = magenta!20, draw = magenta,thick] (5,3) -- ++(-0.5,0.5) -- ++ (0,-1) -- ++(0.5,0.5);
\filldraw[fill = magenta!20, draw = magenta,thick] (5,0) -- ++(-0.5,0.5) -- ++ (0,-1) -- ++(0.5,0.5);
\filldraw[fill = magenta!20, draw = magenta,thick] (1,0) -- ++(-0.5,0.5) -- ++ (0,-1) -- ++(0.5,0.5);
\filldraw[fill = magenta!20, draw = magenta,thick] (1,3) -- ++(-0.5,0.5) -- ++ (0,-1) -- ++(0.5,0.5);

\filldraw[fill = magenta!20, draw = magenta,thick] (0,1) -- ++(-0.5,-0.5) -- ++ (1,0) -- ++(-0.5,0.5);
\filldraw[fill = magenta!20, draw = magenta,thick] (0,4) -- ++(-0.5,-0.5) -- ++ (1,0) -- ++(-0.5,0.5);

\foreach \i in {3,4}
{ 
    \filldraw[fill = orange!20, draw = orange,thick] (\i+0.5,2.5) -- ++(-1,-1)-- ++(1,0) -- ++(0,1);
}
\filldraw[fill = orange!20, draw = orange,thick] (2.5,3.5) -- ++(-1,-1) -- ++(0,1) -- ++(1,0);

\filldraw[fill = orange!20, draw = orange,thick] (2.5,2.5) -- ++(-1,0) -- ++(0,-1) -- ++(1,0) -- ++(0,1);

\filldraw[fill = orange!20, draw = orange,thick] (0.5,2.5) -- ++(-1,-1)-- ++(1,0) -- ++(0,1);

\filldraw[fill = orange!20, draw = orange,thick] (2.5,2.5-2) -- ++(-1,0) -- ++(0,-1) -- ++(1,1);

\foreach \i in {1,2,3,4}
{ 
\foreach \j in {1,2,3} 
{
    \ifnum \i = 1
            \ifnum \j = 1
            \draw[red,very thick] (\i+0.5,\j+0.5) -- ++(-0.5,0.5) -- ++(0,-1) -- ++(1,0) -- ++(-0.5,0.5);
            %\filldraw[fill = orange!20, draw = orange,thick] (\i,\j) -- ++(-1,0) -- ++(1,1
            \else
                \draw[red,very thick] (\i+0.5,\j+0.5) -- ++(-0.5,0.5) -- ++(0,-1) -- ++(0.5,-0.5) -- ++(0,1);
                \draw[red,very thick] (\i+0.5+4,\j+0.5) -- ++(-0.5,0.5) -- ++(0,-1) -- ++(0.5,-0.5) -- ++(0,1);
            \fi
    \else
        \ifnum \i = 4
            \ifnum \j = 1
             \draw[red,very thick] (\i,\j) -- ++(1,0) -- ++ (-0.5,0.5) -- ++ (-1,0);
            \draw[red,very thick] (\i,\j+3) -- ++(1,0) -- ++ (-0.5,0.5) -- ++ (-1,0);
            %\filldraw[fill = orange!20, draw = orange,thick] (\i,\j) -- ++(0.5,0.5) -- ++ (0.5,-0.5) -- ++ (-1,0);
            % \filldraw[fill = magenta!20, draw = magenta,thick] (\i,4) -- ++(-0.5,-0.5) -- ++ (1,0) -- ++(-0.5,0.5);
            % \filldraw[fill = magenta!20, draw = magenta,thick] (\i,1) -- ++(-0.5,-0.5) -- ++ (1,0) -- ++(-0.5,0.5);
            %\filldraw[fill = orange!20, draw = orange,thick] (\i+1,\j) -- ++(1,0) -- ++(-1,1);
        \else
                % \filldraw[fill = magenta!20, draw = magenta,thick] (\i+0.5,\j+0.5) -- ++(0.5,-0.5) -- ++ (-0.5,-0.5) -- ++(0,1);
                %  \filldraw[fill = magenta!20, draw = magenta,thick] (\i+0.5-4,\j+0.5) -- ++(0.5,-0.5) -- ++ (-0.5,-0.5) -- ++(0,1);
        \fi
    \else
            \ifnum \j = 1
                \draw[red,very thick] (\i,\j) -- ++(1,0) -- ++ (-0.5,0.5) -- ++ (-1,0) -- ++(0.5,-0.5);
                \draw[red,very thick] (\i,\j+3) -- ++(1,0) -- ++ (-0.5,0.5) -- ++ (-1,0) -- ++(0.5,-0.5);
                %\filldraw[fill = orange!20, draw = orange,thick] (\i,\j) -- ++(0.5,-0.5) -- ++ (0.5,0.5) -- ++ (-1,0);
            \else
                \ifnum \j = 3
                    % \filldraw[fill = magenta!20, draw = magenta,thick] (\i,\j+1) -- ++(-0.5,-0.5) -- ++ (1,0) -- ++(-0.5,0.5);
                    % \filldraw[fill = magenta!20, draw = magenta,thick] (\i,\j+1-3) -- ++(-0.5,-0.5) -- ++ (1,0) -- ++(-0.5,0.5);
                    %\filldraw[fill = orange!20, draw = orange,thick] (\i,\j+1) -- ++(0.5,0.5) -- ++ (0.5,-0.5) -- ++ (-1,0);
                \fi
            \fi
        \fi
    \fi
}
}

%stuff outside M
%\filldraw[fill = magenta!20, draw = magenta,thick] (4,4) -- ++(-0.5,-0.5) -- ++ (1,0) -- ++(-0.5,0.5);
\draw[magenta,thick] (4+0.5,3+0.5) -- ++(0.5,0.5);
\draw[magenta,thick] (0.5,3+0.5) -- ++(0.5,0.5);
\draw[magenta,thick] (4+0.5,0.5) -- ++(0.5,0.5);

\draw[cyan,very thick] (1.5,3.5) -- ++(0.5,0.5);
\draw[cyan,very thick] (0.5,1.5) -- ++(0.5,0.5);
\draw[cyan,very thick] (0.5,1.5+3) -- ++(0.5,0.5);
\draw[cyan,very thick] (0.5+4,1.5+3) -- ++(0.5,0.5);
\draw[cyan,very thick] (4.5,1.5) -- ++(0.5,0.5);
\draw[cyan,very thick] (1.5,0.5) -- ++(0.5,0.5);
\draw[cyan,very thick] (1.5+4,0.5) -- ++(0.5,0.5);
\draw[cyan,very thick] (1.5+4,0.5+3) -- ++(0.5,0.5);

\draw (0.5+2,0.5) -- ++(0,-1);
\draw (0.5+3,0.5) -- ++(0,-1);
\draw (0.5,0.5+2) -- ++(-1,0);

\draw[red,very thick] (5.5,4.5) -- ++(-0.5,0.5) -- ++(0,-1) -- ++(1,0) -- ++(-0.5,0.5);

\draw[red,very thick] (1.5,4.5) -- ++(-0.5,0.5) -- ++(0,-1) -- ++(1,0) -- ++(-0.5,0.5);

\draw[red,very thick] (5.5,1.5) -- ++(-0.5,0.5) -- ++(0,-1) -- ++(1,0) -- ++(-0.5,0.5);

\draw[red,very thick] (1.5,0.5) -- ++(-0.5,0.5) -- ++(0,-1) -- ++(0.5,-0.5) -- ++(0,1);
\draw[red,very thick] (1.5+4,0.5) -- ++(-0.5,0.5) -- ++(0,-1) -- ++(0.5,-0.5) -- ++(0,1);

\draw[red,very thick] (0,1) -- ++(1,0) -- ++ (-0.5,0.5) -- ++ (-1,0) -- ++(0.5,-0.5);
\draw[red,very thick] (0,1+3) -- ++(1,0) -- ++ (-0.5,0.5) -- ++ (-1,0) -- ++(0.5,-0.5);

\foreach \i in {0,1,2,3,4,5}
{ 
\foreach \j in {0,1,2,3,4} 
{
    \node[circle, draw, fill=black, minimum size=4pt, inner sep=0pt] at (\i+0.5, \j+0.5) { };
}
}

\foreach \i in {0,1,2,3,4}
{ 
\foreach \j in {0,1,2,3} {
    \ifnum \j < 3
        \ifnum \i < 4
        \node at (\i+0.72, \j+0.72) {\footnotesize $\i,\j$};
        \else
            \ifnum \j < 3
            \node at (\i+0.72, \j+0.72) {\footnotesize $0,\j$};
            \fi
        \fi
    \else
        \ifnum \i < 4
        \node at (\i+0.72, \j+0.72) {\footnotesize $\i,0$};
        \fi
    \fi
}
}

\node at (4+0.72, 3+0.72) {\footnotesize $0,0$};

\draw[blue,->,very thick] (0.85,-0.15+0.3) to[out=195, in=345, looseness=1.5] (5+0.15,-0.15+0.3);
\draw[blue,->,very thick] (-0.25,1-0.25) to[out=255, in=105, looseness=1.5] (-0.25,4.25-0.25);

\draw[draw=blue, very thick] (1,1) rectangle (5,4);
\node at (5.25,4.25) {\textcolor{blue}{$M$}};

%dual lattice sites

\foreach \i in {0,1,2,3,4,5}
{
    \node[diamond, draw, fill=red, minimum size=6pt, inner sep=0pt] at (1, \i) {};
    \node[diamond, draw, fill=red, minimum size=6pt, inner sep=0pt] at (5, \i) {};
}

\foreach \i in {0,2,3,4,6}
{
    \node[diamond, draw, fill=red, minimum size=6pt, inner sep=0pt] at (\i,1) {};
    \node[diamond, draw, fill=red, minimum size=6pt, inner sep=0pt] at (\i,4) {};
}

\foreach \i in {2,3} {
    \foreach \j in {2} {
        
        % Draw links (horizontal)
        \ifnum\j<3
            \draw (\i+0.5, \j+0.5) -- ++ (0,1);
        \fi

        % Draw links (vertical)
        \ifnum\i<4
            \draw (\i+0.5, \j+0.5) -- ++ (1,0);
        \fi
    }
}

\draw[magenta, thick] (0.5,0.5) -- ++(0.5,0.5);

\fill[white, opacity=0.7, even odd rule]
  (-1,-1) rectangle (6.2,6)   % outer full-coverage rectangle
  (1,1) rectangle (5,4);    % inner hole (highlight area)

\end{tikzpicture}
}
    \caption{Result of Defect ``Motion" by $\mathcal{U}_{0,1}$. The red figures are interpreted as plaquette terms formed by the physical and ancilla sites ($ZZ \sigma^z \sigma^z$) - these are the first set of ``duality transformed" plaquette terms. Subsequent movement steps form $ZZZZ$ plaquette terms in the bulk from $X$ terms and vice versa by performing the Duality map on layers $L_{\geq 2}$. The cyan lines are the ``tears" in the outer defect line caused by moving the inner defect line.}
    \label{fig:u1_movement_result_torus}
\end{figure}
For completeness, we write down the full defect Hamiltonian after the first movement step. The terms here are color-coded to match Fig.(\ref{fig:u1_movement_result_torus}) with $L_x = 4$ and $L_y = 3$.
\begin{align}
\label{eqn:u1_defect_motion_result}
&(\mathcal{U}_{0,1} \mathcal{U}_s) \tilde H_{\sf XM} (\mathcal{U}_{0,1} \mathcal{U}_s)^{\dagger} = \\
& + \left[ \left( \textcolor{cyan}{g^{-1} \sigma^x_{\bar 1 , \bar 0} X_{1,0}} + \textcolor{red}{g Z_{1,0} Z_{1,L_y-1} \sigma^z_{\bar 0 , \bar L_y-1} \sigma^z_{\bar 0 , \bar 0}}  \right) \nonumber + \left( \textcolor{cyan}{g^{-1} \sigma^x_{\bar 0 , \bar 1} X_{0,1}} + \textcolor{red}{g Z_{0,1} Z_{L_x-1,1} \sigma^z_{\bar L_x-1 , \bar 0} \sigma^z_{\bar 0 , \bar 0}} \right) \right]_{\text{tears + plaquette terms}} \\
&\left( \textcolor{magenta}{g^{-1} \sigma^x_{\bar 0 , \bar 0} Z_{0,0}} + g X_{0,0} \right)_{\text{corner of outer defect line}} \nonumber \\
& + \left[ \left( \textcolor{magenta}{g^{-1} \sigma^x_{\bar 2 , \bar 0} Z_{2,0} Z_{3 \text{mod} L_x,0}} + \textcolor{orange}{g X_{2,0} Z_{1,0} Z_{1,L_y-1}} \right) + \left( \textcolor{magenta}{g^{-1} \sigma^x_{\bar 0 , \bar 2} Z_{0,2} Z_{0,3 \text{mod} L_y}} + \textcolor{orange}{g X_{0,2} Z_{0,1} Z_{L_x-1,1}} \right) \right]_{\text{intersection of defect lines}} \nonumber \\
& + \left[ \sum_{3 \leq k \leq L_x-1} \left( \textcolor{magenta}{g^{-1} \sigma^x_{\bar k , \bar 0} Z_{k,0} Z_{k+1,0}} + g X_{k,0} \right) + \sum_{3 \leq k \leq L_y-1} \left( \textcolor{magenta}{g^{-1} \sigma^x_{\bar 0 , \bar k} Z_{0,k} Z_{0,k+1}} + g X_{0,k} \right) \right]_{\text{edges of outer defect line}} \nonumber \\ 
& + \left[ \left( \textcolor{red}{g^{-1} X_{1,1}} + \textcolor{red}{g Z_{1,1} \sigma^z_{\bar 1, \bar 0} \sigma^z_{\bar 0, \bar 1} \sigma^z_{\bar 0, \bar 0}} \right) + \sum_{q \in L_1 \setminus \{ (0,1),(1,0) \} : \ q_y > q_x = 1} \left( \textcolor{red}{g^{-1} X_q} + \textcolor{red}{g Z_q Z_{q - \hat x} \sigma^z_{\bar{q}- \hat x- \hat y} \sigma^z_{\bar{q}- \hat y} } \right) \right. \nonumber \\
& \left. + \sum_{q \in L_1 \setminus \{ (0,1),(1,0) \} : \ q_x > q_y = 1} \left( \textcolor{red}{g^{-1} X_q} + \textcolor{red}{g Z_q Z_{q - \hat y} \sigma^z_{\bar{q}-\hat x- \hat y} \sigma^z_{\bar{q}- \hat x} } \right) \right]_{\text{``duality-transformed" terms}} \nonumber \\
& + \left( g^{-1} Z_{2,2} Z_{3 \text{mod} L_x,2} Z_{2,3 \text{mod} L_y} Z_{3 \text{mod} L_x,3 \text{mod} L_y}  + \textcolor{orange}{g X_{2,2} Z_{2,1} Z_{1,2} Z_{1,1}} \right)_{\text{corner of inner defect line}} \nonumber \\
& + \left[ \sum_{q \in L_2 \setminus \{ (0,2),(2,0) \} : \ q_y > q_x = 2} \left( g^{-1} Z_q Z_{q+\hat x } Z_{q+\hat y } Z_{q + \hat x + \hat y} + \textcolor{orange}{g X_q Z_{q - \hat x} Z_{q - \hat  x - \hat y}}  \right) \right. \nonumber \\
& \left. + \sum_{q \in L_2 \setminus \{ (0,2),(2,0) \} : \ q_x > q_y = 2} \left( g^{-1} Z_q Z_{q+\hat x } Z_{q+\hat y } Z_{q + \hat x + \hat y} + \textcolor{orange}{g X_q Z_{q - \hat y} Z_{q - \hat  x - \hat y}} \right) \right]_{\text{edges of inner defect line}} + \sum_{3 \leq l < k_{\text{max}}} H_l ~ .\nonumber 
\end{align}

\textit{Remark:} Each tear includes the $\sigma^z$ degrees of freedom \textit{not} involved in making up the (outer) defect term, which maintains unitary equivalence between the intermediate Hamiltonians. After the final movement step the outer line vanishes and the inner line is fully ``torn". We then apply the $\mathcal{U}_f$ operator to 'fix' these tears, the prescription directed by the required Bond-Algebraic Automorphism. This has no direct analogue to the cylindrical case. $\mathcal{U}_f$ here is interpreted as the fusion operator generalized to tears.

\begin{center}
   \textbf{The $\mathcal{U}_f$ operator} 
\end{center}

%\subsection{The $\mathcal{U}_f$ operator}

The motivation behind constructing $\mathcal{U}_f$ is by ensuring the boundary plaquette and transverse field terms get mapped to their correct Bond-Algebraic Duals by ``fixing the tears". The boundary plaquette terms are all mapped to tears upon application of the start and all movement operations : 
\begin{equation}
    \label{eqn:plaq_terms}
    Z_q Z_{q+\hat x } Z_{q+\hat y } Z_{q + \hat x + \hat y} \sigma^{z}_{q} \xrightarrow{\mathcal{U}_{L_y-1,0} \cdots \mathcal{U}_{0,1} \mathcal{U}_s} X_q \sigma^{x}_{q} \ \ , \ \ q \in L_0 
\end{equation}

We want $ X_q \sigma^{x}_{q} \to X_q$ to obtain the required isomorphism. So we first define : 
\begin{equation}
    \label{eqn:uf_1}
    \mathcal{U}^{(1)}_f := \prod_{q \in L_0} C^{\sf x}_{q,\bar q}
\end{equation}

Upon conjugation by $\mathcal{U}^{(1)}_f\mathcal{U}_{L_y} \cdots \mathcal{U}_1 \mathcal{U}_s$ the Hamiltonian $\tilde H_{\sf XM}[g]$ is almost completely duality transformed, barring the couplings involving ancilla sites : 
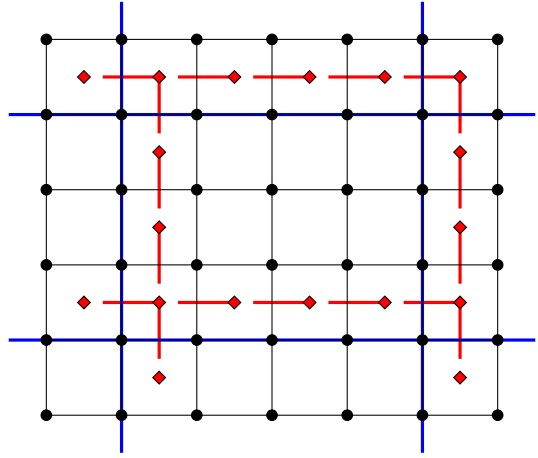
\begin{figure}[ht]
\centering
\begin{tikzpicture}
% Define colors
\definecolor{myblue}{RGB}{0,0,255}

\foreach \i in {2,3,4}
{
\draw[red,very thick] (\i+0.5,1.5) -- ++(-0.75,0);
\draw[red,very thick] (\i+0.5,4.5) -- ++(-0.75,0);
}

\foreach \j in {2,3}
{
\draw[red,very thick] (1.5,\j+0.5) -- ++(0,-0.75);
\draw[red,very thick] (5.5,\j+0.5) -- ++(0,-0.75);
}

\draw[red,very thick] (1.5,1.5) -- ++(-0.75,0); \draw[red,very thick] (1.5,1.5) -- ++(0,-0.75);

\draw[red,very thick] (5.5,1.5) -- ++(-0.75,0); \draw[red,very thick] (5.5,1.5) -- ++(0,-0.75);

\draw[red,very thick] (1.5,4.5) -- ++(-0.75,0); \draw[red,very thick] (1.5,4.5) -- ++(0,-0.75);

\draw[red,very thick] (5.5,4.5) -- ++(-0.75,0); \draw[red,very thick] (5.5,4.5) -- ++(0,-0.75);

%extra sites
\foreach \i in {1,2,3,4,5}
{ 
\node[diamond, draw, fill=red, minimum size=5pt, inner sep=0pt] at (\i+0.5, 1.5){};
\node[diamond, draw, fill=red, minimum size=5pt, inner sep=0pt] at (\i+0.5, 0.5+4){};
}
% Color the leftmost column nodes
%extra sites
\foreach \j in {2,3} {
    \node[diamond, draw, fill=red, minimum size=5pt, inner sep=0pt] at (1.5, \j+0.5) {};
    \node[diamond, draw, fill=red, minimum size=5pt, inner sep=0pt] at (0.5+5, \j+0.5) {};
}

\draw[blue,very thick] (1,1) -- ++(0,3) -- ++(4,0) -- ++(0,-3) -- ++(-4,0);
\draw[blue,very thick] (5,1) -- ++(1.5,0);
\draw[blue,very thick] (5,4) -- ++(1.5,0);
\draw[blue,very thick] (1,4) -- ++(0,1.5);
\draw[blue,very thick] (5,4) -- ++(0,1.5);
\draw[blue,very thick] (1,1) -- ++(0,-1.5);
\draw[blue,very thick] (1,1) -- ++(-1.5,0);
\draw[blue,very thick] (1,4) -- ++(-1.5,0);
\draw[blue,very thick] (5,1) -- ++(0,-1.5);

\node[diamond, draw, fill=red, minimum size=5pt, inner sep=0pt] at (0.5,1.5){};
\node[diamond, draw, fill=red, minimum size=5pt, inner sep=0pt] at (0.5,4.5){};
\node[diamond, draw, fill=red, minimum size=5pt, inner sep=0pt] at (1.5,0.5){};
\node[diamond, draw, fill=red, minimum size=5pt, inner sep=0pt] at (5.5,0.5){};

% Draw the lattice
\foreach \i in {0,1,2,3,4,5,6} 
{
    \foreach \j in {0,1,2,3,4,5}
    {
        % Draw the nodes
        \node[circle, draw, fill=black, minimum size=4pt, inner sep=0pt] at (\i, \j){};
        
        % Draw links (horizontal)
        \ifnum\j<5
            \draw (\i, \j) -- (\i, \j+1);
        \fi

        % Draw links (vertical)
        \ifnum\i<6
            \draw (\i, \j) -- (\i+1, \j);
        \fi
    } 
}
\end{tikzpicture}
    \caption{Couplings of ancilla sites to the boundary plaquettes before the application of $\mathcal{U}^{(2)}_f$. The red lines correspond to the extra site to which they are attached and indicate that this site is coupled to every plaquette in which the line is contained (see Eq.(\ref{eq:ancilla_coupling_b4_Uf_1}),(\ref{eq:ancilla_coupling_b4_Uf_2}),(\ref{eq:ancilla_coupling_b4_Uf_3}) for reference). Upon applying $\mathcal{U}^{(2)}_f$ these lines (couplings) 'disappear'.}
    \label{fig:ancilla_couplings_before_Uf(2)}
\end{figure}
\begin{align}
    & X_{0,0} \xrightarrow{\mathcal{U}^{(1)}_f\mathcal{U}_{L_y-1,0} \cdots \mathcal{U}_{0,1} \mathcal{U}_s} Z_{0,0} Z_{1,0} Z_{0,1} Z_{1,1} \sigma^z_{\bar 0, \bar 0} \sigma^z_{\bar 1 ,\bar 0} \sigma^z_{\bar 0 ,\bar 1} \label{eq:ancilla_coupling_b4_Uf_1}\\
    & X_{k \geq 1,0} \xrightarrow{\mathcal{U}^{(1)}_f\mathcal{U}_{L_y-1,0} \cdots \mathcal{U}_{0,1} \mathcal{U}_s} Z_{k,0} Z_{k+1,0} Z_{k,1} Z_{k+1,1} \sigma^z_{\bar k, \bar 1} \sigma^z_{\bar k+1 ,\bar 0} \label{eq:ancilla_coupling_b4_Uf_2} \\
    & X_{0,k \geq 1} \xrightarrow{\mathcal{U}^{(1)}_f\mathcal{U}_{L_y-1,0} \cdots \mathcal{U}_{0,1} \mathcal{U}_s} Z_{0,k} Z_{0,k+1} Z_{1,k} Z_{1,k+1} \sigma^z_{\bar 0 ,\bar k} \sigma^z_{\bar 0 , \bar k+1}. \label{eq:ancilla_coupling_b4_Uf_3}
\end{align}
We define $\mathcal{U}^{(2)}_f$ supported only on $L_{\sigma}$ to ensure the couplings of ancilla sites to plaquettes matches the Bond-Algebraic assignment. 

\begin{equation}
    \label{eqn:uf_2}
    \mathcal{U}^{(2)}_f := \left( \prod_{q \in L_{\sigma} \setminus \{ (\bar 0, \bar 0) \} } C^{\sf x}_{q,(\bar 0, \bar 0)} \right) \left( \prod_{1 \leq k \leq L_y-1} C^{\sf x}_{(\bar 0 , \bar{k}+1),(\bar 0, \bar k)} \right) \left( \prod_{1 \leq k \leq L_x-1} C^{\sf x}_{(\bar{k}+1,\bar 0),(\bar k , \bar 0)} \right)
\end{equation}

Hence, $\mathcal{U}_f := \mathcal{U}^{(2)}_f \mathcal{U}^{(1)}_f$ and we have 
\begin{equation}
    \label{eqn:final_unitary_xm_torus}
   ( \mathcal{U}_f \mathcal{U}_{L_y-1,0} \cdots \mathcal{U}_{0,1} \mathcal{U}_s ) \tilde H_{\sf XM}[g] ( \mathcal{U}_f \mathcal{U}_{L_y-1,0} \cdots \mathcal{U}_{0,1} \mathcal{U}_s )^{\dagger} = \tilde H_{\sf XM}[g^{-1}]
\end{equation}

as desired.

\subsection{Infinite Cylinder Limit} \label{sec:unitary_infinte_cyl}

The main idea of this procedure is discussed in the End Matter, we supply explicit expressions for the defect Hamiltonians for reference.

\begin{align}
\label{eqn:defect_hamiltonian_0}
& \mathcal{U}'_{s} \tilde H_{c} \mathcal{U'}^{\dagger}_{s} = H^{\mathcal{D}_L,\bar 0 ; \mathcal{D}_R,\bar 0}_{c} = \sum_{q_x=0} \left( g^{-1} \sigma^{x}_{q} + g  X_q \sigma^{z}_{q} \sigma^z_{q - \hat y} \right) \nonumber \\
& + \sum_{q_x = 1} \left( g^{-1} Z_{q}Z_{q+\hat x}Z_{q+\hat y}Z_{q + \hat x + \hat y} + g  X_q \sigma^z_{q - \hat x} \sigma^z_{q - \hat x - \hat y}  \right) + \sum_{2 \leq q_x < L_x} \left( g^{-1} Z_{q}Z_{q+\hat x }Z_{q+\hat y }Z_{q + \hat x + \hat y} + g  X_q  \right)
\end{align}

\begin{align}
\label{eqn:defect_hamiltonian_flipped}
& \mathcal{U}_{\text{flip}} H^{\mathcal{D}_L,\bar 0 ; \mathcal{D}_R,\bar 0}_{c} \mathcal{U}^{\dagger}_{\text{flip}} = H^{\mathcal{D}_R,0 ; \mathcal{D}_R,\bar 0}_{c} = \sum_{q_x=0} \left( g^{-1} Z_q Z_{q+\hat y } \sigma^{x}_{q} + g  X_q \right) + \sum_{q_x = 1} \left( g^{-1} Z_{q}Z_{q+\hat x }Z_{q+\hat y }Z_{q + \hat x + \hat y} + g  X_q \sigma^z_{q - \hat x} \sigma^z_{q - \hat x - \hat y}  \right) \nonumber \\
& + \sum_{2 \leq q_x < L_x} \left( g^{-1} Z_{q}Z_{q+\hat x }Z_{q+\hat y }Z_{q + \hat x + \hat y} + g  X_q  \right)~. 
\end{align}

\begin{align}
\label{eqn:defect_motion_in_hamiltonian}
& \mathcal{U}_{\bar 0,1} H^{\mathcal{D}_R,0;\mathcal{D}_R,\frac{1}{2}}_{c} \mathcal{U}_{\bar 0,1}^{\dagger} = H^{\mathcal{D}_R,0;\mathcal{D}_R,1}_{c} = \sum_{q_x=0} \left( g^{-1} Z_q Z_{q+\hat y } \sigma^{x}_{q} + g  X_q \right) + \sum_{q_x = 1} \left( g^{-1} X_{q} + g \sigma^z_{q - \hat x} \sigma^z_{q - \hat x - \hat y} Z_q Z_{q - \hat y} \right) \nonumber \\
& + \sum_{q_x = 2} \left( g^{-1} Z_{q}Z_{q+\hat x }Z_{q+\hat y }Z_{q + \hat x + \hat y} + g  X_{q} Z_{q - \hat x} Z_{q - \hat  x - \hat y}  \right) + \sum_{3 \leq q_x < L_x} \left( g^{-1} Z_{q}Z_{q+\hat x }Z_{q+\hat y }Z_{q + \hat x + \hat y} + g  X_q  \right)~.
\end{align}
 
\begin{align}
\label{eqn:defect_motion_in_hamiltonian_full_sweep}
& \left( \mathcal{U}_{L_x-2,L_x-1} \cdots \mathcal{U}_{\bar 0,1} \mathcal{U}_{s} \right) \tilde H_{c} \left( \mathcal{U}_{L_x-2,L_x-1} \cdots \mathcal{U}_{\bar 0,1} \mathcal{U}_{s} \right)^{\dagger} = H^{\mathcal{D}_R,0;\mathcal{D}_R,L_x-1}_{c} =
\sum_{q_x = 1} \left( g^{-1} X_{q} + g \sigma^z_{q - \hat x} \sigma^z_{q - \hat x - \hat y} Z_q Z_{q - \hat y} \right) \nonumber \\
& + \sum_{2 \leq q_x \leq L_x-1} \left( g^{-1} X_q + g  Z_{q} Z_{q - \hat y} Z_{q - \hat x} Z_{q - \hat  x - \hat y}  \right) + \sum_{q_x=0} \left( g^{-1} Z_q Z_{q+\hat y } \sigma^{x}_{q} + g  X_{q} Z_{q - \hat x} Z_{q - \hat  x - \hat y} \right).
\end{align}

\section{General Defect Configurations} \label{sec:gauging}
To illustrate the procedure of gauging the subsystem symmetries and its connection to the self-duality, we first review how this self-duality can be realized by gauging all subsystem symmetries, then projecting into an equivalent, non-trivial sector which hosts the dual degrees of freedom. Subsequently, we discuss the concept of partial gauging to generate defect configurations corresponding to performing a ``local" duality transformation within a sub-region of the lattice - which comes at the cost of a defect line on the boundaries between the duality-transformed and original regions of the lattice. 
\\
\\
Unlike the case of the $1+1$ D TFIC, these defect lines are not freely deformable as the Hilbert space dimension needed to host them depends on the perimeter of the defect line. However there still exists a sequence of unitary transformations that can be performed to map the \textit{minimally gauged} theory to a duality transformed one. Being invertible maps, they cannot ``shrink" one defect configuration into another with a smaller perimeter and we explicitly see this from the ``tears" (which contain the extra degrees of freedom) in the outer defect line after each movement step of the inner line.
\subsection{Duality via Full Gauging} \label{sec:full_gauging}
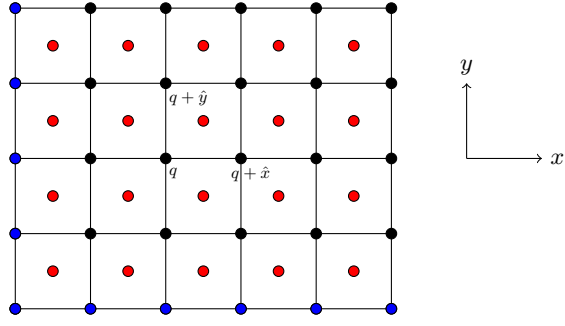
\begin{figure}[h!]
\centering
\begin{tikzpicture}

% Define colors
\definecolor{myblue}{RGB}{0,0,255}

% Draw the lattice
\foreach \i in {0,1,2,3,4,5} {
    \foreach \j in {0,1,2,3,4} {
        % Draw the nodes
        \node[circle, draw, fill=black, minimum size=4pt, inner sep=0pt] at (\i, \j) {};
        
        % Draw links (horizontal)
        \ifnum\j<4
            \draw (\i, \j) -- (\i, \j+1);
        \fi

        % Draw links (vertical)
        \ifnum\i<5
            \draw (\i, \j) -- (\i+1, \j);
        \fi
    }
}

% Color the bottommost row nodes
\foreach \i in {0,1,2,3,4,5} {
    \node[circle, draw, fill=myblue, minimum size=4pt, inner sep=0pt] at (\i, 0){};
}
\foreach \j in {0,1,2,3,4} {
    \node[circle, draw, fill=myblue, minimum size=4pt, inner sep=0pt] at (0, \j){};
}
%extra sites
\foreach \i in {1,2,3,4,5}
{ 
\foreach \j in {0,1,2,3} {
    \node[circle, draw, fill=red, minimum size=4pt, inner sep=0pt] at (\i-0.5, \j+0.5) {};
}
}

\draw node[scale=0.7] at (3.13,1.8) {$q+\hat x $};
\draw node[scale=0.7] at (2.3,2.8) {$q+\hat y $};
\draw node[scale=0.7] at (2.1,1.8) {$q$};

\draw[->] (6,2) -- ++ (1,0) node[right] {$x$};
  \draw[->] (6,2) -- ++ (0,1) node[above] {$y$};

\end{tikzpicture}
\caption{Extra sites introduced at the centers of all lattice sites. As before, identify blue sites with opposite ends when pasting into a torus. Similarly, the orange sites are identified with the corresponding red sites at opposite ends.}
\label{fig:full_gauging_xm_extra_sites}
\end{figure}

\noindent
The couplings are taken to be : 
\begin{equation}
\label{eqn:xu_moore_full_gauging_hamiltonian}
H'_{L_x L_y} = - \sum_{q} \left( g^{-1} Z_{q} Z_{q+\hat x } Z_{q+\hat y } Z_{q + \hat x + \hat y} \sigma^{x}_{q} + g X_{q} \right)
\end{equation}
With a Hilbert space $\mathcal{H} \otimes \mathcal{H}_{\text{ancilla}}$. The matter fields act non-trivially on $\mathcal{H}$ while the ancilla $\sigma$ fields act on $\mathcal{H}_{\text{ancilla}}$. 
\\
\\
By periodic boundary conditions, we have $\dim(\mathcal{H}) = \dim(\mathcal{H}_{\text{ancilla}}) = 2^{L_x L_y}$~.
What were previously subsystem symmetries corresponding to the product $\prod X$ along each horizontal and vertical loop of sites is now 'gauged' into local symmetries of the enlarged Hamiltonian $H'_{L_x L_y}$ : 
\begin{equation}
\label{eqn:gauge_symmetries}
    G_q := \sigma^{z}_{q} \sigma^z_{q - \hat x} \sigma^z_{q - \hat x - \hat y} \sigma^z_{q - \hat y} X_q
\end{equation}
Evidently, 
\begin{equation}
    \label{eqn:gauge_syms_commute_w_gauged_hamiltonian}
    [G_q , H'_{L_x L_y}] = 0  \ \ \forall \ q
\end{equation}
as $G_q$ can anticommute with either 2 or 0 Paulis in any local term of $H'_{L_x L_y}$. Thus, $G_q$ commutes with every local term in the Hamiltonian.
\\
\\
All $G_q$'s can be set \textit{independently} to $\pm1$. There are $2^{L_x L_y}$ gauge sectors corresponding to every possibility : projecting into any one sector leaves us in a $$\frac{2^{2L_x L_y}}{2^{L_x L_y}} = 2^{L_x L_y}$$ dimensional subspace, which is the size of the Hilbert space of the original problem. The effective Hamiltonian in any sector is conveniently represented by a set of $2 L_x L_y$ operators which :
\begin{enumerate}
    \item Obey the Pauli algebra.
    \item Commute with every $G_q$~. This condition ensures that each of the $G_q = \pm 1$ sectors of $\mathcal{H} \otimes \mathcal{H}_{\text{ancilla}}$ is an invariant subspace of these new Pauli operators.
\end{enumerate}
Define : 
   \begin{align}
        \tilde X_{\bar q} &= Z_{q + \hat x + \hat y} Z_{q+\hat x } Z_{q+\hat y } Z_{q} \sigma^{x}_{q}  \nonumber \\ 
        \tilde Z_{\bar q} &= \sigma^{z}_{q}  
        \label{eqn:new_paulis_full_gauging_xm}
    \end{align}

These operators satisfy both requirements. We can hence project into the $G_q = 1 ~ \forall ~ q$ sector, which gives us the ``Gauss laws" :
\begin{equation}
\label{eqn:gauss_laws_p1}
G_q = 1 =  \sigma^{z}_{q} \sigma^z_{q - \hat x} \sigma^z_{q - \hat x - \hat y} \sigma^z_{q - \hat y} X_q 
\end{equation}
From Eq.(\ref{eqn:new_paulis_full_gauging_xm}) and the Gauss Law
\begin{equation}
\label{eqn:gauss_law_p2}
X_q = \tilde Z_{q} \tilde Z_{q - \hat x} \tilde Z_{q - \hat  x - \hat y} \tilde Z_{q - \hat y}
\end{equation}
\textit{which is only true within the chosen Gauge sector, and is well defined because the RHS is a legitimate operator that acts in this sector.}
\\
From Eq.(\ref{eqn:gauss_law_p2}) and Eq.(\ref{eqn:new_paulis_full_gauging_xm}) the enlarged Hamiltonian $H'_{L_x L_y}$ projected into this subspace acts as : 
\begin{equation}
\label{eqn:full_gauged_hamitonian_xm_p1}
\Tilde{H}'_{L_x L_y} = \sum_{q} \left( g^{-1} \tilde X_{q} + g \tilde Z_{q} \tilde Z_{q - \hat x} \tilde Z_{q - \hat  x - \hat y} \tilde Z_{q - \hat y} \right)
\end{equation}
Relabeling $q' : = \bar q$ we have 
\begin{equation}
\label{eqn:full_gauged_hamitonian_xm_p2}
\Tilde{H}'_{L_x L_y} = \sum_{q'} \left( g^{-1} \tilde X_{q'} + g \tilde Z_{q'- \hat x- \hat y} \tilde Z_{q'- \hat x} \tilde Z_{q'- \hat y} \tilde Z_{q'} \right) 
\end{equation}

Which implements the duality.
\subsection{Arbritrary Defect Sectors via ``Partial Gauging"} \label{sec:partial_gauging}
Consider the situation where a \textit{rectangular subregion} of the lattice is coupled to ancillary fields. Let $P^{\sigma}_q$ denote the plaquette term coupled to the ancilla $\sigma^z$ field at $\bar q \equiv \bar q$. If the plaquette term is not coupled to an ancilla, we denote it by $B_q$. We call the rectangular region of lattice sites which make up the $P^{\sigma}$ plaquettes by : 
\begin{equation}
    \label{eqn:twisted_subregion}
    R := \{ q : 0 \leq (q-Q)_{x/y} \leq S_{x/y} \}.
\end{equation}
The complement of $R$ is denoted by $\bar R$. The boundary $\partial R$ is defined naturally to be all the sites located in the perimeter : 
\begin{equation}
    \label{eqn:bdy_of_twisted_subregion}
    \partial R := \{ q : (q-Q)_{x/y} = 0,S_{x/y} \}.
\end{equation}

\begin{figure}[h!]
\centering
\begin{tikzpicture}

% Define colors
\definecolor{myblue}{RGB}{0,0,255}

% Draw the lattice
\foreach \i in {0,1,2,3,4,5,6} {
    \foreach \j in {0,1,2,3,4,5} {
        % Draw the nodes
        \node[circle, draw, fill=black, minimum size=4pt, inner sep=0pt] at (\i, \j) {};
        
        % Draw links (horizontal)
        \ifnum\j<5
            \draw (\i, \j) -- (\i, \j+1);
        \fi

        % Draw links (vertical)
        \ifnum\i<6
            \draw (\i, \j) -- (\i+1, \j);
        \fi
    }
}

\draw[magenta,thick] (0.8+1,4.2+1) -- ++ (4.4,0) -- ++(0,-3.4) -- ++(-4.4,0) -- ++(0,3.4);
\node at (3-0.4,5.4){\textcolor{magenta}{$R$}};

% Color the bottommost row nodes
\foreach \i in {0,1,2,3,4,5,6} {
    \node[circle, draw, fill=myblue, minimum size=4pt, inner sep=0pt] at (\i, 0){};
}
\foreach \j in {0,1,2,3,4,5} {
    \node[circle, draw, fill=myblue, minimum size=4pt, inner sep=0pt] at (0, \j){};
}
%extra sites
\foreach \i in {1,2,3,4}
{ 
\foreach \j in {1,2,3} {
    \node[circle, draw, fill=red, minimum size=4pt, inner sep=0pt] at (\i+0.5+1, \j+0.5+1) {};
}
}

\draw node[scale=0.7] at (3.13-1,1.8-1) {$q+\hat x $};
\draw node[scale=0.7] at (2.3-1,2.8-1) {$q+\hat y $};
\draw node[scale=0.7] at (2.1-1,1.8-1) {$q$};

\draw[->] (7,2) -- ++ (1,0) node[right] {$x$};
  \draw[->] (7,2) -- ++ (0,1) node[above] {$y$};

\end{tikzpicture}
\caption{Partial Gauging of a rectangular sub-region '$R$' of the Xu-Moore lattice}
\label{fig:partial_gauging_xm_model}
\end{figure}
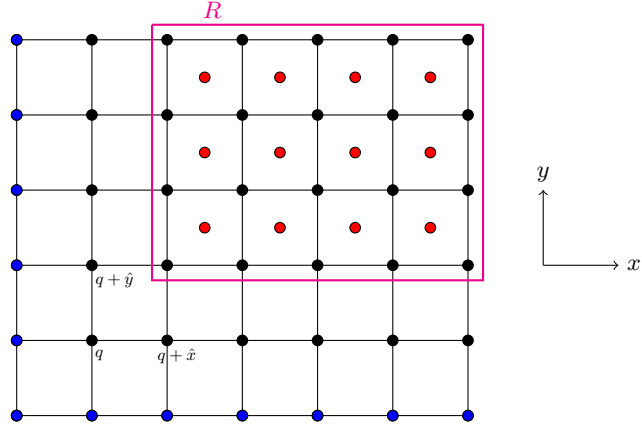

\begin{figure}[h!]
\centering
\begin{tikzpicture}

% Define colors
\definecolor{myblue}{RGB}{0,0,255}

% Draw the lattice
\foreach \i in {0,1,2,3,4,5,6} {
    \foreach \j in {0,1,2,3,4,5} {
        % Draw the nodes
        \node[circle, draw, fill=black, minimum size=4pt, inner sep=0pt] at (\i, \j) {};
        
        % Draw links (horizontal)
        \ifnum\j<5
            \draw (\i, \j) -- (\i, \j+1);
        \fi

        % Draw links (vertical)
        \ifnum\i<6
            \draw (\i, \j) -- (\i+1, \j);
        \fi
    }
}

\draw[magenta,thick] (0.8+1,4.2+1) -- ++ (4.4,0) -- ++(0,-3.4) -- ++(-4.4,0) -- ++(0,3.4);
\node at (3-0.4,5.4){\textcolor{magenta}{$R$}};

% Color the bottommost row nodes
\foreach \i in {0,1,2,3,4,5,6} {
    \node[circle, draw, fill=myblue, minimum size=4pt, inner sep=0pt] at (\i, 0){};
}
\foreach \j in {0,1,2,3,4,5} {
    \node[circle, draw, fill=myblue, minimum size=4pt, inner sep=0pt] at (0, \j){};
}

%dual lattice edges

\foreach \i in {1,2,3,4} {
    \foreach \j in {1,2,3} {
        
        % Draw links (horizontal)
        \ifnum\j<3
            \draw[red] (\i+0.5+1, \j+0.5+1) -- ++ (0,1);
        \fi

        % Draw links (vertical)
        \ifnum\i<4
            \draw[red] (\i+0.5+1, \j+0.5+1) -- ++ (1,0);
        \fi
    }
}

%arrows to indicate the structure gauge-invariant X variables

\foreach \i in {2,3,4,5}
{ 
\foreach \j in {2,3,4} {

    \ifnum \i = 2
        \ifnum \j = 2
            \draw[blue,thick,->] (\i+0.5,\j+0.5) -- ++(0.4,0.4);
        \else
            \ifnum \j=4
                \draw[blue,thick,->] (\i+0.5,\j+0.5) -- ++(0.4,-0.4);
            \else
                \draw[blue,thick,->] (\i+0.5,\j+0.5) -- ++(0.4,0.4);
                \draw[blue,thick,->] (\i+0.5,\j+0.5) -- ++(0.4,-0.4);
            \fi
        \fi
    \else
        \ifnum \i = 5
            \ifnum \j = 2
            \draw[blue,thick,->] (\i+0.5,\j+0.5) -- ++(-0.4,0.4);
            \else
                \ifnum \j=4
                    \draw[blue,thick,->] (\i+0.5,\j+0.5) -- ++(-0.4,-0.4);
                 \else
                    \draw[blue,thick,->] (\i+0.5,\j+0.5) -- ++(-0.4,0.4);
                    \draw[blue,thick,->] (\i+0.5,\j+0.5) -- ++(-0.4,-0.4);
                \fi
            \fi 
        \else
            \ifnum \j = 2
                \draw[blue,thick,->] (\i+0.5,\j+0.5) -- ++(0.4,0.4);
                \draw[blue,thick,->] (\i+0.5,\j+0.5) -- ++(-0.4,0.4);
            \else
                \ifnum \j=4
                    \draw[blue,thick,->] (\i+0.5,\j+0.5) -- ++(0.4,-0.4);
                    \draw[blue,thick,->] (\i+0.5,\j+0.5) -- ++(-0.4,-0.4);
                \else
                    \draw[blue,thick,->] (\i+0.5,\j+0.5) -- ++(0.4,0.4);
                    \draw[blue,thick,->] (\i+0.5,\j+0.5) -- ++(0.4,-0.4);
                    \draw[blue,thick,->] (\i+0.5,\j+0.5) -- ++(-0.4,0.4);
                    \draw[blue,thick,->] (\i+0.5,\j+0.5) -- ++(-0.4,-0.4);
                \fi
            \fi
        \fi
    \fi

}
}

%dual lattice (extra) sites
\foreach \i in {2,3,4,5}
{ 
\foreach \j in {2,3,4} {
    \node[circle, draw, fill=red, minimum size=4pt, inner sep=0pt] at (\i+0.5, \j+0.5) {};
}
}
\end{tikzpicture}
\caption{The blue directed arrows indicate the $Z$ matter field terms which are to be multiplied to the corresponding ancilla $\sigma^x$ to define the gauge invariant $\tilde X$ at that dual lattice site.}
\label{fig:gauge_invariant_vars}
\end{figure}
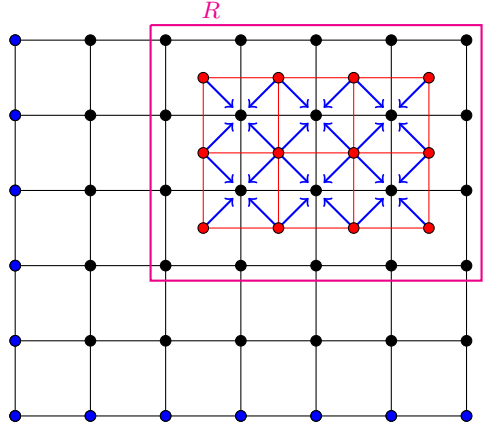

% For later convenience, we make the distinction between the top right and bottom right boundaries $\partial R_t$ and $\partial R_b$ respectively : 
% \begin{align}
%     &\label{eqn:top_bdy_of_twisted_subregion}
%     \partial R_t := \{ q : (q-Q)_{x/y} = \ S_{x/y} \} \\
%     &\label{eqn:bottom_bdy_of_twisted_subregion}
%     \partial R_b := \{ q : (q-Q)_{x/y} = \ 0 \}.
% \end{align}
While the region $R$ considered here is rectangular, the generalization to arbitrary shapes is obvious. The corresponding Hamiltonian is :
\begin{equation}
    \label{eqn:rectangular_region_hamiltonian}
    \tilde H_{S_x S_y} = \sum_{P^{\sigma}_q} \left( g^{-1} P^{\sigma}_q \sigma^{x}_{q} + g X_q \right) + \sum_{B_q} \left( g^{-1} B_q  + g X_q \right)
\end{equation}
The Hilbert space of this theory is $\mathcal{H} \otimes \mathcal{H}_{\text{ancilla}}$ where $\dim \left( \mathcal{H}_{\text{ancilla}} \right) = 2^{S_x S_y}$.
\\
\\
The non trivial gauge sectors are given by : 
\begin{equation}
    G_{q} = \left( \prod_{a,b = \pm1} \sigma^z_{q + a \frac{x}{2} + b \frac{y}{2}} \right) X_q \ \ \forall  \ q \in R \setminus \partial R
\end{equation}
Where we have emphasized the convention of placing extra sites on plaquette centers for clarity. Due to the loss of periodicity we have only $(S_x - 1)(S_y - 1)$ gauge symmetries, short of the number of ancilla degrees of freedom by $S_x + S_y - 1$. The dimension of the ``gauge reduced" Hilbert space upon projecting into the $G_q = 1$ sector is 
\begin{equation}
\label{eqn:duality_defect_dimension}
 \dim \left( P_{G_q=1} \left( \mathcal{H} \otimes \mathcal{H}_{\text{ancilla}} \right) \right) = \dim \left( \mathcal{H}_{\text{gauged}} \right) = \frac{2^{L_xL_y + S_x S_y}}{2^{(S_x-1)(S_y-1)}} = 2^{L_x L_y + S_x + S_y - 1}.
\end{equation}
We infer that the Hilbert space dimension \textit{is a function of the perimeter of a defect term one desires to host}. Hence these defects are \textbf{in general rigid and non-deformable on the finite lattice}, in contrast to what one finds from gauging the 1D TFIC as in \cite{seiberg-LSM}.
\\
\\
For all $0 \leq q_{x/y} - Q_{x/y} \leq S_{x/y}$, we choose new Pauli operators that are well defined in the $G_q$ gauge sectors (i.e. commute with $G_q \ \forall \ q$) : 
\begin{align}
    &\tilde Z_{q} = \sigma^{z}_{q} \\
    & \tilde X_{q} = \sigma^{x}_{q} \left( \prod_{\langle q', \bar q \rangle , \ q ' \in R \setminus \partial R} Z_{q'} \right)
\end{align}
where $\langle \ , \ \rangle$ indicates nearest neighbors. The gauge invariant $\tilde X$ variables are defined by dressing the dual $\sigma^x$ fields with the appropriate $Z_q$ terms for $q \in R \setminus \partial R$ such that they commute with every gauge sector $G_{q'}$. For example, if $q - Q = (S_x - 1 , S_y - 1)$ we have 
$\tilde X_{q} = \sigma^{x}_{q} Z_q $ and if $q - Q = (S_x - 2 , S_y - 1)$ we have $\tilde X_{q} = \sigma^{x}_{q} Z_q Z_{q+\hat x } $ (see Fig.(\ref{fig:gauge_invariant_vars})). 

Imposing the gauge constraint $G_q = 1$ to project into $\mathcal{H}_{\text{gauged}}$ : 
\begin{align}
    \label{eqn:gauged_rectangular_region_hamiltonian}
    & P_{G_q=1} \left( H^{\text{twisted}}_{S_x S_y} \right) = H^{\text{gauged}}_{S_x S_y} = \sum_{q \ : \ P^{\sigma}_q \subset R \setminus \partial R} g^{-1} \tilde X_{\bar q} + \sum_{q \ : \ P^{\sigma}_q \cap \partial R \neq \emptyset} \textcolor{orange}{ g^{-1} \tilde X_{\bar q} \left( \prod_{\langle q', \bar q \rangle \ ; \ q' \in \partial R} Z_{q'} \right)} \\
    & + \sum_{q \in R \setminus \partial R}  g \left( \prod_{a,b = 0,1} \tilde Z_{\bar q - ax - by} \right) + \sum_{B_q} \left( g^{-1} B_q  + g X_q \right) \nonumber
\end{align}

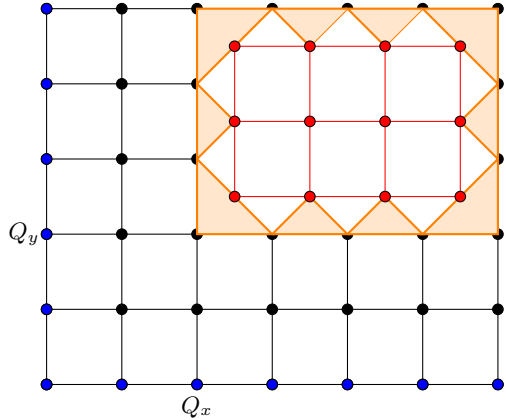
\begin{figure}[h!]
\centering
\begin{tikzpicture}

% Define colors
\definecolor{myblue}{RGB}{0,0,255}

% Draw the lattice
\foreach \i in {0,1,2,3,4,5,6} {
    \foreach \j in {0,1,2,3,4,5} {
        % Draw the nodes
        \node[circle, draw, fill=black, minimum size=4pt, inner sep=0pt] at (\i, \j) {};
        
        % Draw links (horizontal)
        \ifnum\j<5
            \draw (\i, \j) -- (\i, \j+1);
        \fi

        % Draw links (vertical)
        \ifnum\i<6
            \draw (\i, \j) -- (\i+1, \j);
        \fi
    }
}

\filldraw[fill=white, draw=white] (2,2) rectangle (6,5);

% Color the bottommost row nodes
\foreach \i in {0,1,2,3,4,5,6} {
    \node[circle, draw, fill=myblue, minimum size=4pt, inner sep=0pt] at (\i, 0){};
}
\foreach \j in {0,1,2,3,4,5} {
    \node[circle, draw, fill=myblue, minimum size=4pt, inner sep=0pt] at (0, \j){};
}

%dual lattice edges

\foreach \i in {1,2,3,4} {
    \foreach \j in {1,2,3} {
        
        % Draw links (horizontal)
        \ifnum\j<3
            \draw[red] (\i+0.5+1, \j+0.5+1) -- ++ (0,1);
        \fi

        % Draw links (vertical)
        \ifnum\i<4
            \draw[red] (\i+0.5+1, \j+0.5+1) -- ++ (1,0);
        \fi
    }
}

%arrows to indicate the structure gauge-invariant X variables

\foreach \i in {2,3,4,5}
{ 
\foreach \j in {2,3,4} 
{
    \ifnum \i = 2
        \ifnum \j = 2
            \filldraw[fill = orange!20, draw = orange,thick] (\i+0.5,\j+0.5) -- ++(-0.5,0.5) -- ++(0,-1) -- ++(1,0) -- ++(-0.5,0.5);
        \else
            \ifnum \j=4
                \filldraw[fill = orange!20, draw = orange,thick] (\i+0.5,\j+0.5) -- ++(0.5,0.5) -- ++(-1,0) -- ++(0,-1) -- ++(0.5,0.5);
            \else
                \filldraw[fill = orange!20, draw = orange,thick] (\i+0.5,\j+0.5) -- ++(-0.5,0.5) -- ++(0,-1) -- ++(0.5,0.5);
            \fi
        \fi
    \else
        \ifnum \i = 5
            \ifnum \j = 2
            \filldraw[fill = orange!20, draw = orange,thick] (\i+0.5,\j+0.5) -- ++(0.5,0.5) -- ++(0,-1) -- ++(-1,0) -- ++(0.5,0.5);
        \else
            \ifnum \j=4
                \filldraw[fill = orange!20, draw = orange,thick] (\i+0.5,\j+0.5) -- ++(-0.5,0.5) -- ++(1,0) -- ++(0,-1) -- ++(-0.5,0.5);
            \else
                \filldraw[fill = orange!20, draw = orange,thick] (\i+0.5,\j+0.5) -- ++(0.5,0.5) -- ++(0,-1) -- ++(-0.5,0.5);
            \fi
        \fi
        \else
            \ifnum \j = 2
                \filldraw[fill = orange!20, draw = orange,thick] (\i+0.5,\j+0.5) -- ++(0.5,-0.5) -- ++ (-1,0);
                \filldraw[fill = orange!20, draw = orange,thick] (\i+0.5,\j+0.5) -- ++(-0.5,-0.5);
            \else
                \ifnum \j = 4
                    \filldraw[fill = orange!20, draw = orange,thick] (\i+0.5,\j+0.5) -- ++(0.5,0.5);
                    \filldraw[fill = orange!20, draw = orange,thick] (\i+0.5,\j+0.5) -- ++(-0.5,0.5) -- ++(1,0);
                \fi
            \fi
        \fi
    \fi
}
}
%dual lattice (extra) sites
\foreach \i in {2,3,4,5}
{ 
\foreach \j in {2,3,4} {
    \node[circle, draw, fill=red, minimum size=4pt, inner sep=0pt] at (\i+0.5, \j+0.5) {};
}
}

\node at (2,-0.3) {$Q_x$};
\node at (-0.3,2) {$Q_y$};

\end{tikzpicture}
\caption{The partially gauged Hamiltonian projected into the sector with the duality defect line. The colors correspond to their respective terms in Eq.(\ref{eqn:gauged_rectangular_region_hamiltonian_SHIFTED_ORIGIN}).}
    \label{fig:gauged_hamiltonian_xm_torus}
\end{figure}

Making the choice of origin and the rectangular shape explicit as in Fig.(\ref{fig:gauged_hamiltonian_xm_torus}), Eq.(\ref{eqn:gauged_rectangular_region_hamiltonian}) takes the following form (see Fig.(\ref{fig:gauged_hamiltonian_xm_torus}) to match the corresponding color-coded terms) :
\begin{align}
    \label{eqn:gauged_rectangular_region_hamiltonian_SHIFTED_ORIGIN}
    & H^{\text{gauged}}_{S_x S_y} = \sum_{0 \leq q_{x} < Q_x \ \textbf{or} \ 0 \leq q_{y} < Q_y} \left( g^{-1} Z_q Z_{q+\hat x } Z_{q+\hat y } Z_{q + \hat x + \hat y}  + g X_q \right) + \nonumber \\
    & \textcolor{red}{\sum_{L_{x/y}-1 > q_{x/y} \geq Q_{x/y}}  \left( g \tilde Z_{\bar q - \hat  x - \hat y} \tilde Z_{\bar q - \hat x} \tilde Z_{\bar q - \hat y} \tilde Z_{\bar q}  + g^{-1} \tilde X_{\bar q - \hat  x - \hat y} \right)} 
    \\
    & + \textcolor{red}{\sum_{q_{x/y} > Q_{x/y}
    \ ; \ q_x = L_x-1 \ \textbf{or} \ q_y = L_y-1} g^{-1} \tilde X_{\bar q}} + \textcolor{orange}{\Delta_{\text{c}} H_{(Q_x,Q_y)}} \nonumber
\end{align}
where 
\begin{equation}
    \label{eqn:delta_H_QxQy}
    \Delta_{\text{c}} H_{Q_x , Q_y} = \sum_{q_{x/y} = Q_{x/y} \ , \ Q_{y/x} \leq q_{y/x} \leq L_{y/x}-1} g^{-1} \tilde X_{\bar q} \left( \prod_{\langle q', \bar q \rangle \ ; \ q' \in \partial R} Z_{q'} \right)
\end{equation}
is the defect line. Contrasting this with Eq.(\ref{eq.D-defect}), the defect line here is closed (hence the subscript in $\Delta_{c}$) because it is supported on the boundary of a finite region which is subjected to partial gauging and gauge-reduction. An open defect line $\Delta H_{Q_x, Q_y}$ can be thought of as just the bottom left corner of the closed defect line $\Delta_c H_{Q_x, Q_y}$, which are equivalent in the thermodynamic limit $L_{x/y} \to \infty$ :
\begin{align}
    \label{eqn:delta_H_QxQy_open}
    & \Delta H_{Q_x , Q_y} = \sum_{q_{x/y} = Q_{x/y} \ , \ Q_{y/x} \leq q_{y/x}} g^{-1} \tilde X_{\bar q} \left( \prod_{\langle q', \bar q \rangle \ ; \ q' \in \partial R} Z_{q'} \right) \\
& =g^{-1} \tilde X_{\bar Q_x, \bar Q_y} Z_{Q_x,Q_y}Z_{Q_x+1,Q_y}Z_{Q_x,Q_y+1} 
+ \sum_{j > Q_x} \tilde X_{\bar j, \bar Q_y} Z_{j, Q_y} Z_{j+1, Q_y} +  \sum_{j > Q_y} \tilde X_{\bar Q_x, \bar j + 1} Z_{Q_x, j} Z_{Q_x, j+1} \nonumber
\end{align}
\subsection{Fusion Rule Nuances and Mobility Problems}
On the torus, we consider the situation with a defect on lines $x=L_x$,$y=L_y$, along the boundary of the model. By similar arguments to the infinite cylinder, we see that the pure single fracton defect Hamiltonian. 

\begin{equation}  
    H^{f;q_f}_{t} =  \sum_{q \neq q_f} \left( (-1)^{\delta(q,q_f)} g^{-1}Z_{q} Z_{q+\hat x } Z_{q+\hat y } Z_{q + \hat x + \hat y} + g X_q \right)
\end{equation}has no unitary connection to a bare Xu-Moore Hamiltonian $H_{\sf XM,torus}^{f}$.  The fusion operator of a single fracton with a duality defect is

\begin{eqnarray}
    \lambda({\mathcal{D}_{R;(L_x,L_y)},f;(i,j))} = (\prod_{k> i,}\prod_{l>  j} X_{k,l} )\sigma^z_{(\bar i, 1/2)}.
\end{eqnarray} 

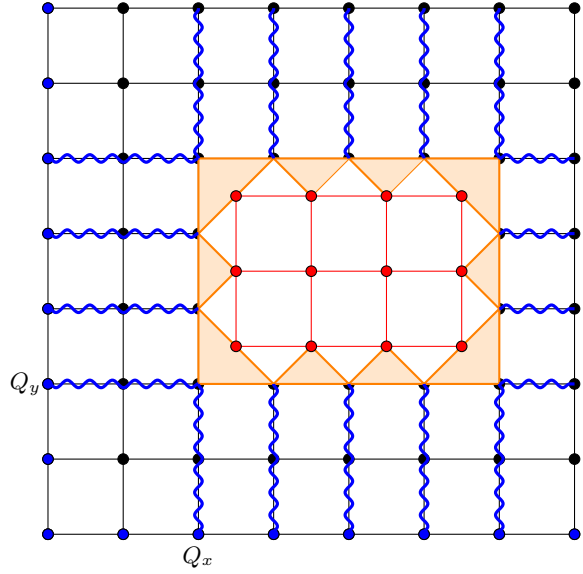
\begin{figure}[h!]
\centering
\begin{tikzpicture}

% Define colors
\definecolor{myblue}{RGB}{0,0,255}

% --- Base lattice: 8x8 (0..7 in both x and y) ---
\foreach \i in {0,...,7} {
    \foreach \j in {0,...,7} {
        % Draw the nodes
        \node[circle, draw, fill=black, minimum size=4pt, inner sep=0pt] at (\i, \j) {};
        
        % Vertical links
        \ifnum\j<7
            \draw (\i, \j) -- (\i, \j+1);
        \fi

        % Horizontal links
        \ifnum\i<7
            \draw (\i, \j) -- (\i+1, \j);
        \fi
    }
}

% White rectangle mask (still in the same place)
\filldraw[fill=white, draw=white] (2,2) rectangle (6,5);

% Color the bottommost row nodes (blue)
\foreach \i in {0,...,7} {
    \node[circle, draw, fill=myblue, minimum size=4pt, inner sep=0pt] at (\i, 0){};
}
% Color the leftmost column nodes (blue)
\foreach \j in {0,...,7} {
    \node[circle, draw, fill=myblue, minimum size=4pt, inner sep=0pt] at (0, \j){};
}

% --- Dual lattice edges (red) ---
\foreach \i in {1,2,3,4} {
    \foreach \j in {1,2,3} {
        % Vertical links
        \ifnum\j<3
            \draw[red] (\i+0.5+1, \j+0.5+1) -- ++ (0,1);
        \fi
        % Horizontal links
        \ifnum\i<4
            \draw[red] (\i+0.5+1, \j+0.5+1) -- ++ (1,0);
        \fi
    }
}

% --- Orange decorations ---
\foreach \i in {2,3,4,5}
{ 
\foreach \j in {2,3,4} 
{
    \ifnum \i = 2
        \ifnum \j = 2
            \filldraw[fill = orange!20, draw = orange,thick] (\i+0.5,\j+0.5) -- ++(-0.5,0.5) -- ++(0,-1) -- ++(1,0) -- ++(-0.5,0.5);
        \else
            \ifnum \j=4
                \filldraw[fill = orange!20, draw = orange,thick] (\i+0.5,\j+0.5) -- ++(0.5,0.5) -- ++(-1,0) -- ++(0,-1) -- ++(0.5,0.5);
            \else
                \filldraw[fill = orange!20, draw = orange,thick] (\i+0.5,\j+0.5) -- ++(-0.5,0.5) -- ++(0,-1) -- ++(0.5,0.5);
            \fi
        \fi
    \else
        \ifnum \i = 5
            \ifnum \j = 2
                \filldraw[fill = orange!20, draw = orange,thick] (\i+0.5,\j+0.5) -- ++(0.5,0.5) -- ++(0,-1) -- ++(-1,0) -- ++(0.5,0.5);
            \else
                \ifnum \j=4
                    \filldraw[fill = orange!20, draw = orange,thick] (\i+0.5,\j+0.5) -- ++(-0.5,0.5) -- ++(1,0) -- ++(0,-1) -- ++(-0.5,0.5);
                \else
                    \filldraw[fill = orange!20, draw = orange,thick] (\i+0.5,\j+0.5) -- ++(0.5,0.5) -- ++(0,-1) -- ++(-0.5,0.5);
                \fi
            \fi
        \else
            \ifnum \j = 2
                \filldraw[fill = orange!20, draw = orange,thick] (\i+0.5,\j+0.5) -- ++(0.5,-0.5) -- ++ (-1,0);
                \filldraw[fill = orange!20, draw = orange,thick] (\i+0.5,\j+0.5) -- ++(-0.5,-0.5);
            \else
                \ifnum \j = 4
                    \filldraw[fill = orange!20, draw = orange,thick] (\i+0.5,\j+0.5) -- ++(0.5,0.5);
                    \filldraw[fill = orange!20, draw = orange,thick] (\i+0.5,\j+0.5) -- ++(-0.5,0.5) -- ++(1,0);
                \fi
            \fi
        \fi
    \fi
}
}

% --- Wiggly blue lines from boundary of a rectangle ---
\def\xmin{2}
\def\ymin{2}
\def\xmax{6}
\def\ymax{5}
\def\Lsize{7} % lattice goes from 0..7

% Loop over boundary sites of the rectangle
\foreach \i in {\xmin,...,\xmax} {
  \foreach \j in {\ymin,...,\ymax} {
    % Only boundary sites
    \ifnum\i=\xmin
      % left edge
      \draw[myblue, very thick, decorate, decoration={snake, amplitude=0.5mm, segment length=3mm}]
        (\i,\j) -- (0,\j);
    \fi
    \ifnum\i=\xmax
      % right edge
      \draw[myblue, very thick, decorate, decoration={snake, amplitude=0.5mm, segment length=3mm}]
        (\i,\j) -- (\Lsize,\j);
    \fi
    \ifnum\j=\ymin
      % bottom edge
      \draw[myblue, very thick, decorate, decoration={snake, amplitude=0.5mm, segment length=3mm}]
        (\i,\j) -- (\i,0);
    \fi
    \ifnum\j=\ymax
      % top edge
      \draw[myblue, very thick, decorate, decoration={snake, amplitude=0.5mm, segment length=3mm}]
        (\i,\j) -- (\i,\Lsize);
    \fi
  }
}

% Dual lattice (extra) sites
\foreach \i in {2,3,4,5}
{ 
\foreach \j in {2,3,4} {
    \node[circle, draw, fill=red, minimum size=4pt, inner sep=0pt] at (\i+0.5, \j+0.5) {};
}
}

\node at (2,-0.3) {$Q_x$};
\node at (-0.3,2) {$Q_y$};

\end{tikzpicture}
\caption{A rectangular duality defect and the corresponding subset of $\eta$ operators which have expressible unitary fusion operators with the duality defect.}
\label{fig:defect-subalgebra-fusion}
\end{figure}

This represents an exchange of the fracton and \textit{two} local field rotation defects $\tilde{Z}$. By equivalent argument, this reproduces the defect fusion rule of $\cal{D}$ with $\eta$ defects. An essential subtlety arises from the constrained mobility of these fractons and $\eta$'s. As depicted in Fig. \ref{fig:defect-subalgebra-fusion}, only $\eta$ defects whose associated defect lines can be rigidly extended to the defect surface can fuse with $\cal{D}$. 

In Fig. \ref{fig:defect-subalgebra-fusion} we also show the symmetry which arises from the presence of the duality defect. It invokes a product of each vertical and horizontal $\eta$ which makes contact with the duality domain wall. As such this defect may only ``fuse" with the $\eta$ defects pictured. In general, for such $\eta$'s as the ones pictured, we can write an exact fusion operator 

\begin{eqnarray}\label{eqn:EtaDualityFusionOps}
\lambda(\eta^{\text{row}}_{i,j}, \mathcal{D}_Q) = \big(\prod_{k=i+1/2}^{Q_x} X_{k,j} \big)\sigma_{Q_x,k}\sigma^{z}_{Q_x,k-1}\\\nonumber
\lambda(\eta^{\text{col}}_{i,j}, \mathcal{D}_Q)= \big(\prod_{k=j+1/2}^{Q_x} X_{k,j} \big)\sigma^z_{i-1,Q_x}\sigma^{z}_{i,Q_x}.
\end{eqnarray}

The action of this fusion operator on the defect Hamiltonian of both an $\eta$ defect and a $\cal{D}$ defect is

\begin{eqnarray}
   \lambda(\eta^{\text{row}}_{\bar{i},j}, \mathcal{D}_Q)H_{(\bar{i},j) ,Q}^{row,(\cal{D})}\lambda(\eta^{\text{row}}_{\bar{i},j}, \mathcal{D}_Q)=H^{(D)}_Q.
\end{eqnarray}
\newpage
\subsection{Infinite Cylinder Limit} \label{sec:defects_infinte_cyl}

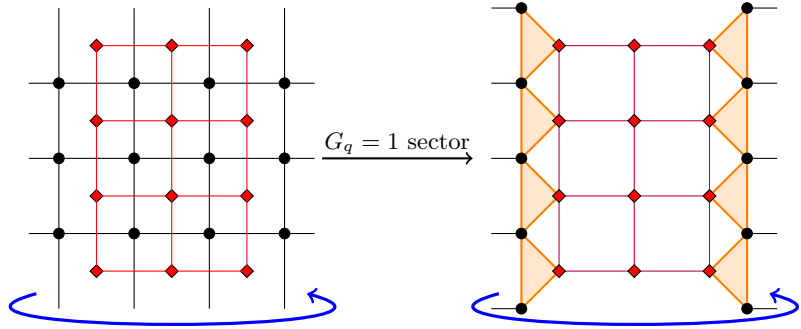
\begin{figure}[h!]
    \centering
    \begin{tikzpicture}
        \def\cols{3}
        \def\rows{2}
        % Add vertex bulges (small circles)
        \foreach \x in {0,...,\cols} 
        {
            \foreach \y in {0,...,\rows} 
            {
                \node[draw=black, fill=black, circle, minimum size=4pt, inner sep=0pt] at (\x,\y){ };
            }
        }
        %Add extra sites
         \foreach \x in {0,...,2} {
            \foreach \y in {-1,...,\rows} {
                \node[draw=black, fill=red, diamond, minimum size=5pt, inner sep=0pt] 
                    at (\x+0.5,\y+0.5){ };
            }
        }

        \foreach \x in {0,...,\cols} 
        {
            \foreach \y in {-1,...,\rows} 
            {
                \draw (\x,\y) -- (\x,\y+1);
                
                \ifnum \x<\cols
                    \ifnum \y > -1
                    \draw (\x,\y) -- (\x+1,\y);
                    \fi
                \fi
            }
        }

        \draw[blue,->,very thick] (-0.3,-0.8) to[out=195, in=345, looseness=1.5] (0.3+\cols,-0.8);

       \foreach \x in {0,...,2}
       {
            \foreach \y in {-1,...,2}
            {
                \ifnum \x < 2
                \draw[red] (\x+0.5,\y+0.5) -- ++ (1,0);
                \fi
                
                \ifnum \y < 2 
                \draw[red] (\x+0.5,\y+0.5) -- ++ (0,1);
                \fi
            }
       }
       \foreach \y in {0,...,2}
            {
                \draw (0,\y) -- ++(-0.4,0);
                \draw (3,\y) -- ++(0.4,0);
            }
        \draw[->,black,thick] (3.5,1)--(5.5,1);
        \node at (4.5,1.2) {$G_q=1$ sector};
\end{tikzpicture}
\hspace{-2cm}
\begin{tikzpicture}
        \def\cols{3}
        \def\rows{2}
       
       \foreach \x in {0,...,2}
       {
            \foreach \y in {-1,...,2}
            {
                \ifnum \x < 2
                \draw[purple] (\x+0.5,\y+0.5) -- ++ (1,0);
                \fi
                
                \ifnum \y < 2 
                \draw[purple] (\x+0.5,\y+0.5) -- ++ (0,1);
                \fi

                \ifnum \x=0
                   \filldraw[fill = orange!20, draw = orange,thick] (\x+0.5,\y+0.5) -- ++ (-0.5,0.5) -- ++(0,-1) -- ++(0.5,0.5);
                \fi

                \ifnum \x=2
                    \filldraw[fill = orange!20, draw = orange,thick] (\x+0.5,\y+0.5) -- ++ (0.5,0.5) -- ++(0,-1) -- ++(-0.5,0.5);
                \fi
            }
       }
        %Add extra sites
            \foreach \x in {0,...,2} 
            {
                \foreach \y in {-1,...,\rows} 
                {
                    \node[draw=black, fill=red, diamond, minimum size=5pt, inner sep=0pt] 
                    at (\x+0.5,\y+0.5){ };
                }
            }

            \foreach \y in {-1,...,3}
            {
                \node[draw=black, fill=black, circle, minimum size=4pt, inner sep=0pt] at (0,\y) { };
                \draw (0,\y) -- ++(-0.4,0);
                \node[draw=black, fill=black, circle, minimum size=4pt, inner sep=0pt] at (3,\y) { };
                \draw (3,\y) -- ++(0.4,0);
            }
            \draw[blue,->,very thick] (-0.3,-0.8) to[out=195, in=345, looseness=1.5] (0.3+\cols,-0.8);

\end{tikzpicture}
    \caption{Partial Gauging to unveil a line-defect structure}
    \label{fig:xm_cylinder_partial_gauging_defects}
\end{figure}

We now consider the situation where $H_{c}$ has plaquettes coupled to ancillas in the sub-region $R$ defined by : 
\begin{equation}
   \{ q : \ 0\leq q_x < S_x \ ; -\infty< q_y <\infty \}
\end{equation}
The gauging procedure can be performed in the bulk of this region, but the boundaries ($q_x = 0,S_x-1$) require a careful choice of gauge-invariant variables due to a loss of horizontal translational symmetry. This choice unveils the structure of the defect lines (Fig.(\ref{fig:xm_cylinder_partial_gauging_defects})).

The Hamiltonian under consideration : 
\begin{align}
\label{eqn:cylinder_hamiltonian_xm_full_gauged}
& H^{S_x}_{\text{partial gauged}, c} = \sum_{0 \leq q_x < S_x  \ ; \ -\infty < q_y < \infty} \left( g^{-1} Z_{q} Z_{q+\hat x } Z_{q+\hat y } Z_{q + \hat x + \hat y} \sigma^x_{\bar q } + g X_q \right) \\
& + \sum_{S_x \leq q_x \leq L_x-1  \ ; \ -\infty < q_y < \infty} \left( g^{-1} Z_{q} Z_{q+\hat x } Z_{q+\hat y } Z_{q + \hat x + \hat y} + g X_q \right) \nonumber
\end{align}

The system is partially gauged, possessing the following local symmetries: 
\begin{equation}
\label{eqn:xm_cylinder_partial_gauge_sectors}
G_q = \sigma^{z}_{q} \sigma^z_{q - \hat x} \sigma^z_{q - \hat x - \hat y} \sigma^z_{q - \hat y} X_q \ \ \text{for} \ \ 1 \leq q_x \leq S_x-1 \ , \ -\infty < q_y < \infty
\end{equation}

Focusing our attention to one column, we see that there are $S_x$ ancilla sites but only $S_x - 1$ gauge degrees of freedom. Projecting into each of the $G_q=1$ sectors still leaves the Hilbert space dimension increased by one column of ancilla sites. I.e.
\begin{equation}
\label{eqn:ancilla_site_number_calc_xm_cyl_partial_gauging}
 \mathcal{H}^{\text{partial gauged}}_{c} \xrightarrow{G_q = 1} \mathcal{H}_{c} \bigotimes_{-\infty < q_y < \infty} \mathbb{C}^2 
\end{equation}

We now choose the $G_q$ Gauge Invariant matter fields $\tilde X , \tilde Z$ : 
\begin{align}
    \label{eqn:gauge_invariant_matter_fields_xm_cyl_p.g}
    & \tilde X_q = Z_{q} Z_{q+\hat x } Z_{q+\hat y } Z_{q + \hat x + \hat y} \sigma^x_{\bar q } \ ; \ \tilde Z_q = \sigma^z_{\bar q } & 1\leq q \leq S_x-1 \\
    & \tilde X_q = Z_{q+\hat x } Z_{q + \hat x + \hat y} \sigma^x_{\bar q } \ ; \ \tilde Z_q = \sigma^z_{\bar q } & q=0 \\
    & \tilde X_q = Z_{q} Z_{q+\hat y } \sigma^x_{\bar q } \ ; \ \tilde Z_q = \sigma^z_{\bar q } & q=S_x 
\end{align}

Enforcing the Gauss Laws by projecting into $G_q = 1$ gives us Eq.(\ref{eqn:gauss_law_p2}) for $1 \leq q_x \leq S_x-1$, and the effective Hamiltonian (Fig.(\ref{fig:xm_cylinder_partial_gauging_defects})) is therefore given by ($-\infty < q_y <\infty $ in all summations) : 
\begin{align}
    \label{eqn:defect_hamiltonian_xm_cyl_gauss_laws_LEFT_RIGHT}
    &H^{\mathcal{D}_R,0; \mathcal{D}_L,S_x}_{c} =\sum_{q_x=0 \ ; \ q_y} \left( g^{-1} Z_q Z_{q+\hat y } \tilde X_{\bar q} + g X_q \right) +   \sum_{1 \leq q_x < S_x-1 \ ; \ q_y} \left( g^{-1} \tilde X_{\bar q} + g \tilde Z_{\bar q - \hat  x - \hat y} \tilde Z_{\bar q - \hat x} \tilde Z_{\bar q - \hat y} \tilde Z_{\bar q} \right) \nonumber \\
    & + \sum_{q_x=S_x-1 \ ; \ q_y} \left( g^{-1} \tilde X_{q+\frac{x}{2}+ \frac{y}{2}} Z_{q + \hat x + \hat y} Z_{q+\hat x } + g \tilde Z_{\bar q - \hat  x - \hat y} \tilde Z_{\bar q - \hat x} \tilde Z_{\bar q - \hat y} \tilde Z_{\bar q} \right) \\
    & +  \sum_{S_x \leq q_x < L_x \ ; \ q_y} \left( g^{-1} Z_q Z_{q+\hat y } Z_{q+\hat x } Z_{q + \hat x + \hat y} + g X_{q} \right) \nonumber
\end{align}
The defects are drawn as triangular in Fig.(\ref{fig:xm_cylinder_partial_gauging_defects}) with one of the vertices belonging to the dual lattice ($\tilde X$) and the remaining vertices being a part of the original lattice ($Z$). We denote a Hamiltonian possessing a left(right) pointing defect by the subscript $\mathcal{D}_{L(R)}$. The index $q_x$ of the $Z$ terms is used to label the location of the defect.
\\
\\
Thus, the $\mathcal{D}_{R}$ and $\mathcal{D}_{L}$ superscripts on $H_{XM,cyl}$ in Eq.(\ref{eqn:defect_hamiltonian_xm_cyl_gauss_laws_LEFT_RIGHT}) indicate the presence of defect terms of the respective forms (upto overall factors) : 
\begin{equation}
\label{eqn:defect_terms_xm_cyl_LEFT_RIGHT}
\mathcal{D}_R : \sum_{-\infty < q_y <\infty}  X_{q} Z_{q - \hat x} Z_{q - \hat  x - \hat y} \ \ \ \text{or} \ \ \ \mathcal{D}_L : \sum_{-\infty < q_y <\infty}  X_{q} Z_{q+\hat x } Z_{q+\hat x -y}
\end{equation}

To study the mobility of defects, it is better to fix a uniform direction of the defect. We do this by acting on $H^{\mathcal{D}_R,0; \mathcal{D}_L,S_x}_{XM,cyl}$ with a Unitary to ``flip" one of these around, say $\mathcal{D}_L \to \mathcal{D}_R$

\begin{equation}
    \label{eqn:flipping_DL_to_DR}
    \mathcal{U}^{L \to R}_{\text{flip}} = \left( \prod_{q_x = S_x-1 ,; q_y} C^{\sf z}_{q+\frac{x}{2}+ \frac{y}{2} , q+\hat x } C^{\sf z}_{q+\frac{x}{2}+ \frac{y}{2} , q+\hat x  + y} \right)
\end{equation}

Thus,
\begin{align}
\label{eqn:defect_hamiltonian_xm_cyl_gauss_laws_RIGHT}
  & \mathcal{U}^{L \to R}_{\text{flip}} H^{\mathcal{D}_R,0; \mathcal{D}_L,S_x}_{c} {(\mathcal{U}^{L \to R}_{\text{flip}} )}^{-1} \nonumber \\
  & =\sum_{q_x=0 \ ; \ q_y} \left( \textcolor{orange}{g^{-1} Z_q Z_{q+\hat y } \tilde X_{\bar q}} + g X_q \right) +   \sum_{1 \leq q_x \leq S_x-1 \ ; \ q_y} \left( g^{-1} \tilde X_{\bar q} + g \tilde Z_{\bar q - \hat  x - \hat y} \tilde Z_{\bar q - \hat x} \tilde Z_{\bar q - \hat y} \tilde Z_{\bar q} \right) \nonumber \\
    & + \sum_{q_x = S_x \ ; \ q_y} \left( g^{-1} Z_q Z_{q+\hat y } Z_{q+\hat x } Z_{q + \hat x + \hat y} + \textcolor{orange}{g \tilde Z_{\bar q - \hat x} \tilde Z_{\bar q - \hat  x - \hat y} X_{q}} \right) \\
    & +  \sum_{S_x < q_x < L_x \ ; \ q_y} \left( g^{-1} Z_q Z_{q+\hat y } Z_{q+\hat x } Z_{q + \hat x + \hat y} + g X_{q} \right) \nonumber \\
    & = H^{\mathcal{D}_R,0; \mathcal{D}_R,S_x - \frac{1}{2}}_{XM,cyl} \nonumber
\end{align}

\end{document}